\documentclass[11pt,a4paper]{article}
\usepackage{lmodern}
\pdfoutput=1
\usepackage{jheppub}
\usepackage{rotating}
\usepackage{array}
\usepackage{amsmath}
\usepackage[normalem]{ulem}
\usepackage{slashed}
\usepackage{booktabs}
\usepackage[pdftex,table]{xcolor}
\usepackage{enumerate}
\usepackage{parskip} 
\usepackage{upgreek}
\usepackage{layouts}
\usepackage{appendix}
\usepackage{siunitx}
\usepackage{paralist}
\usepackage{multirow}
\usepackage{xspace}
\usepackage{tikz}
\usepackage{hyperref}
\usepackage[compat=1.1.0]{tikz-feynman}
\usepackage{mathrsfs}
\usepackage{pgfplots}
\usepackage{fontawesome5}
\usepackage{silence}
\usepackage{footmisc}
\pgfplotsset{compat=1.15}
\usetikzlibrary{arrows, decorations.pathmorphing}
\tikzset{snake it/.style={decorate, decoration=snake}}
\tikzfeynmanset{warn luatex=false}
\DeclareSIUnit\barn{b}

\newcommand{\pyth}{\textsc{Pythia}\xspace}
\newcommand{\alpi}{\textsc{ALPiNIST}\xspace}

\newcounter{rowcntr}[table]
\renewcommand{\therowcntr}{\roman{rowcntr}}
\newcolumntype{N}{>{(\refstepcounter{rowcntr}\therowcntr)}r}
\preprint{MPP-2024-129}

\title{On the impact of heavy meson production spectra on searches for heavy neutral leptons}

\author[1,2]{Jonathan L. Schubert}
\author[2]{Babette D{\"obrich}}
\author[2]{Jan Jerhot}
\author[3]{Tommaso Spadaro}

\affiliation[1]{Technical University of Munich, TUM School of Natural Sciences, Physics Department, Chair for Data Science in Physics, 85748 Garching, Germany}
\affiliation[2]{Max-Planck-Institut für Physik (Werner-Heisenberg-Institut), Boltzmannstr. 8, 85748 Garching bei München, Germany}
\affiliation[3]{Laboratori Nazionali di Frascati dell’INFN, Via E. Fermi, 54, 00044 Frascati Italy}

\emailAdd{babette@mpp.mpg.de}
\emailAdd{jan.jerhot@cern.ch}
\emailAdd{jonathan.schubert@cern.ch}
\emailAdd{tommaso.spadaro@cern.ch}

\abstract{Feebly Interacting Particles are a commonly considered extension to the Standard Model of Particle Physics. 
In many theoretical frameworks these particles can explain observed physical phenomena which are in tension with the current model. 
\textsc{ALPiNIST} is a simplified Monte Carlo framework aimed at evaluating past, present, and future, short and long baseline experiments for their sensitivities to different models of Axion-Like Particles. 
We present the extension of this framework to accommodate new classes of Feebly Interacting Particles with emphasis on Heavy Neutral Leptons. 
This extension is especially well motivated, solving multiple of the standing issues with the Standard Model at the same time.
The fundamental importance of inputs on the resulting parameter sensitivity, and thus the need for a unified simulation set-up, is highlighted.}


\begin{document}

\maketitle
\flushbottom

\section{Introduction} 
\label{sec:introduction}

    %about fips 
    Unlike their equivalent for Beyond Standard Model extensions at very large mass scales, which are accessible in the lab only at colliders, searches for Feebly Interacting Particles (FIPs, typically below the GeV scale) do not necessarily rely on high interaction energies~\cite{Beacham:2019nyx,Antel:2023hkf}.
    Instead, due to the suppressed nature of their Standard Model (SM) couplings, high intensities are required to facilitate such searches.
    Therefore, the FIP paradigm can be studied in a multitude of ways, including nuclear recoil for FIPs of cosmological or solar origin~\cite{babyiaxo,Dutta:2024kuj}, (far) displaced detection at particle colliders~\cite{FASER:2018eoc, Gorkavenko:2023nbk, MATHUSLA:2018bqv}, and fixed-target production.\footnote{By referring to fixed-target experiments in this work, we refer to beam-dump~\cite{SHiP2023, SHADOWS2023, HIKE2023, Giffin:2022rei}, neutrino physics~\cite{CHARM:1980yym, CCFRNuTeV:1996vbm, DUNE:2020fgq} and light meson decay-in-flight experiments~\cite{NA62:2017rwk, PIONEER:2022yag}.}

    One of the arguably most well-motivated FIPs are Heavy Neutral Leptons (HNLs), whose discovery could help to understand the origin and scale of neutrino masses~\cite{Minkowski:1977sc,Gell-Mann:1979vob,Shaposhnikov:2006nn},
    the Dark Matter problem~\cite{Dodelson:1993je,Shi:1998km,Abazajian:2001nj,Asaka:2006nq,Boyarsky:2009ix}, and the Baryon asymmetry of the universe~\cite{Fukugita:1986hr,Akhmedov:1998qx,Asaka:2005pn,Klaric:2021cpi,Drewes:2021nqr}. 
    For a comprehensive overview for constraints and searches of HNLs in a wide mass region, see for example~\cite{HNL}.
    In our study we are scrutinising the phenomenology and prospects of HNL searches in proton fixed-target experiments.
    
    %about (heavy) mesons
    In proton fixed-target experiments, a predominant production process of FIPs, and also of HNLs, is in the decay of secondary mesons originating in the primary proton shower. 
    Specifically, it is the most boosted forward component of those mesons which is almost exclusively relevant in the computation of the FIP yield. 
    However, since the signal from this component is typically dominated by beam induced backgrounds in fixed-target set-ups, and is thus least well known, there are uncertainties to be accounted for when modelling the production of FIP production from mesons. 
    A detailed modelling is needed though in order to prove or disprove the existence of HNLs in data. 
    While in decay-in-flight experiments, for example with Kaons, there is good control over the yield and kinematic distributions, the validation of the heavy meson flux is more challenging due to typically much shorter lifetimes. 
    One of the standard tools used to model heavy meson production~\cite{Ruf:2115534,Gorkavenko:2021mpj} in this context is the event generator \pyth, other options are discussed below.
    
    In the case of lighter mesons, an attempt~\cite{Dobrich:2019dxc} has been made to validate existing experimental literature against \pyth, such that uncertainties in shape and yield can be accounted for.
    For heavier mesons, this endeavour is much more difficult due to statistics~\cite{Lourenco:2006vw} -- a fact which has triggered projects to overcome this problem~\cite{SHiP:2024oua}.

    Given the heavy meson spectrum, the production, propagation and decay of HNLs can be modelled for a specific experimental set-up (c.f. figure \ref{fig:bd_schematic}). 
    A public, unified framework is then relevant in order to reliably compare different FIP parameter reaches for different acceptance parameters (geometry, acceptance cuts...).
    One such tool is \alpi~\cite{Jerhot:2022chi, Afik:2023mhj}, originally developed for the specific case of axion-like particles. 
    Software with similar functionality has been made available including \textsc{MadDump}~\cite{Buonocore:2018xjk}, and more recently \textsc{SensCalc}~\cite{Ovchynnikov:2023cry}, and \textsc{DisplacedDecayCounter}~\cite{Domingo:2023dew}.
    Such software is particularly useful for FIPs that can couple to multiple SM particles, such as HNLs, which can couple to all SM leptons. 
    However, the relative strength of these couplings depends on the exact model at hand. 
    Often, for simplicity, benchmark cases where the HNL couples to only one lepton are shown. 
    With light-weight Monte Carlos (MCs) such as the ones mentioned above, a plethora of other cases can be studied easily.
    %Such software is particularly useful because physics is richer than the PBC benchmarks.
    
    In this paper we  introduce a version of \alpi that can handle the full production and decay chains for HNLs. In addition, we compare in detail the experimental knowledge of heavy meson production for energies relevant in proton fixed-target experiments. 
    We compare those measurements to simulations and \emph{scrutinise predictions for the possible discovery or exclusion of HNLs  in a number of experiments}.
    The main focus of this version of \alpi is to facilitate a public MC simulation tool to estimate (proposed) experiments for their sensitivities to the benchmark cases proposed by physics beyond colliders~\cite{Beacham:2019nyx,Antel:2023hkf}.\footnote{As for ALPs, the HNL implementation also allows for non-trivial combinations thereof.}

    This paper is structured as follows: In section~\ref{sec:FIP_BD_pheno} we review the concepts of FIP searches in beam-dumps and highlight the variables interesting for the case of HNL searches. 
    In section~\ref{sec:distributions}, we discuss the experimental status of heavy meson spectra in the relevant energy regimes for HNL production at beam-dumps. 
    We compare this to state-of-the-art simulation tools.
    In section~\ref{sec:ALPINIST} we review the MC simulation framework \alpi which will be employed to obtain sensitivity projections for a number of scenarios. 
    Finally, we summarise the results of our studies in section~\ref{sec:res} and discuss the impact of our study for past and future HNL searches. 
    Lastly, we summarise our findings and discuss the relevance to other FIP cases in section~\ref{sec:conc}. 

\section{Fixed-target phenomenology of feebly interacting particles}\label{sec:FIP_BD_pheno}

    \subsection{Proton beam-dump simulation of FIP searches}

    Employing a proton beam-dump in Feebly Interacting Particle searches has gained renewed interest after the last such endeavours in the 1980s. The rationale is that weakly coupled particles at MeV-GeV mass scales for a certain range of decay lengths/couplings can be accessed most sensitively.\footnote{A number of subtleties have to be considered here~\cite{Dobrich:2024ajq}, the primary one being the proper modelling of the FIPs' production, as is the main point of this article.}
    
\begin{figure}
        \centering
        \begin{flushleft}
\begin{footnotesize}
\vspace*{-14em}
\begin{tikzpicture}[line cap=round,line join=round,>=triangle 45,x=0.5cm,y=0.5cm]
\clip(0.6225995992514683,-5.692672824868455) rectangle (28.698864219790572,16.00510061003318);
\fill[line width=2pt,fill=black,fill opacity=0.18] (3.9797910449283718,6) -- (7.979791044928372,6) -- (7.979791044928372,2) -- (3.9797910449283718,2) -- cycle;
\fill[line width=2pt,fill=black,fill opacity=0.04] (14,6) -- (22,6) -- (22,2) -- (14,2) -- cycle;
\fill[line width=2pt,fill=black,fill opacity=0.1] (22,6) -- (26.5,6) -- (26.5,2) -- (22,2) -- cycle;
\fill[line width=2pt,fill=black,fill opacity=0.36] (26.5,6) -- (27,6) -- (27,2) -- (26.5,2) -- cycle;
\draw (3.9797910449283718,6)-- (7.979791044928372,6) -- (7.979791044928372,2);
\draw (7.979791044928372,2)-- (3.9797910449283718,2) -- (3.9797910449283718,6);
% \draw (14,6);
\draw (22,2)-- (14,2) -- (14,6) -- (22,6) ;
\draw (4.7,5.8) node[anchor=north west] {\textbf{dump}};
\draw (14.2,5.35) node[anchor=north west] {\textbf{decay volume}};
% \draw (22,6);
\draw (27,2)-- (22,2)-- (22,6)-- (27,6);
\draw (24.5,2)-- (24.5,6);
\draw [dotted] (22.75,2)-- (22.75,6);
\draw [dotted] (23.75,2)-- (23.75,6);
\draw (27,6)-- (27,2);
\draw (26.5,2)-- (26.5,6);
\draw (22.5,6)-- (22.5,2);
\draw (24,6)-- (24,2);
\draw (21.5,7.7) node[anchor=north west] {\textbf{magnetic}};
\draw (20.8,7.05) node[anchor=north west] {\textbf{spectrometer}};
\draw (23,2.7) node[anchor=north west] {\scriptsize{$B$}};
\draw (22.75,3.2) node[anchor=north west] {$\odot$};
\draw (22.75,4.25) node[anchor=north west] {$\odot$};
\draw (22.75,5.3) node[anchor=north west] {$\odot$};
\draw (24.5,1.9) node[anchor=north west] {\textbf{calorimeter}};
\draw [line width=1pt,->] (2.5,4)-- (4.3,4);
\draw [dashed] (4,4.009014219606395)-- (18.813737693380236,3.593318071739856);
\draw [line width=1pt] (4.505960836178257,3.995727721369871) -- (5.62592932844057,4.187722320043407);
\draw [line width=1pt] (4.793952734188567,3.9797281714804096) -- (6.057917175456033,3.5477403244649546);
\draw [line width=1pt] (5,4) -- (6.02591807567711,3.8677313222541807);
\draw [line width=1pt] (4.905949583414798,3.6597371736911835) -- (4.249968037946872,3.995727721369871);
\draw [line width=1pt] (5.135550934516929,3.982523806431687) -- (6.195822695009405,4.251272260124695);
\draw (0.6,5.2) node[anchor=north west] {$p/e$ beam};
\draw (9.7,4.7) node[anchor=north west] {$X$};
\draw (19.9,5.3) node[anchor=north west] {$\ell^+$};
\draw (20.7,4.5) node[anchor=north west] {$\nu$};
\draw (20.0,3.6) node[anchor=north west] {$\ell^-$};
\draw [line width=1pt, style=dotted ] (18.813737693380236,3.593318071739856) -- (30.5,4.3); %neutrino line 26.5 cal endpoint
\draw [line width=1pt] (18.69632107966186,3.596897058466403)-- (23.25,5.348473881106331) -- (26.5,5);
\draw [line width=1pt] (18.69632107966186,3.596897058466403)-- (23.25,3.133813973558165) -- (26.5,3.1685494892055472);
\end{tikzpicture}
\end{footnotesize}
\end{flushleft}
\vspace*{-10em}
% \hspace*{-4em}
\hspace*{-4em}
        \caption{Schematic layout of a typical beam-dump experiment.}
        \label{fig:bd_schematic}
    \end{figure}
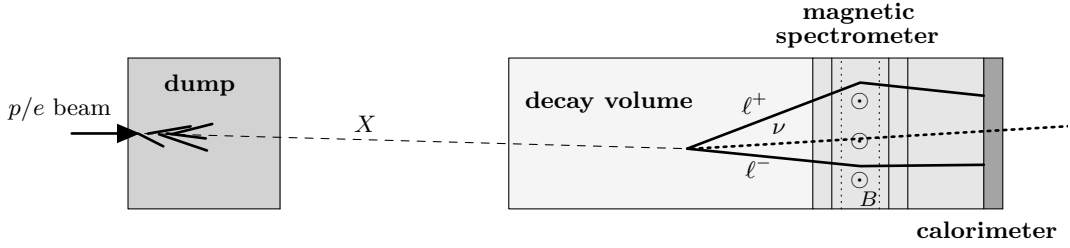

    The general principle of a fixed-target experiment is presented schematically in figure~\ref{fig:bd_schematic} and can be summarised as follows:
    \begin{enumerate}
        \item A highly energetic beam particle impinges on a stationary target (dump) and scatters off the material. 
        \item A feebly interacting particle $X$ is produced in primary or secondary interactions of the beam particle with the dump material with a probability $\chi_X$
        \item Due to its inherently limited interaction strength, $X$ is assumed to propagate freely up to the point of its eventual decay. 
        \item With a given probability $P_\mathrm{FV}$, the decay of $X$ occurs in a fiducial decay volume. 
        \item With a given probability $P_\mathrm{acc}$, the SM decay products of $X$ are within the acceptance of a detector apparatus. 
    \end{enumerate}
    For a generic FIP with mass $m_X$ and SM coupling $g_X$, the number of detectable events with a given final state can thus be estimated as 
    \begin{equation}\label{eq:N_detections}
        N_\mathrm{evt}(m_X, g_X) = N_\mathrm{PoT}\chi^{}_X(g_X,m_X) P_\mathrm{FV}(m_X, \Gamma_X) \epsilon_\mathrm{det}P_\mathrm{acc}(m_X) \text{,}
    \end{equation}
    where $N_\mathrm{PoT}$ is the number of beam particles on target, $\Gamma_X(g_X,m_X)$ is the width of $X$, and $\epsilon_\mathrm{det}$ quantifies the detection probability of the final states crossing the detectors.
    As nearly all beam particles will interact with the dense material of the dump, beam-dump experiments typically feature a very high luminosity compared to collider experiments, albeit at an inherently lower centre-of-mass energy. 
    This makes them the ideal class of experiments to look for FIPs, as the FIPs' very small interaction cross sections can be compensated by the high luminosity. 
    
    \subsection{The Heavy Neutral Lepton} \label{sec:hnl_pheno}

        In this section we will outline the steps in simplifying the fixed-target phenomenology of the HNL into a tabulation-compatible framework.
        The HNL is commonly introduced as part of the type-I seesaw, where $\mathcal N$ new right chiral singlet fermions $\nu_{R I}$ featuring a Majorana mass $M_I$ are added to the content of the SM Lagrangian~\cite{Minkowski:1977sc,Gell-Mann:1979vob}.
        These states can couple to the SM `active' neutrinos $\nu_{L\alpha}$ of lepton generation $\alpha$ via a Yukawa-like term which after electroweak symmetry breaking yields
        \begin{equation}\label{eq:Lagrangian_seesaw}
            \mathcal{L}^{\nu_L \nu_R}_{\alpha I} = -\left(F_{\alpha I}^{\nu}\right)^*\frac{v}{\sqrt{2}}\nu_{L\alpha}^{\dagger}\nu_{R I}+ h.c.
        \end{equation}
        Here, $F_{\alpha I}$ denotes the Yukawa coupling of the $I$-th right chiral singlet to lepton generation $\alpha$, and $v$ is the Higgs' vacuum expectation value.
        Diagonalising in the mass basis yields $\mathcal N$ heavy Majorana states (HNLs) $\mathrm{N}_I$ with masses $m_{\mathrm{N}_I}$.\footnote{Additionally, this yields three light Majorana neutrinos with masses $m_{\nu i}$ determined by the seesaw relation $\Bigl( V^\dagger_\mathrm{PMNS} \mathrm{diag}(m_{\nu_1},m_{\nu_2},m_{\nu_3}) V^{}_\mathrm{PMNS} \Bigr)_{\alpha\beta} \simeq - v^2\sum_I \frac{F_{\alpha I}F_{\beta I}}{M_I^2} m_{\mathrm{N}_I}$.}
        Consequently the HNL inherits couplings to SM fields from its left chiral component, namely 
        \begin{equation}\label{eq:Lagrangian_HNL_int}
            \mathcal L _\mathrm{int} = \frac g {2\sqrt 2} W_\mu^+\sum_{\alpha, I} \theta_{\alpha I}^* \bar{N^c_I}\gamma^\mu (1-\gamma_5) \ell^-_\alpha + \frac g {4\cos\theta_W}Z_\mu \sum_{\alpha, I} \theta_{\alpha I}^* \bar{N^c_I}\gamma^\mu (1-\gamma_5) \nu_\alpha + h.c.
        \end{equation}
        where the interactions are suppressed by the mixing angles
        \begin{equation}
            \theta_{\alpha I} \equiv \frac{F_{\alpha I }v}{\sqrt 2 M_I}.
        \end{equation}
        For simplicity in phenomenological studies, the HNLs are commonly approximated by a single fermion $\mathrm{N}$ with mass $m_\mathrm{N}$, and mixing angles $\theta_\alpha$ which is either Majorana or Dirac in nature.
        It is further convenient to define the \emph{coupling suppression} $U_\alpha^2 = |\theta_\alpha|^2$, which in the following we will often refer to simply as the coupling. 
        This allows us to simplify the parameter space from $7\mathcal N -3 $ free parameters in the full HNL theory to the HNL mass, the set of $(U_e, U_\mu, U_\tau)$ and the choice between Majorana or Dirac type HNL. 
        For recent reviews, please see~\cite{Abdullahi:2022jlv, Antel:2023hkf} and references therein.

        In accordance with eq.~(\ref{eq:Lagrangian_HNL_int}), HNLs can appear and decay through weak interactions at a suppressed rate (determined by $U_\alpha^2$). 
        Thus, possible production mechanisms at beam-dump facilities include i) deep inelastic and coherent proton-nucleus scattering processes, and ii) the (semi-) leptonic decay of secondary mesons and $\tau$ leptons. 
        In this work we will limit ourselves to the production through ii), as this contribution typically outweighs mechanism i) by several orders of magnitude for the typical centre-of-mass energies of fixed-target experiments~\cite{Bondarenko:2018ptm}.
        We further assume that the HNL does not couple to lighter dark sector particles, so that its width is purely determined by the decays allowed by the interactions prescribed by eq.~(\ref{eq:Lagrangian_HNL_int}). 

        The theoretical description of heavy neutrinos regarding their production in meson decays and decay channels has long been established in literature~\cite{Shrock:1980ct, Shrock:1981wq, Johnson:1997cj} and is mostly consistent with more recent reviews dedicated to HNLs~\cite{Gorbunov:2007ak, Atre:2009rg}. 
        A crucial difference however, are the contributions of neutral current interaction to some final states of the HNL decay, which can lead to significant discrepancies in terms of phenomenological implications~\cite{Ruchayskiy:2011aa,PhysRevD.104.095019}. 
        We consider width contributions to the HNL with a mass up to the $B_s$ mass as listed in the most recent review on the subject~\cite{Bondarenko:2018ptm} with updated underlying theory inputs as listed in appendix~\ref{sec:HNL_interactions}.
        This leads to slightly different branching ratios for three-body decays of beauty mesons which are shown for an electronphilic HNL in figure~\ref{fig:BC6_heavyProdBrs}.
        A brief summary of the production and decay widths are listed for the reader's convenience in appendix~\ref{sec:HNL_interactions}.
        
\section{Heavy meson distributions at fixed-target experiments} 
\label{sec:distributions}
    
    In the $\SI{100}{MeV}$ to $\mathrm{few}\,\si{GeV}$ mass range, where beam-dump searches dominate, HNLs are most abundantly produced in the decays of charm and beauty mesons (see section~\ref{sec:hnl_pheno} and appendix~\ref{sec:HNL_interactions}). 
    Thus, it is paramount to accurately describe the meson production in order to evaluate the resulting HNL distributions and eventual detector responses. 
    This not only extends to the production probability of the meson, but importantly also to their kinematic spectra, as we will see in this section.
    
    The simulation package \pyth is one of the tools most used in the community to determine the expected yield of mesons produced by proton interactions, e.g.~\cite{CERN-SHiP-NOTE-2015-009,Moghaddam:2022tac,Fieg:2023kld}. 
    
    \subsection{Production cross section}\label{sec:sigma_qq}

    The most straightforward way in which meson properties affect the production of Feebly Interacting Particles is through the meson production cross section, determining how many mesons and therefore potential decays into FIPs we observe.
    For heavy mesons this is commonly quantified by the pair production cross section of the relevant heavy valence quark $\sigma_{q\bar q}$.
    The production rate of a given meson is then determined by the convolution of the quark level cross section and fragmentation functions subject to the QCD environment around the heavy quarks~\cite{Lourenco:2006vw}. 

    These cross sections can be calculated from the theory side. 
    \pyth provides an internally generated estimate for $\sigma_{q\bar q}$. 
    However, these are known to underestimate the measured cross sections, prompting often a naive rescaling of the estimate by a factor $k_{q \bar q}$ typically of the order $\mathrm{few}$~\cite{Lourenco:2006vw, CERN-SHiP-NOTE-2015-009}.
    Alternatively, the cross sections can be calculated directly in various schemes using pQCD (e.g. massive quarks at NLO~\cite{Vogt:2007aw}), with one of the most widely used schemes being the fixed order next-to-leading logarithm (FONLL) approach~\cite{Cacciari:1998it, Cacciari:2001td}.
    To calculate the production cross section, we employ the FONLL software interfaced with LHAPDF6~\cite{Buckley:2014ana} to use the NNPDF4.0 proton parton distribution functions~\cite{NNPDF:2021njg}.

    On the experimental side, the open charmed (or $D$) mesons feature a variety of relatively clean decay signatures, and a sizeable production cross section even at typical energies for beam-dump experiments. 
    Notably, charm decays affect significantly the energy spectrum of neutrinos emitted at these centre-of-mass energies, which gave a motivation for several experiments in the past to accurately measure the charm production cross section (see e.g.~\cite{Frixione:1994nb} for a review).
    For open beauty (or $B$) mesons, however, the production cross section at centre-of-mass energies around $\SI{30}{GeV}$ is smaller than for charmed mesons by three orders of magnitude. 
    No experimental data is available for centre-of-masses below $\SI{40}{GeV}$. 
    
\begin{figure}
        \centering
        \includegraphics[width = 0.995\textwidth]{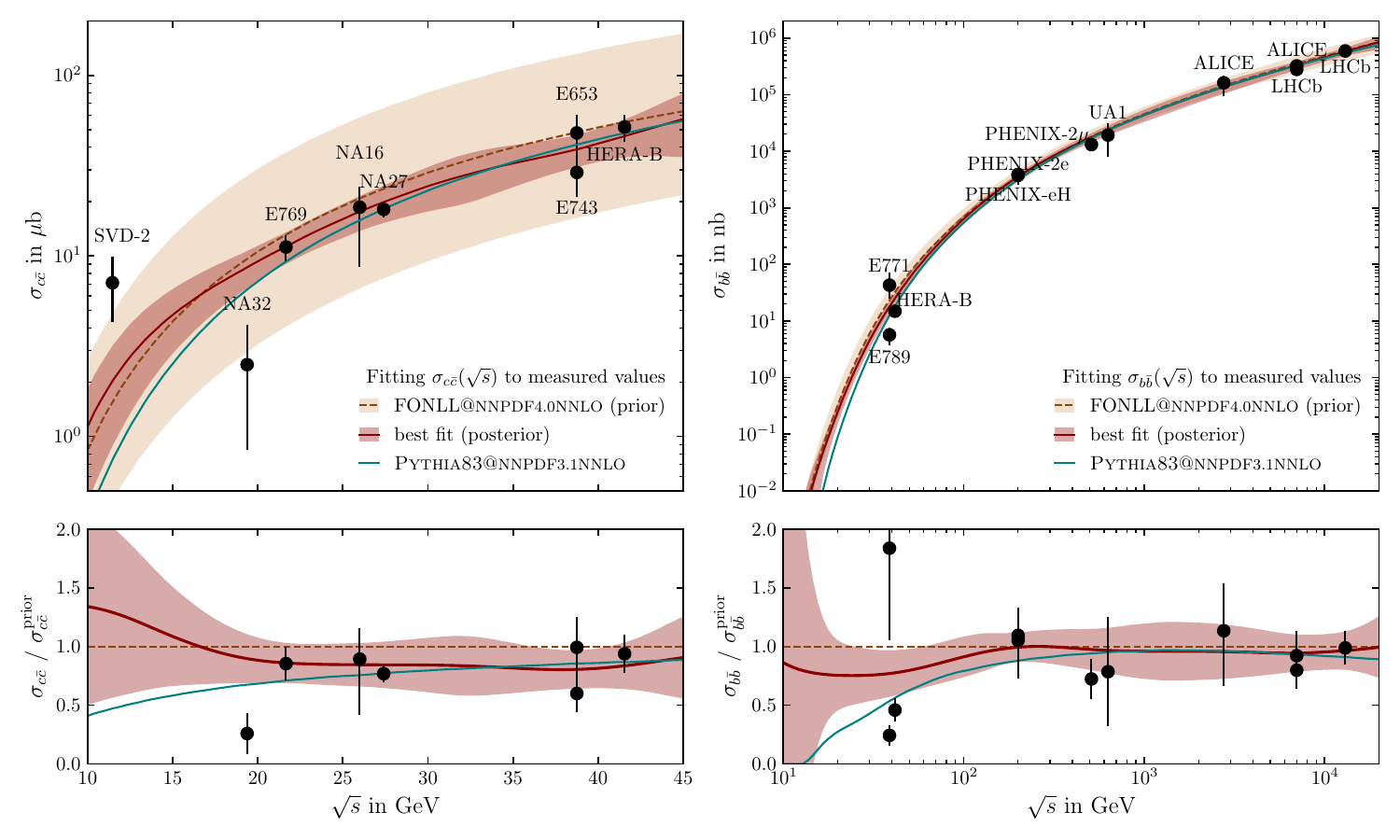}
        \caption{Charm (left) and beauty (right) pair production cross sections as a function of the centre-of-mass energy of the $pp$ collision.
        The FONLL calculation~\cite{Cacciari:1998it,Cacciari:2001td} (dashed brown curve with orange filled area showing the uncertainties from scale and quark mass variation), the Bayesian fit to the data~\cite{SVD-2:2017ovj,ACCMOR:1988pxc,E769:1996jqf,Adamovich:1992cv,LEBC-EHS:1983llu,LEBC-EHS:1988oic,Ammar:1988ta,FermilabE653:1991vmo,HERA-B:2007rfd,%Dcitations
        E789:1994nhc,Jansen:1994bz,HERA-B:2005tnp,PHENIX:2018dwt,PHENIX:2020iaj,UA1:1990vvp, ALICE:2012acz,LHCb:2010wqx,LHCb:2016qpe%B citations
        }
        (red curve with light-red filled area showing the $68\%,\mathrm{CL}$), and the \pyth calculation (teal curve), scaled by a factor $k_{q\bar q}$ obtained through an ODR fit to the data are compared.
        The lower panels show the production cross section normalised to the central value predicted by the FONLL calculation.
        For further details see text.}
        \label{fig:qqbar_production_crosssection}
    \end{figure}

     We use the PyMC Gaussian Process~\cite{AbrilPla2023} to perform a Bayesian fit to the data (see appendix~\ref{sec:fit_details} for details). 
     The posterior distribution is calculated as a marginal likelihood using the central value and the estimated error of the FONLL calculation as a prior.
     The resulting posterior shows a good agreement with the FONLL calculations within the respective uncertainties, as shown in figure~\ref{fig:qqbar_production_crosssection}. 
     The fit results for relevant beam energies of different beam-dump facilities are summarised in table~\ref{tab:qqbar_production_crosssection}, with further information about the fit in appendix~\ref{sec:fit_details}.
    
    In figure~\ref{fig:qqbar_production_crosssection} we also compare the Bayesian fit to the \pyth result\footnote{This result was derived using the most up to date NNPDF proton parton distribution function available in \pyth~8.3, the NNPDF3.1 NNLO+LUXQED~\cite{NNPDF:2021njg}.} scaled by a multiplicative constant $k_{q \bar q}$ determined by an orthogonal distance regression (ODR) fit to the data. We find that at the higher end of the centre-of-mass energies presented, the scaled \pyth curve agrees reasonably well with both the fit result and the FONLL calculation. 
    Especially at the lower end of the $\sqrt{s}$ spectrum, the \pyth curve significantly underestimates the production cross sections for both charmed and beauty quarks.
    %In the absence of data at lower centre of mass energies the $\sigma_{b\bar b}$ fit relies on extrapolation to the low energy regime.
    %Encouraged by the good agreement between fit and data in this regime in the $\sigma_{c\bar c}$ case, we decided to rely on the FONLL values as a prior for the extrapolation.  
    For $\sigma_{b\bar b}$, data points at much higher energy scales are included to achieve a more stable extrapolation and the fit is performed on $\log(\sqrt{s})$.
    Nonetheless, the diverging uncertainty at $\sqrt{s}\to0$ reflects the absence of data and cannot be avoided.
    
    Unless otherwise stated, the production cross sections for the various beam-dump experiments considered in this work are the fit results listed in table~\ref{tab:qqbar_production_crosssection}.   

    \begin{table}[h]
        \centering
        \begin{tabular}{ccc}
             $E_\mathrm{beam}$ in $\mathrm{GeV}$& $\sigma_{c\bar c}$ in $\mathrm{\mu b}$ & $\sigma_{b\bar b}$ in $\mathrm{nb}$ \\
            \hline
            $70$	&$2.2\pm1.2$	& -- \\
            $120$	&$4.6\pm1.8$	&$0.02\pm0.02$ \\
            $400$	&$20.0\pm4.5$	&$2.63\pm0.76$ \\
            $800$	&$39.5\pm7.7$	&$18.6\pm4.6$ \\
            \hline
        \end{tabular}
        \caption{Best fit results for quark anti-quark pair production cross sections $\sigma_{q\bar q}$ for typical beam energies $E_\mathrm{beam}$ at beam-dump facilities.}
        \label{tab:qqbar_production_crosssection}
    \end{table}

    \subsection{Differential production cross sections}\label{sec:dsigma_qq}

    Another important aspect about mesons regards their kinematic spectra, as these greatly impact the kinematic spectra of their eventual FIP daughters.
    The meson spectra are usually given in terms of the Feynman parameter $x_F$ quantifying the centre-of-mass momentum transfer of the beam to the meson ($x_F=p_z^\mathrm{CM}p_{z,\max}^\mathrm{CM\ -1} $) and $p_T$, the momentum orthogonal to the incident beam axis' modulus.
    A common way to parameterise the differential cross section of heavy mesons in a fixed-target experiment is given by~\cite{Lourenco:2006vw}
    \begin{equation}\label{eq:diff_meson_parametrisation}
        \frac{\partial^2 \sigma}{ \partial{x_F}\partial{p_T^2}} \propto (1-|x_F|)^n e^{-(ap_T + b p_T^2)},
    \end{equation}
    which we will use in the following to summarise different measurement results.
    It is important to note that heavy meson decay lengths are typically orders of magnitude below the typical material interaction lengths, justifying a prompt decay assumption.\footnote{The longest living heavy meson is the $B^\pm$ meson with a life time equivalent to $c\tau_{B^\pm} = \SI{491\pm 1}{\micro\metre}$. 
    Meanwhile, the nuclear interaction lengths in copper ($\SI{153.2}{mm}$) or lead ($\SI{175.9}{mm}$) are orders of magnitude longer.~\cite{ParticleDataGroup:2022pth}}
    Therefore, the following considerations are valid on the nucleon level irrespective of the target material.
    
    \subsubsection{Open charmed mesons}
    
    Many of the experiments that measured the total production cross section also made an effort to measure their differential distributions.
    A summary of the experimental information is listed in table~\ref{tab:diffxs_cc}. 
    Large uncertainties and significantly differing power laws are obtained by different experiments. 
    As we will see in section~\ref{sec:Impact_CHARM_HNL_spectra}, such differences are relevant for the sensitivities of searches for HNLs. 
    Using the data presented in table~\ref{tab:diffxs_cc}, we employ a similar Bayesian fitting procedure as presented in section~\ref{sec:sigma_qq}.
    As a theory motivated prior for these fits, we use \pyth~8.3 to generate meson spectra to which we then fit eq.~(\ref{eq:diff_meson_parametrisation}) using orthogonal distance regression. 
    As most of the experimental data in table~\ref{tab:diffxs_cc} is presented with $a=0$ and $b$ as a free parameter to fit the measured $p_T$ distribution, we follow this convention.
    The $b$ fit results on the \pyth generated data consistently underestimate the experimental results. 
    Therefore, we decided to scale the prior $b$ value by a global factor of $2$.
    In the absence of a theory motivated uncertainty, we apply a prior $68\%\,\mathrm{CL}$ region as $\pm 50\%$ with respect to the central prior value.
    The resulting band covers almost all of the data uncertainty ranges.\footnote{This is not the case for 2 of the 5 data points for the $b_{c \bar c}$ fit, resulting in the poor expected log point wise predictive density score presented in table~\ref{tab:fit_details}.}
    
    \begin{table}[h]
        \centering
        \begin{tabular}{cccccc}
             experiment  & target & $E_\mathrm{beam}$& $n$ & $a$  & $b$  \\
             &material&[$\si{GeV}$]& & [$\si{\per\giga\electronvolt}$] & [$\si{\per\giga\electronvolt\squared}$] \\
             \hline
             ACCMOR(NA32)   &$\mathrm{Si}$   & $200$ & $5.5^{+2.1}_{-1.8}$ & -- & $1.4^{+0.6}_{-0.4}$ \\
             WA82           &$\mathrm{Cu}$   & $370$ & $6\pm 0.3$          & -- & $0.93\pm0.09$ \\
             LEBC-EHS(NA27) &$\mathrm{H_2}$  & $400$ & $4.9\pm0.5$         & -- & $0.99\pm0.09$ \\
             CHARM          &$\mathrm{Cu}$   & $400$ & $ 6.2^{+0.4}_{-0.5}\pm 0.4$ & $2.0 \pm 0.3$ & -- \\
             LEBC-MPS(E743) &$\mathrm{H_2}$  & $800$ & $8.6\pm2.0$         & -- & $1.1 \pm0.3$ \\
             E653           &emulsion        & $800$ & $6.9^{+1.9}_{-1.8}$ & -- & $0.84^{+0.10}_{-0.08}$ \\
             HERA-B         &$\mathrm{C},\mathrm{Ti},\mathrm{W}$ & $920$ & $7.5 \pm 3.2$ & -- & -- \\
            \hline
        \end{tabular}
        \caption{Experimental results~\cite{ACCMOR:1988pxc, Adamovich:1992cv, LEBC-EHS:1987evz, Bergsma:1987br, Ammar:1988ta, FermilabE653:1991vmo, HERA-B:2007rfd} for the differential cross section measurements of open charmed meson production in proton beam fixed-target interactions as parameterised by eq.~(\ref{eq:diff_meson_parametrisation}).}
        \label{tab:diffxs_cc}
    \end{table}

\begin{figure}
        \centering
        \includegraphics[width = 0.995\textwidth]{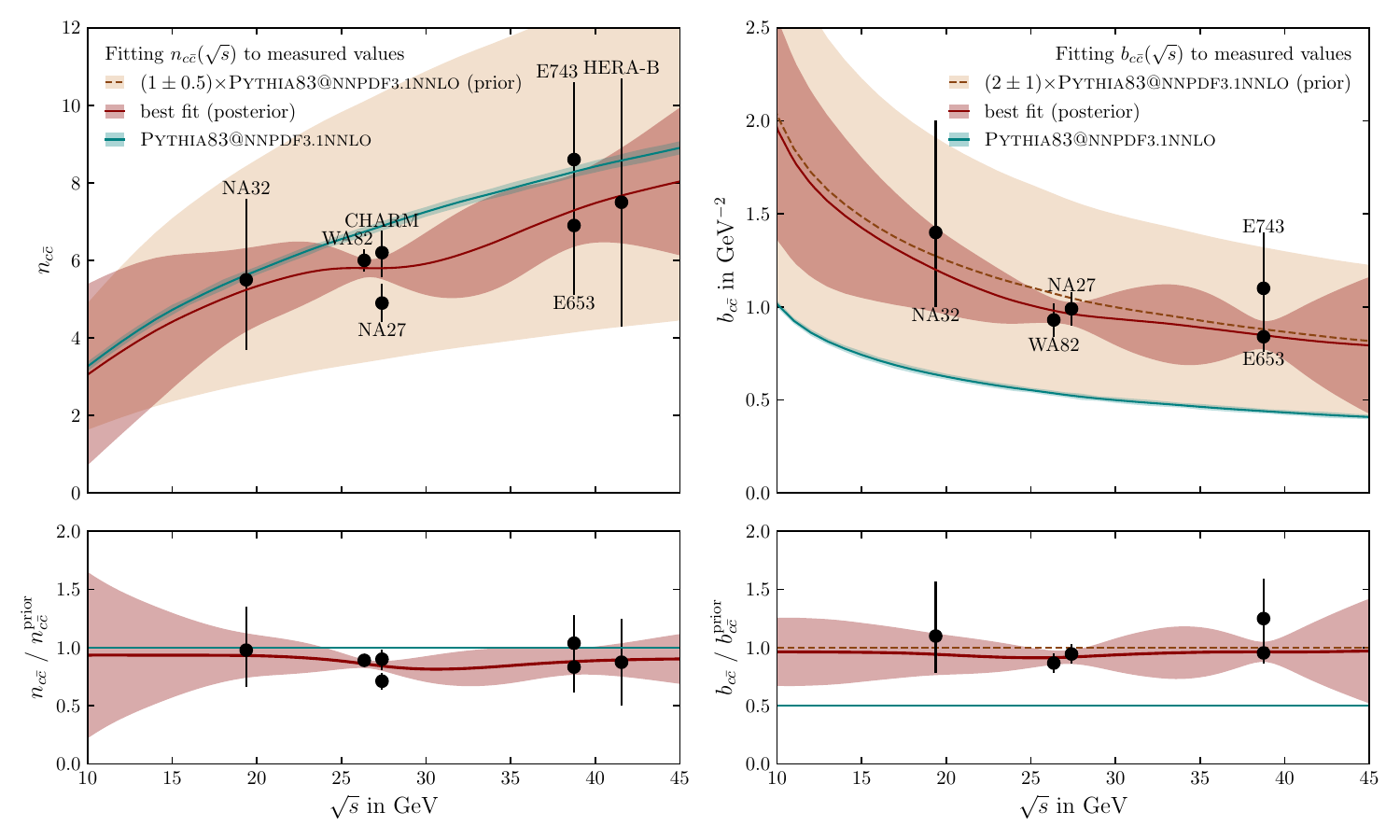}
        \caption{Charm differential cross section parameters of forward (left) and transversal (right) momentum as parameterised by eq.~(\ref{eq:diff_meson_parametrisation}) at $a=0$ pair production cross sections as a function of the centre-of-mass energy of the $pp$ collision.
        The \pyth results are derived using an ODR fit of eq.~(\ref{eq:diff_meson_parametrisation}) at $a=0$ to the spectra generated with \pyth~8.3.
        The scaled \pyth calculation (dashed brown curve with orange filled area showing assumed prior uncertainty), the Bayesian fit to the data~\cite{ACCMOR:1988pxc, Adamovich:1992cv, LEBC-EHS:1987evz, Bergsma:1987br, Ammar:1988ta, FermilabE653:1991vmo, HERA-B:2007rfd}
        (red curve with light-red filled area showing the $68\%,\mathrm{CL}$), and the unscaled \pyth (teal curve with teal area showing the fit uncertainty).
        The lower panels show the respective parameters normalised to the scaled central values derived from the \pyth spectra (prior).
        For further details see text.}
        \label{fig:DX_cc_fit}
    \end{figure}

    The results of this fit are shown in figure~\ref{fig:DX_cc_fit} and summarised for common beam-dump energies in table~\ref{tab:DX_cc_fit}.
    We find that $n_{c\bar c}^\pyth$ (ODR fit results of the \pyth spectra) agrees reasonably well with the data. 
    The fact, that $b_{c\bar c}^\pyth$ consistently underestimates the data points means that \pyth overestimates the multiplicity of events featuring large $p_T$.
    For FIP searches with far away on-axis detectors this would result in underestimating the signal, due to disproportionately many FIPs inheriting the larger $p_T$ tendency and going out-of-acceptance. 
    
    \begin{table}[h]
        \centering
        \begin{tabular}{ccc}
             $E_\mathrm{beam}$ in $\mathrm{GeV}$& $n_{c\bar c}$ in $\mathrm{\mu b}$ & $b_{c\bar c}$ in $\mathrm{GeV^{-2}}$ \\
            \hline
            $70$	&$3.6\pm2.0$	&$1.66\pm0.54$ \\
            $120$	&$4.4\pm1.6$	&$1.41\pm0.40$ \\
            $400$	&$5.81\pm0.28$	&$0.96\pm0.06$ \\
            $800$	&$7.29\pm0.95$	&$0.85\pm0.08$ \\
            \hline
        \end{tabular}
        \caption{Best fit results for differential cross section as parameterised by eq.~(\ref{eq:diff_meson_parametrisation}) at $a=0$ for typical beam energies $E_\mathrm{beam}$ at beam-dump facilities.}
        \label{tab:DX_cc_fit}
    \end{table}

    In the above, we have considered all open charmed mesons to follow the same differential distributions. 
    While most of the experiments do not consider a unique differential cross section for the different possible charm flavours ($D^+$, $D^-$, $D^0$, ...), LEBC-EHS (NA27) attempted to identify different curves for the different flavours. 
    This is motivated by an expected contribution to the production cross sections of the \emph{leading particle effect}, possibly including ``$2\to 3$" parton processes (heavy recombination). 
    The underlying principle of this phenomenon is that spectra of secondary particles sharing a valence quark with the beam particle are harder with respect to their charge conjugate counterparts which do not have such a valence quark in common.

    The different meson spectra would consequently not only impact the distribution of FIPs by virtue of their production mechanism being meson dependent (i.e. decay channels of $D^\pm$ do not necessarily exist for $D^0$ with the same kinematics), but also yield different spectra for (anti)-HNLs in case of Dirac-like HNLs.\footnote{Furthermore, this would greatly impact the oscillation patterns which could be observed for example in the pseudo-Dirac-limit for two HNLs.~\cite{Boyanovsky:2014una,Tastet:2019nqj,Mikulenko:2023iqq}} 
    Our present understanding is that the real impact on the differential cross section of the leading particle effect in proton-nucleon collisions at the energy scales of beam-dump experiments is as of yet experimentally inconclusive. 
    However, the interested reader can find a brief summary in appendix~\ref{sec:leading_particle_effect}. 
    Nevertheless, the fit presented in this section can be seen as an envelope for the individual distributions of open charmed mesons.

    \subsubsection{Open beauty mesons}\label{sec:dsigma_bb_experimental}
        %a word on B mesons
        For beauty mesons the experimental landscape in terms of production (differential) cross sections for fixed-targets in the relevant energy range is even more sparse than for their charmed counterparts.
        Therefore, no attempt was made at validating the spectra generated with \pyth~8.309.
        Nonetheless, the interested reader can find a brief summary of the impact of possible future \pyth extensions in appendix~\ref{sec:leading_particle_bb}.

\section{The ALP{\footnotesize I}NIST framework} \label{sec:ALPINIST}

    In order to study the effects of a multitude of different input assumptions on the expected sensitivities of various past, present, and future experiments to Feebly Interacting Particles, we make use of the Axion-Like Particle in Numerous Interactions Simulated and Tabulated (\alpi) simulation framework~\cite{jan_jerhot_2022_5844011, Jerhot:2022chi}.
    In this section we will give a brief overview of the framework (section~\ref{sec:ALPINIST_recap}), introduce the experiments that are already implemented for simple simulations (section~\ref{sec:ALPINIST_expeirments}), and briefly outline a validation of the set up (section~\ref{sec:ALPINIST_validation}).

    \subsection{Axion-Like Particles in Numerous Interactions Simulated and Tabulated}\label{sec:ALPINIST_recap}

    The \alpi framework was originally implemented to facilitate the study of parameter reach of different fixed-target experiments (beam-dumps) to Axion-Like Particles with various coupling structures~\cite{Jerhot:2022chi, Afik:2023mhj}.
    The updated version, published together with this work\footnote{Available on Github under \href{https://github.com/jjerhot/ALPINIST}{\faGithubSquare~github.com/jjerhot/ALPINIST}.}, has been extended by several different models of Feebly Interacting Particles, covering all the PBC benchmark cases~\cite{Beacham:2019nyx,Antel:2023hkf}, which can now be studied in a unified MC simulation set-up. 
    The general layout of the simulation framework (shown in figure~\ref{fig:code_schematic} and discussed in section~\ref{sec:implementation}) is closely related to the general layout of beam-dump experiments (see section~\ref{sec:FIP_BD_pheno}) separating production and detection of the FIP.
    The central idea of this simulation workflow, is that the kinematic and model-dependent part can, to a large extent, be separated (see section~\ref{sec:factorisation}).
    Using common input and theory assumptions, and levels of abstraction, experiments can then be compared on an even footing.
    
    \subsection{Experiments} \label{sec:ALPINIST_expeirments}
    
        The \alpi framework features an abstraction scheme of fixed-target experiments, classifying them by common features. 
        These defining features are
        \begin{itemize}
            \setlength\itemsep{-0.4em}
            \setlength{\leftmargin}{2em}
            \setlength{\labelwidth}{4em}
            \addtolength{\itemindent}{\leftmargin}
            \item[$E_\mathrm{beam}$] the beam energy
            \item[$N_\mathrm{PoT}$] the number of beam particles on target
            \item[$\mathrm{M}$] the target material and associated material constants
            \item[$l_\mathrm{DV}$] the length of the decay volumes (for neutral $l_n$ and charged $l_c$ final states)
            \item[$z_\mathrm{DV}$] the distance of the decay volume from the target
            \item[$\theta_\mathrm{off}$] the angular offset with respect to the beam axis
            \item[$z_\mathrm{Ecal}$] the distance of the electromagnetic calorimeter from the target
        \end{itemize}
        For some experiments, the quantities $z_\mathrm{Spect}$ (the distance of an eventual spectrometer from the target) and $z_\mathrm{MuV}$ (the distance of an eventual muon ID system from the target) are also relevant.
        Using $z_\mathrm{Ecal}$ and the calorimeter's orthogonal geometry, one can determine the solid angular coverage $\Omega_\mathrm{cov}$ which gives a figure of merit for an experiment's geometric acceptance. 
        These values are listed in table~\ref{tab:experiment_geometries} for experiments already implemented in the \alpi framework.
        This highlights the versatility of the \alpi approach, where extremely near (DarkQuest) and long baseline (NuTeV) experiments, and different beam energies spanning more than an order of magnitude can be simulated in a unified set-up. 

        The implementation of the NuCal, NA62/HIKE, DUNE ND, DarkQuest, and KOTO  experiments were already described in detail in previous works~\cite{Jerhot:2022chi, Afik:2023mhj}.
        The geometries of the SHiP and SHADOWS experiments were updated in order to match the design changes as presented in the respective technical proposals~\cite{SHADOWS2023, SHiP2023}, which brought significant changes to the SHiP experimental layout in order to adjust to its new housing facility (CERN cavern ECN3, see also appendix~\ref{sec:SHiP_validation}).
        The newly added BEBC and NuTeV experiments will be introduced in more detail below.
        \subsubsection{BEBC} \label{sec:BEBC_experiment}
\begin{figure}
                \centering
                \includegraphics[width = .695\textwidth]{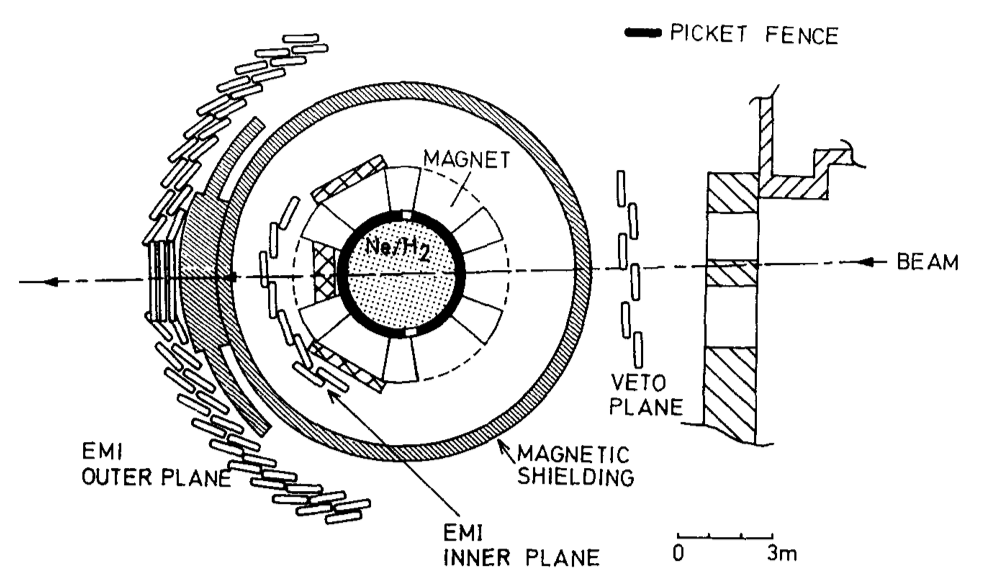}
                \caption{The BEBC chamber and detector array as adapted from~\cite{BEBCWA66:1986err}.}
                \label{fig:BEBC_detectors}
            \end{figure}

            The Big European Bubble Chamber (BEBC) was a $\SI{35}{m^3}$ hydrogen bubble chamber set up in the CERN West Area, hosting a wealth of different experiments~\cite{Harigel:160549}.
            %\footnote{For different experiment set ups, the hydrogen was mixed with other gases. For the 1982 dump run, for example, this was $\SI{74}{\mol \%}$ Neon. }
            In this context we refer to the WA66 experiment, which took data during the 1982 CERN SPS beam-dump run~\cite{BEBCWA66:1986err}.
            The centre of the chamber was aligned with the beam axis and located $\SI{405.9}{m}$ downstream from the front of the dump.
            The detector set-up is shown in figure~\ref{fig:BEBC_detectors} and consists of a veto plane reducing backgrounds due to upstream muons, the chamber in the centre which acts as both decay volume and tracker simultaneously, and the external muon identifier inner and outer plane. 
            The chamber was, further, surrounded by proportional tubes and equipped with two solenoids in Helmholtz arrangement generating a central magnetic field of $\SI{3.5}{T}$~\cite{FOETH1980203,Wittgenstein:1972zz}.
            These were neglected in the \alpi model of BEBC, as they have negligible impact on the cuts applied on true MC events.
            Changes to the decay probability due to the $\mathrm{Ne}/\mathrm{H_2}$ mixture in the chamber were also neglected.
            To `optimise the event rate and the measurability of tracks', the fiducial volume was limited to a $\SI{16.6}{m^3}$ $\SI{6.2}{mrad} \times \SI{8.8}{mrad}$.
            In \alpi we approximate this by a $(r=\SI{1.25}{m},\, h= \SI{3.55}{m})$ cylinder which is concentric and coaxial with the bubble chamber.
            
        \subsubsection{NuTeV} \label{sec:NuTeV_experiment}
\begin{figure}
                \centering
                \includegraphics[width = .595\textwidth]{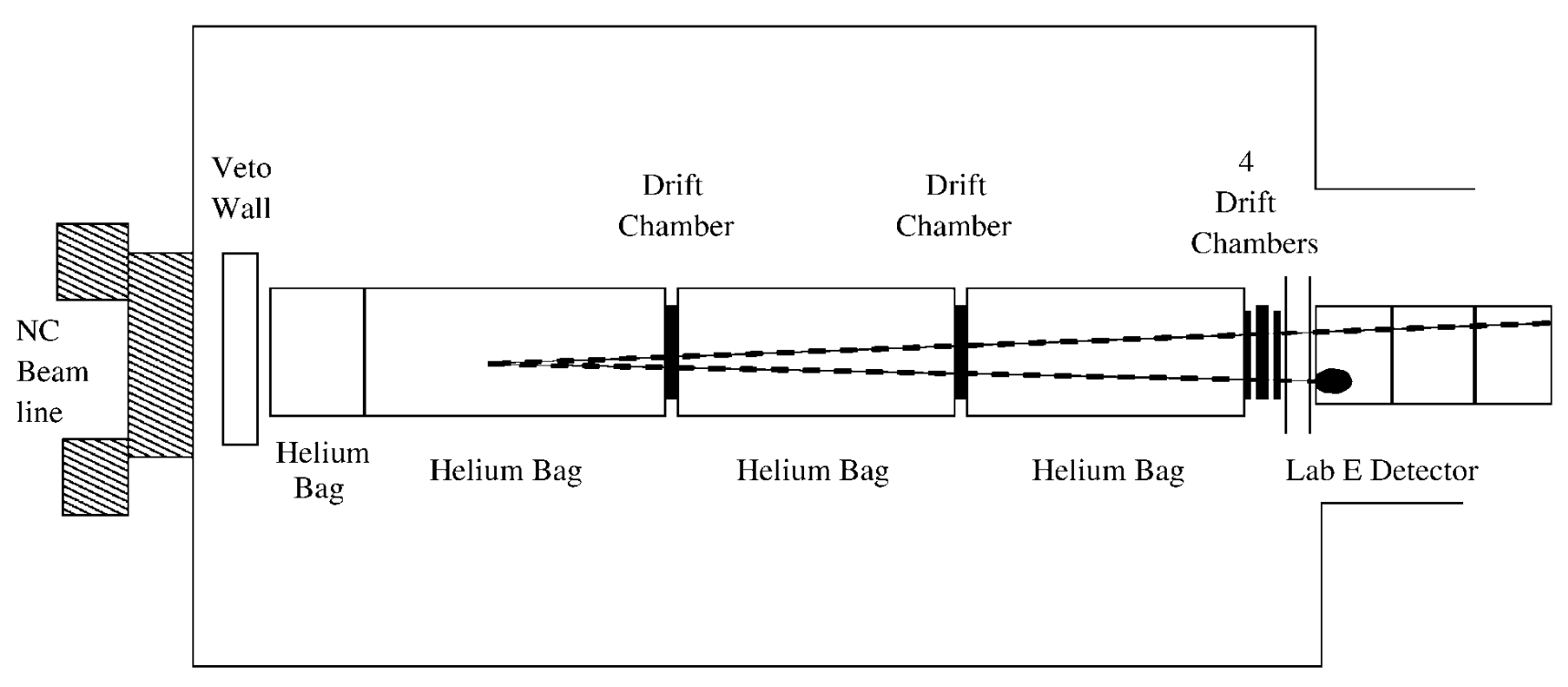}
                \includegraphics[width = .395\textwidth]{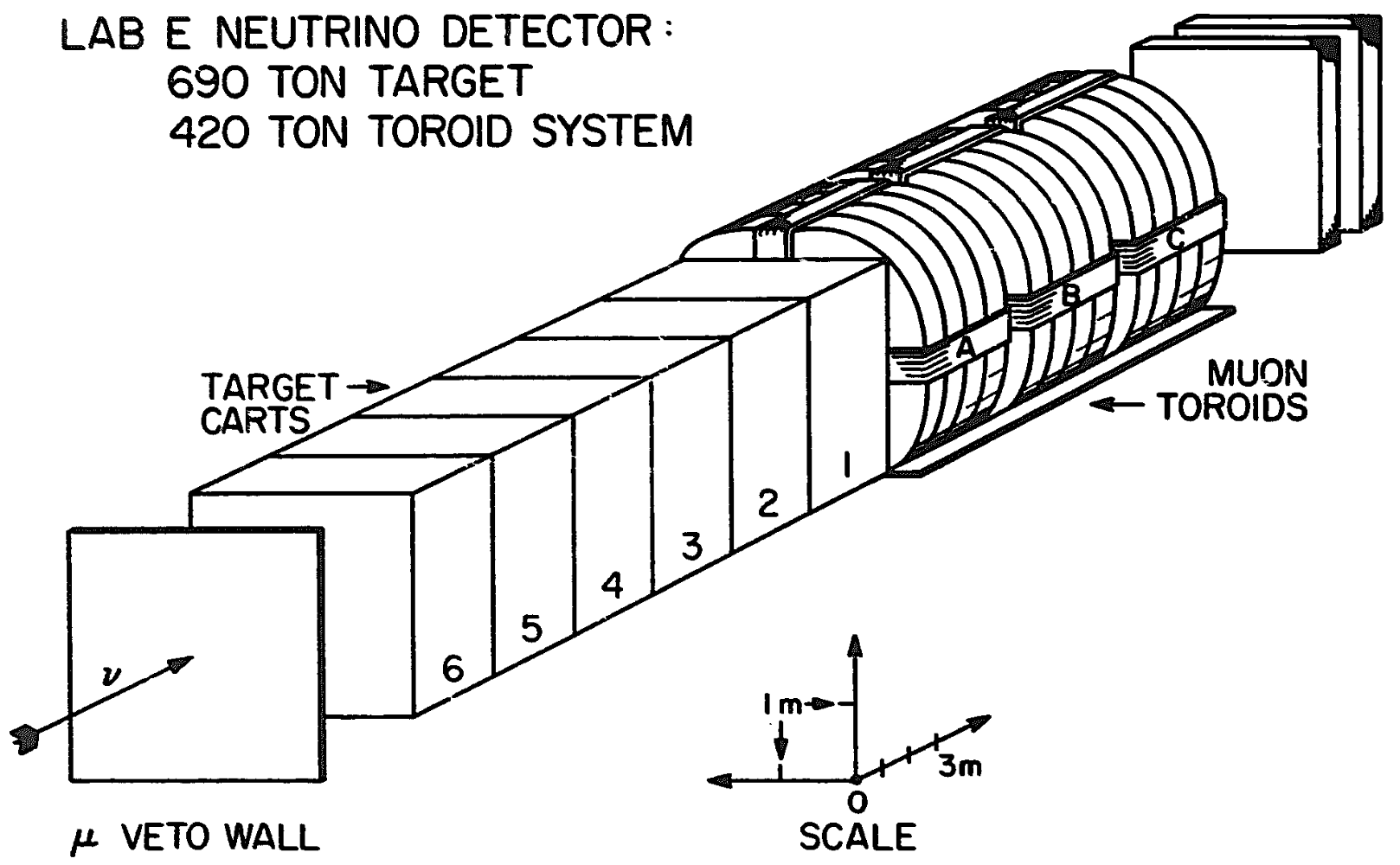}
                \caption{The NuTeV decay and tracking (left) and particle identification (right) array as adapted from~\cite{NuTeV:1999kej,Sakumoto:1990py}.}
                \label{fig:NuTeV_detectors}
            \end{figure}

            The Neutrinos at TeVatron (NuTeV) or E815 experiment was located at the Fermilab National Accelerator Laboratory's Tevatron and dedicated to the study of neutrinos.
            During the fixed-target run, $2.54\times 10^{18}$ protons were dumped into a $\mathrm{BeO}$ target at an energy of $\SI{800}{GeV}$.
            At $\SI{1.4}{km}$ downstream from the target, a detector set-up as depicted in figure~\ref{fig:NuTeV_detectors} was used to study the produced neutrinos and their interactions, but also provided excellent sensitivity to muonphlic HNLs.
            This array consisted of a veto wall to reduce upstream backgrounds, several drift chambers interspaced in the decay chamber facilitating the tracking of charged decay or interaction products, and a 690-ton iron-scintillator sampling calorimeter followed by a toroidal muon spectrometer~\cite{Sakumoto:1990py,King:1991gs}.
            The decay volume was further filled with helium bags which were not included in the \alpi model.
    
        \begin{table}[h]
        \centering
        \begin{tabular}{ccccccccccc}
                Experiment  & Status    & $E_\mathrm{beam}$ & $N_\mathrm{PoT}$  & Target & $l_\mathrm{n,DV}$ & $l_\mathrm{c,DV}$ & $z_\mathrm{DV}$ & $\theta_\mathrm{off}$ & $\Omega_\mathrm{cov}$\\
                            &           & [GeV]             & [$10^{18}$]       &        & [m]               & [m]               & [m]             & [mrad]                & [$\SI{}{\micro sr}$]\\
                \hline
                CHARM       & completed             & 400 & 2.4  & Cu & 35  & 35  & 480 & 10 & 34   \\
                BEBC        & completed             & 400 & 2.72 & Cu & 2.5 & 2.5 & 404 & 0  & 53   \\
                NuCal       & completed             & 70  & 1.7  & Fe & 23  & 23  & 64  & 0  & 700  \\
                NuTeV       & completed             & 800 & 2.54 & BeO & 34 & 34  & 1400& 0  & 4.3  \\
                NA62        & running               & 400 & 1    & Cu & 139 & 81  & 82  & 0  & 84   \\
                KOTO        & running               & 30  & 2.2  & Au & 3.1  & 3.1 &23.9&280 & 4300 \\
                DarkQuest-I  & in prep.~\cite{gori} & 120 & 1.44 & Fe & 13.5& 1 & 5  & 0  & 12000\\
                DarkQuest-II & in prep.~\cite{gori} & 120 & 100  & Fe & 13.5& 7 & 12 & 0  & 12000\\
                DUNE ND     & in prep.       & 120 & 1100 & C  & 10  & 10  & 575 & 0  & 36   \\
                SHiP        & in prep.       & 400 & 600  & Mo & 60.5& 50  & 33.5& 0  & 3900 \\
                HIKE        & rejected~\cite{vallee}& 400 & 15   & Cu & 139 & 81  & 82  & 0  & 84   \\
                SHADOWS     & rejected~\cite{vallee}& 400 & 15   & Cu & 23  & 20  & 10 &  69 & 4800 \\
                \hline
        \end{tabular}
        \caption{Simplified geometries of several experiments assumed in \alpi, as well as their respective statuses.}
        \label{tab:experiment_geometries}
    \end{table}
    
    \subsection{Validation} \label{sec:ALPINIST_validation}

        The kinematic treatment of the \alpi simulation framework has already been studied extensively in previous works~\cite{Jerhot:2022chi,Afik:2023mhj}.
        A natural way to validate that the set-up is compatible with the simulation of HNLs is a comparison to previously published data, which will be the main focus of this section.

    \subsubsection{Comparison to BEBC}\label{sec:BEBC_comparison}

        One of the most stringent limits on HNL mixing parameters in the mass regime around $m_\mathrm{N}\simeq\SI{1}{GeV}$ comes from the WA66 experiment at the Big European Bubble Chamber using the CERN west area beam-dump~\cite{WA66:1985mfx}. 
        The original analysis searched for electronphilic and muonphilic HNLs produced in the two-body decays of $D^\pm$ mesons, and decaying only via charged current interactions, and was later recast as limits for tauphilic HNLs decaying also in neutral current interactions~\cite{Barouki:2022bkt}.
        Using the simplified geometry as described in section~\ref{sec:BEBC_experiment}, the BEBC results were interpreted in the \alpi framework.  
        
        In an effort to recreate the published exclusion limits, we assumed HNL production from the decays $D^\pm\to\ell^\pm\mathrm{N}$ with a $D$ meson spectrum parameterised by eq.~(\ref{eq:diff_meson_parametrisation}) with $n=4,\,a=2,\,b=0$.
        In the original search, the $D$ meson spectrum was normalised to the measured neutrino flux. 
        As the numerical value of the normalisation was not published, we use instead the $\sigma_{c\bar c}=\SI{15.1}{\micro\barn}$ production cross section as measured by the LEBC-EHS collaboration~\cite{LEBC-EHS:1988oic}, including a cascade production factor of $2.3$ in accordance with cascade evaluations by the SHiP collaboration~\cite{CERN-SHiP-NOTE-2015-009}. 
        Consequently, the produced HNLs within geometric acceptance of BEBC were allowed to decay in charged current interactions.
        On the resulting final products the kinematic cuts as performed by the WA66 search were applied.
        These can be summarised in natural units as 
        \begin{enumerate}
            \setlength\itemsep{0em}
            \setlength{\parskip}{0pt}
            \item $\SI{1}{GeV}< \sum p_\mathrm{track}$
            \item $\SI{0.8}{GeV}<p^e_\mathrm{track}$, and  $\SI{3}{GeV}<p^\mu_\mathrm{track}$
            \item $p_T + \sqrt{p_T^2 + M_\mathrm{vis}^2} < m_{D^\pm} - m_\ell$,
        \end{enumerate}
        where $p_\mathrm{track}$ is the total momentum associated with a charged track, $c$ the speed of light, $M_\mathrm{vis}$ is the invariant mass of the charged tracks and $p_T$ the modulus of their combined momentum transverse to the beam direction.
        The limit determined by the $D^\pm$ ($m_{D^\pm}$), and lepton mass ($m_\ell$) are chosen depending on which lepton family the HNL couples to.
        Requiring the minimum momenta for the charged tracks results in combined detection and identification efficiencies of $\SI{89}{\%}$ for final states including two muons, and else $\SI{88}{\%}$.

        \begin{figure}
            \centering
            \includegraphics[width=.95\textwidth]{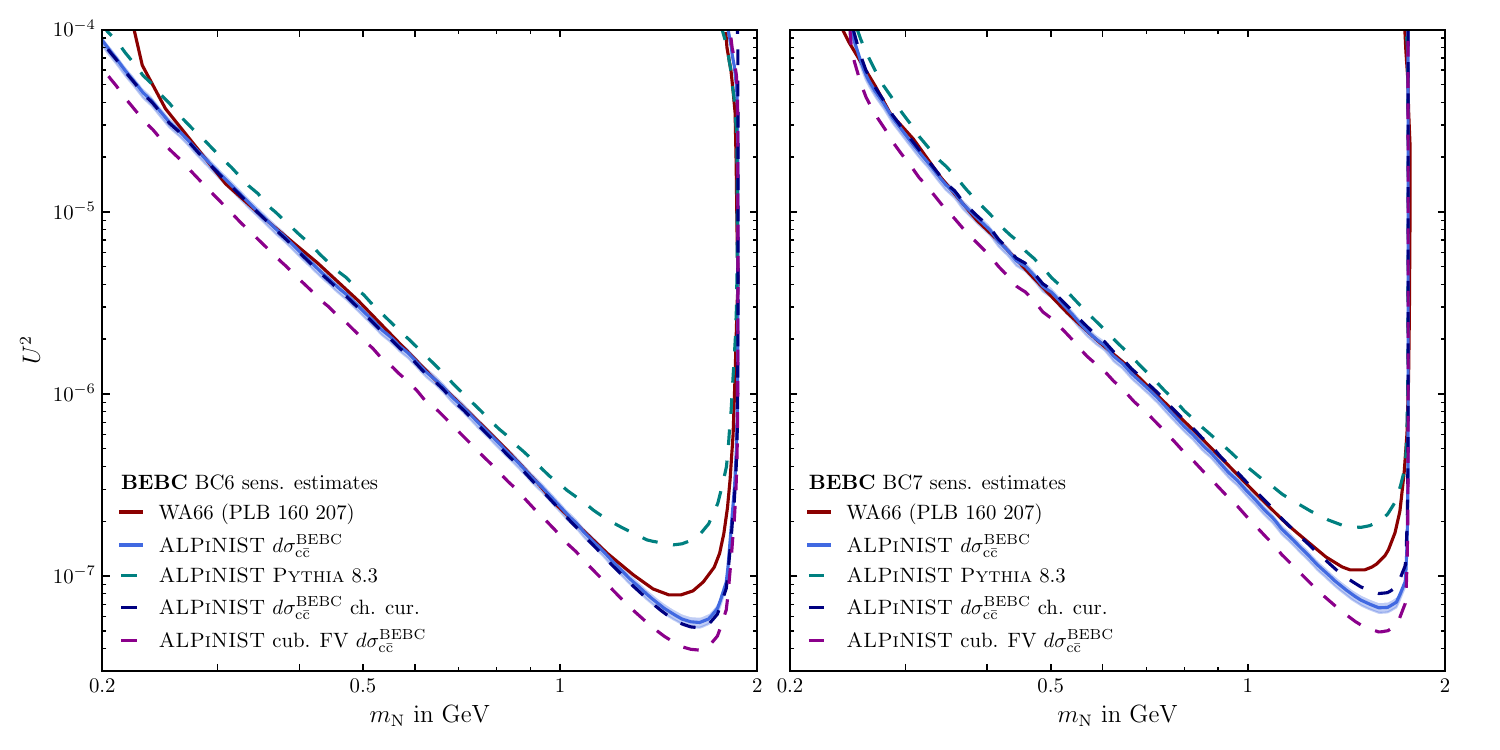}
            \caption{BEBC published (red) and \alpi recast exclusion bounds for an electronphilic (left) and muonphilic (right) HNL produced in $D^\pm \to \mathrm{N}\ell^\pm$, and decaying to $\nu \ell\ell,\,\nu \ell\ell^\prime,\text{ and }\pi\ell$. 
            The blue curves were generated using charmed meson distributions quoted by WA66 as input.
            The dark blue dashed curve shows an HNL only decaying in charged currents, while for the light blue also neutral current interactions are permitted.
            The teal dashed curve uses meson inputs generated with \pyth's \texttt{HardQCD:hardccbar} in standard settings. 
            The magenta dashed curve uses a cuboid fiducial volume, as proposed in a recent recast of the BEBC results~\cite{Barouki:2022bkt}.
            For further assumptions please find the surrounding text.
            }
            \label{fig:BEBC_comparison}
        \end{figure}

        Figure~\ref{fig:BEBC_comparison} compares the published data by the WA66 collaboration~\cite{WA66:1985mfx} to the recast performed with \alpi in terms of an electronphilic (left) and muonphilic (right) HNL produced only in $D^\pm \to \mathrm{N}\ell^\pm$, and decaying only to $\nu \ell\bar\ell,\,\nu \ell\bar\ell^\prime,\text{ and }\pi\ell$ and charge conjugate processes, where $\ell$ is a light lepton ($e$ and $\mu$).
        The curves correspond to a parameter exclusion at $\SI{90}{\%\,CL}$, where the confidence level is evaluated with one event having been observed after cuts with $0.6$ expected background events for the muonphilic HNL, while for the electronphilic HNL no such event was observed.
        The recast reproduces the published curves very well for most of the covered parameter space. 
        Disagreements can be observed for both benchmark cases at the upper mass end of the kinetically allowed region, where \alpi shows a slightly more optimistic sensitivity.
        For the electronphilic case, the two curves also diverge at the lowest masses presented, where the published sensitivity drops significantly when approaching the dimuon threshold. 
        This is unexpected, as this threshold should not correspond to a significant kinematic limit for the electronphilic case.
        In both cases, using \pyth generated kinematic distributions as the $D^\pm$ meson inputs (teal dashed curve) leads to a less restrictive exclusion bound.
        The magenta dashed curve represents the same assumptions as the blue curve, but instead employs a cuboid fiducial volume, as proposed in a recent recast of the BEBC results~\cite{Barouki:2022bkt}.
        Lacking information about the uncertainties of the meson distribution used by BEBC, we varied only the overall scaling according to the neutrino flux uncertainty of $\SI{14}{\%}$ in accordance with values cited by the CHARM experiment~\cite{CHARM:1985nku} which shared the same target resulting in the blue band.
        We see that the cuboid approximation yields a more optimistic parameter reach compared to the smaller cylindrical fiducial volume implemented in \alpi.
        
    \subsubsection{Comparison to NuTeV} \label{sec:NuTeV_comparison}
    
        The most recent search for HNLs at a beam-dump facility comes from the NuTeV collaboration~\cite{NuTeV:1999kej}, which focused on HNLs mixing with the SM muon neutrino. 
        This analysis was implemented in \alpi using the simplified geometry described in section \ref{sec:NuTeV_experiment}.
        To reproduce their findings we assume the production cross section ($\sigma_{c\bar c}=\SI{76}{\micro \barn}$) and meson distributions (parameterised by eq.~(\ref{eq:diff_meson_parametrisation}) with $n=6.8,\,a=0,\,b=0.81$) as measured by the E653 experiment~\cite{FermilabE653:1991vmo}.\footnote{The discrepancy between the cross sections measured in the emulsion~\cite{FermilabE653:1991vmo} and bubble chamber~\cite{Ammar:1988ta} experiments may highlight the importance of including cascade effects to the production cross section as presented in ref.~\cite{CERN-SHiP-NOTE-2015-009}.}
        We further assumed the same kinematic distribution for $D_s$ mesons using the $D_s(D^++D^0)^{-1}$ ratio of $0.2$ as given by the cited theory reference~\cite{Frixione:1994nb}. 
        Finally, we apply the full kinematic cuts on the final decay products as listed by the NuTeV search~\cite{NuTeV:1999kej}. 
        For this we define, $E_\mathrm{track}$ as the energy of final state particles, the visible final states' invariant mass $M_\mathrm{vis}$ and total transverse momentum $p_T$, the missing energy $E_\mathrm{miss}$, and the momentum transfer to the hardest final state muon $Q_\mathrm{vis}$.
        The cuts can then be summarised as
        \begin{enumerate}
            \setlength\itemsep{0em}
            \setlength{\parskip}{0pt} 
            \item $\SI{2}{GeV}<E^\mu_\mathrm{track}$, and  $\SI{10}{GeV} < E^e_\mathrm{track}, E^\pi_\mathrm{track}$
            \item $M_\mathrm{vis}^2 \times (2m_pQ_\mathrm{vis})^{-1} < 0.1$
            \item $m_p^2 + 2m_pE_\mathrm{miss} - Q_\mathrm{vis}^2 > \SI{4.0}{GeV^2}$
            \item $p_T + \sqrt{p_T^2 + M_\mathrm{vis}^2} < \SI{3.0}{GeV}$,
        \end{enumerate}
        in natural units, where $m_p$ is the proton mass.
        
        \begin{figure}
            \centering
            \includegraphics[width=.65\textwidth]{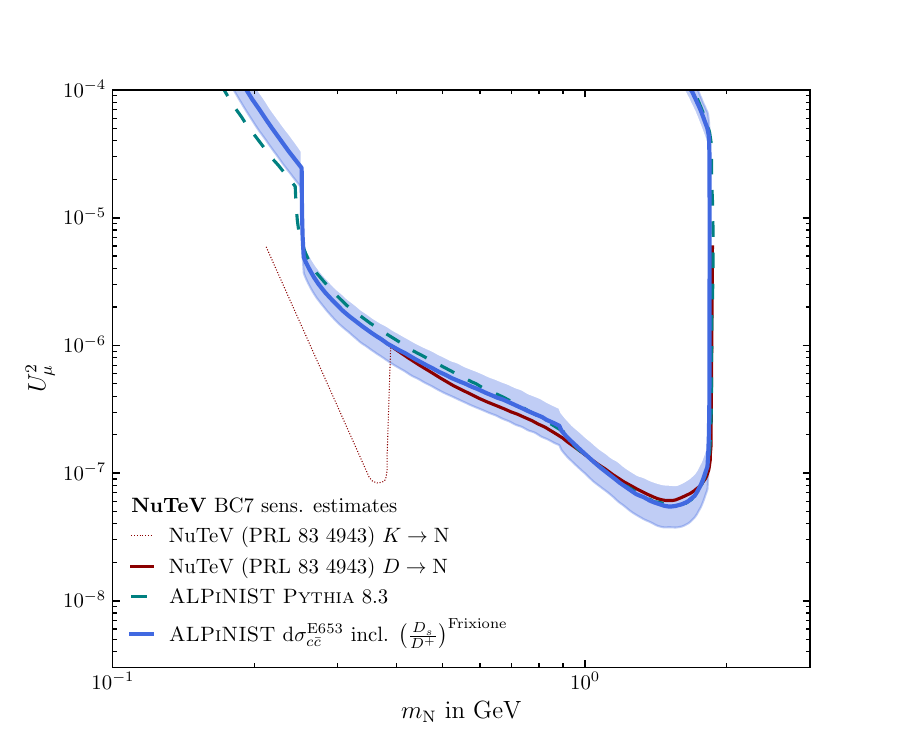}
            \caption{NuTeV published (red) and \alpi recast exclusion bounds for a muonphilic HNL. 
            The blue curve was generated using E653 charmed meson distributions~\cite{FermilabE653:1991vmo} as input, and varying the relevant values within their $1\sigma$ bounds.
            The teal dashed curve uses meson inputs generated with \pyth's \texttt{HardQCD:hardccbar} in standard settings. 
            }
            \label{fig:NuTeV_comparison}
        \end{figure}
        
        Figure~\ref{fig:NuTeV_comparison} shows the comparison of the published data by the NuTeV collaboration~\cite{NuTeV:1999kej} to the recast performed with \alpi in terms of a muonphilic HNL, which was evaluated for the final states $\nu e\mu,\nu\mu\mu,\pi\mu\text{, and }\rho\mu$.
        The curves show the HNL parameter exclusion at $\SI{90}{\%CL}$. 
        In accordance with the NuTeV results, we assume no data event after cuts with an expected background of $0.57$ events.
        Below $\sim\SI{400}{MeV}$, the NuTeV sensitivity is dominated by HNLs produced in Kaon decays. 
        As described in appendix~\ref{sec:HNL_interactions}, we do not consider this production channel, which is why the curves diverge. 
        Above this threshold, however, the curves agree well. 
        We also present the uncertainties of the experimental inputs by varying the input parameters in the listed confidence levels, shown as the blue band. 
        Lastly, we ran the same simulation set-up with inputs coming from \pyth's \texttt{HardQCD:hardccbar}. 
        This is shown as the green dashed curve and agrees well with the published curve near the kinematic limit.
        However, towards lower masses we observe a weaker sensitivity than reported. 
        We note that the discrepancy is well within the bounds coming from input uncertainties as presented for the curve based on empirical inputs.
    
\section{Results} \label{sec:res}

    \subsection{Impact on Feebly Interacting Particle sensitivities} \label{sec:Impact_CHARM_HNL_spectra}

        Combining all the items covered in section~\ref{sec:distributions} (and also appendix~\ref{sec:leading_particle_effect}), we have a plethora of different parameters to play with when describing heavy meson distributions at beam-dump experiments.
        As several Feebly Interacting Particles could be produced in the decays of these heavy mesons, the choice of input parameters has a direct impact on the expected sensitivity of a given experiment to these FIPs.
        Here, we would like to give a brief idea on the impact this choice has on the resulting sensitivity. 
        For this, we compare the sensitivity of the CHARM experiment~\cite{BERGSMA1983361,CHARM:1985nku} to a muonphilic HNL.
        The experiment had received data corresponding to $2.4\times 10^{18}$ dumped protons, the interaction products of which were analysed by a detector array ca. $\SI{480}{m}$ downstream from the CERN WA beam-dump (see appendix~\ref{sec:CHARM_validation} for further details).
        It was mostly sensitive to HNLs produced in open charmed meson decays, making it an ideal candidate for an isolated impact study of charmed meson distributions. 
        
        To evaluate the sensitivity, and consequently the parameter space excluded by the absence of a signal, we use the \alpi framework (see section~\ref{sec:ALPINIST}).
        We compare different meson distributions generated with \pyth namely
        \begin{enumerate}[\roman{enumi}.]
            \setlength{\itemsep}{0pt}
            \setlength{\parskip}{0pt}
            \item \pyth~8.309 in standard settings (\texttt{hardQCD:hardccbar}),
            \item a statistical combination of tables generated with \pyth~8.3's \texttt{hardQCD:hardccbar} and \texttt{hardQCD:3parton} (c.f. text around figure~\ref{fig:23cc_ppScatter}),
            \item the distributions from the cascade production simulation by the SHiP collaboration~\cite{CERN-SHiP-NOTE-2015-009}.
        \end{enumerate}
        These are contrasted by results using the empirical distribution given in eq.~(\ref{eq:diff_meson_parametrisation}) with parametrisations as proposed by the CHARM collaboration\footnote{Note that this parametrisation ($n=5,\,a=2,\,b=0$) differs from the dedicated analysis conducted later~\cite{Bergsma:1987br} (see table~\ref{tab:diffxs_cc}), but was quoted as the underlying distribution for the final HNL search~\cite{CHARM:1985nku}.} and distributions measured by the LEBC-EHS collaboration~\cite{LEBC-EHS:1988oic} (see table~\ref{tab:NA27_results}).
        These differential cross sections are normalised to match the respective production rates $\chi_{pp\to c\bar c}=\sigma_{c\bar c}^{}\sigma_{pp}^{-1}$.
        To account for the averaged proton-nucleon cross section we apply an enhancement of $A_\mathrm{Cu}^{1/3}$~\cite{Lourenco:2006vw, Carvalho:2003pza}.
        For standard \pyth~8.3 this is $\chi^{2\to 2}_{pp\to c\bar c} = 4.0\times 10^{-4}$ which increases to  $\chi^{2\to 2,3}_{pp\to c\bar c} = 7.2\times 10^{-4}$  when combining it with \texttt{hardQCD:3parton} production.
        Even though the SHiP cascade production is also based on \pyth, we use the commonly referenced value of $\chi^\mathrm{SHiP}_{pp\to c\bar c}=4\times 10^{-3}$~\cite{PhysRevD.104.095019,Ovchynnikov:2023cry}.
        We previously found that the flux of neutrinos at the CERN WA beam-dump was consistent with $\chi_{pp\to c\bar c} = 3.5\times 10^{-3}$ (see section~\ref{sec:BEBC_comparison}). 
        As the original CHARM search gives no numerical value for the charm production cross section or rate~\cite{CHARM:1985nku}, we will adopt this value as the $\chi_{pp\to c\bar c}^\mathrm{CHARM}$ rate.
        For the LEBC-EHS distributions, we have $\chi^\mathrm{LEBC-EHS}_{pp\to c\bar c} = 1.5\times 10^{-3}$~\cite{LEBC-EHS:1988oic, ROPP2022}.
        
        \begin{figure}
            \centering
            \includegraphics[width = 0.95\textwidth]{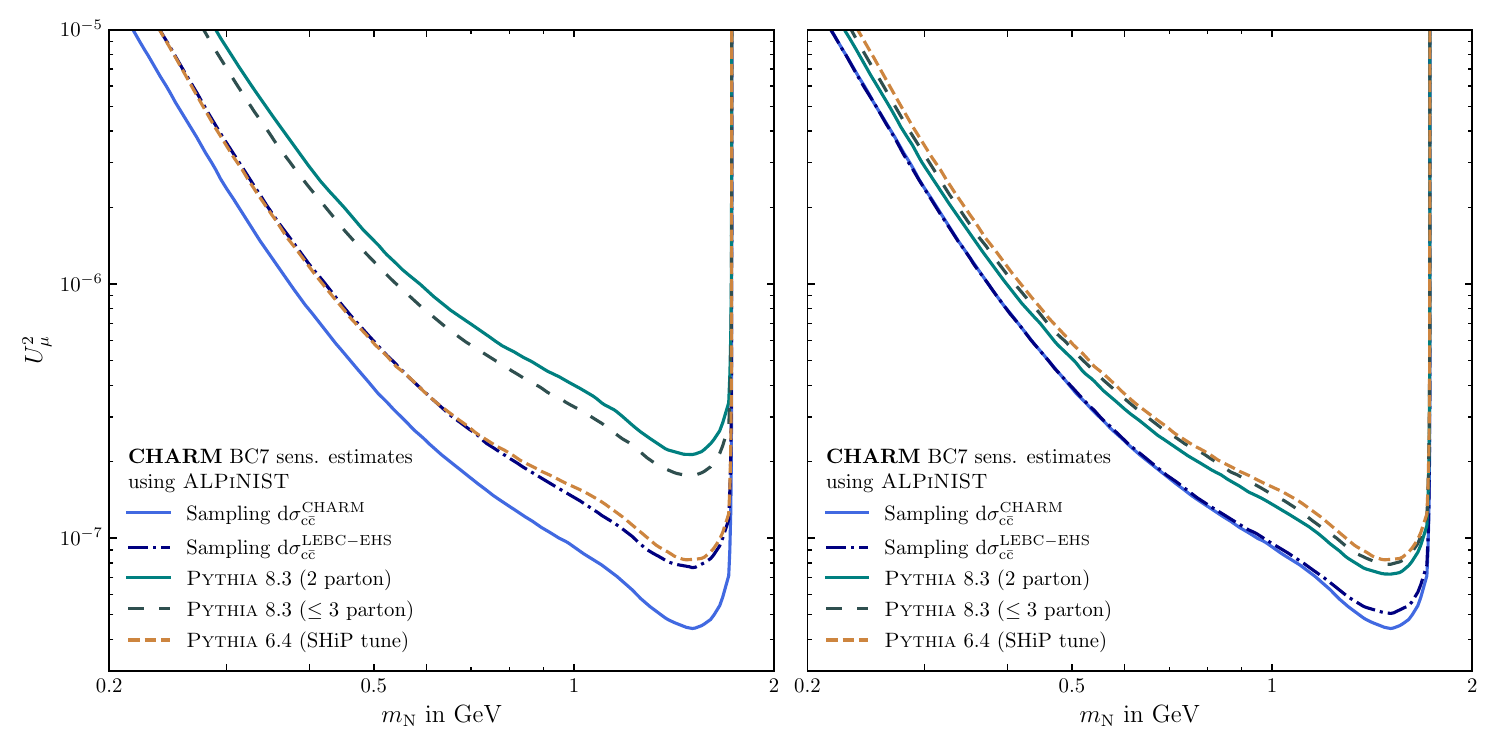}
            \caption{CHARM experiment sensitivity ($90\,\%\,\mathrm{CL}$ excl. bound) to a muonphilic HNL evaluated for different charmed meson distribution inputs.
            The case of employing the respective production rates associated with the distribution (left) is compared to the case of a common normalised production rate (right). 
            For further information see text.}
            \label{fig:CHARM_BC7_impacts}
        \end{figure}

        The resulting sensitivity estimates for a muonphilic HNL at the CHARM experiment are shown in figure~\ref{fig:CHARM_BC7_impacts}.
        The left panel shows the results for meson spectra generated as described above, while the right tile shows the results of a normalised production rate of charmed mesons of $\chi_{pp\to c\bar c} = 3.5\times 10^{-3}$ for all spectra.
        Including the differences in integrated cross section $\sigma_\mathrm{c\bar c}$ (figure~\ref{fig:CHARM_BC7_impacts} left), we observe differences in the expected sensitivity up almost an order of magnitude.
        The differences due to production cross section are a constant factor (with respect to $m_\mathrm{N}$) between the sensitivity curves.
        When normalising $\chi_{pp\to c\bar c}$ (figure~\ref{fig:CHARM_BC7_impacts} right), the differences between curves represent only the impact of the underlying $D$ meson kinematics.\footnote{The maximum difference between the fragmentation functions was $\sim17\,\%$, resulting in a subleading difference in sensitivity of $\sim 8\,\%$.}
        For $m_\mathrm{N}<\SI{0.5}{GeV}$, the difference between the curves is of the order of $2$. 
        When approaching the kinematic limit, the gap between \pyth generated mesons and sampled from empirical distributions widens reaching a maximum of around a factor of $4$.
        This shows, that while the integrated production cross section has an obvious impact on the sensitivity estimate, the substantial contribution of the associated kinematic distributions must not be overlooked.
        
    \subsection{Beyond benchmark cases} \label{sec:ternary_comparisons}
        
\begin{figure}
                \centering
                \includegraphics[width=.95\textwidth]{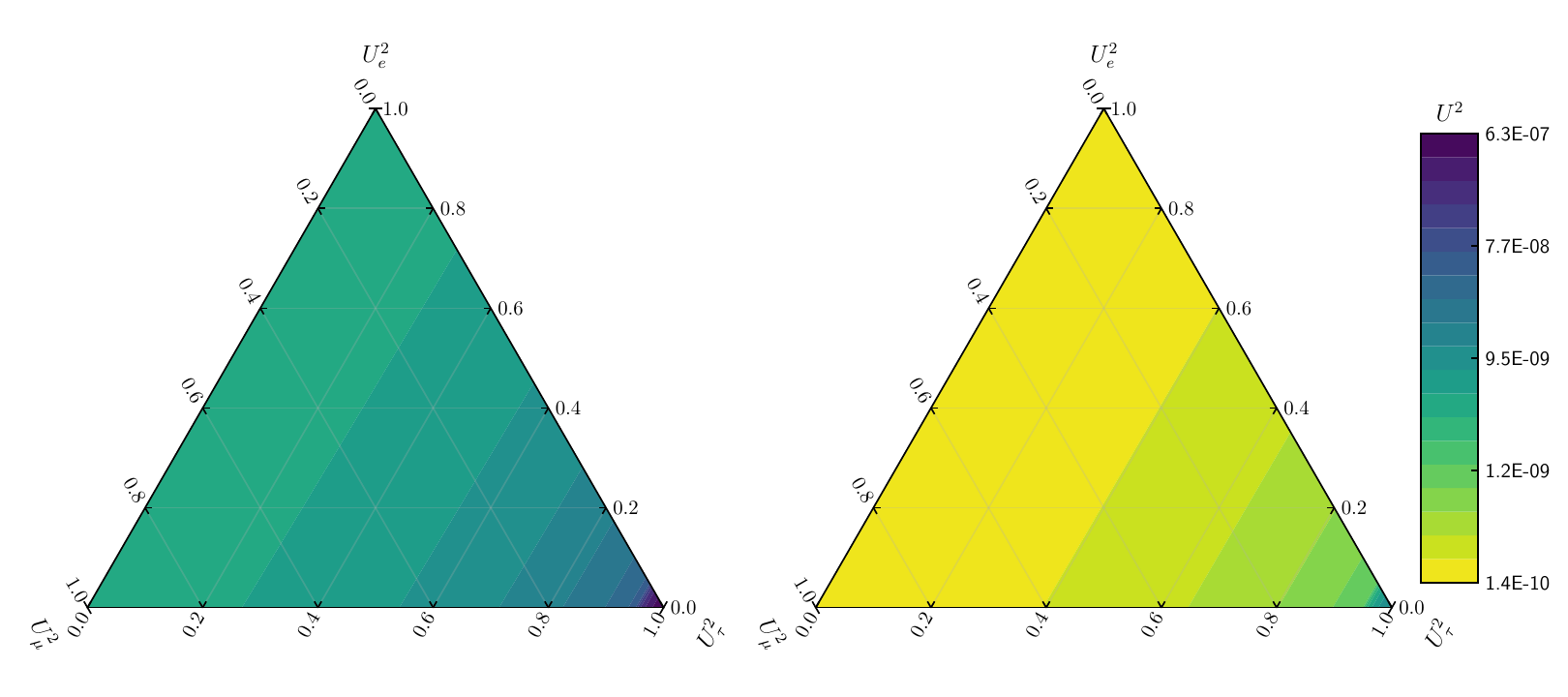}
                \caption{
                The DUNE-ND (left) and SHiP (right) predicted HNL sensitivities ($90\,\%\,\mathrm{CL}$ excl. lower bound) for different coupling combinations assuming $m_\mathrm{N}=\SI{1.5}{GeV}$ and charmed meson distributions as given by \pyth~8.3.
                }
                \label{fig:ternary}
            \end{figure}

        A common way to present HNL sensitivities without having to resort to benchmark case scenarios is the ternary representation at a fixed mass.
        The three axes grid indicates ratios between the individual leptonic couplings $U_\alpha^2$, while the maximal sensitivity to $U^2=U_e^2+U_\mu^2+U_\tau^2$ can be shown in colour scale. 
        Figure~\ref{fig:ternary} shows the expected $U^2$ $\SI{90}{\%\,CL}$ lower bound sensitivity limit for the future DUNE Near Detector~\cite{Berryman:2019dme,DUNE:2021tad} and SHiP~\cite{SHiP2023} experiments in such a representation for an HNL with mass $m_\mathrm{N}=\SI{1.5}{GeV}$, with logarithmic colour scale.
        The predicted sensitivities for both experiments follow a similar pattern, where in the case of coupling to light leptons ($e$ and $\mu$ along the top left axis) the sensitivity is roughly similar, while it strongly decreases as the $\tau$ coupling component becomes more dominant (bottom right corner). 
            
\begin{figure}
                \centering
                \includegraphics[width=.495\textwidth]{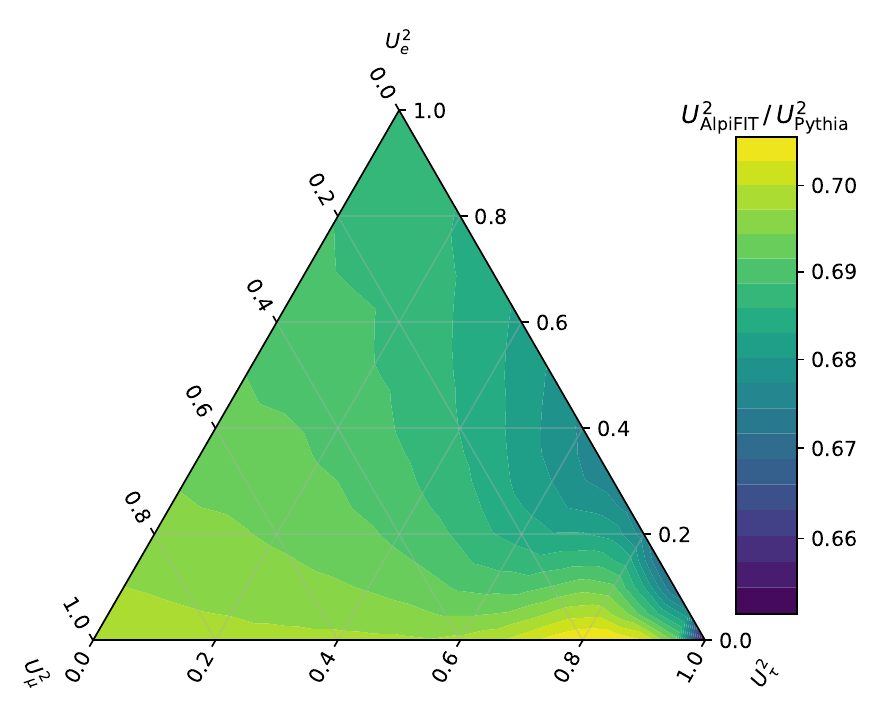}
                \includegraphics[width=.495\textwidth]{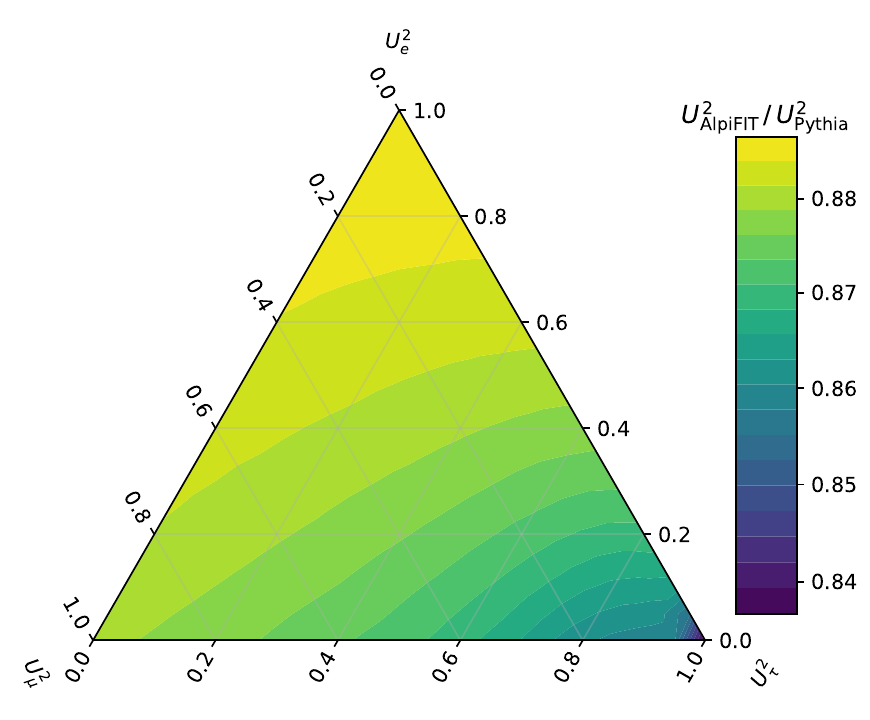}
                \caption{
                Impact on the underlying meson distribution choice on the DUNE-ND (left) and SHiP (right) predicted HNL sensitivities ($90\,\%\,\mathrm{CL}$ excl. lower bound) for different coupling combinations assuming $m_\mathrm{N}=\SI{1.5}{GeV}$. 
                The sensitivity calculated with meson distributions as given by the central fit values of the parametrisation presented in section~\ref{sec:dsigma_qq} is normalised to the sensitivity based on \pyth~8.3 inputs shown in figure~\ref{fig:ternary}. 
                }
                \label{fig:ternary_comparison}
            \end{figure}

        This representation also allows us to investigate the impact of meson distribution choice for different coupling combinations.
        For this, we calculate expected sensitivities based on the fit result  central values for the parametrisation of the differential meson production cross section described by eq.~(\ref{eq:diff_meson_parametrisation}) found in section~\ref{sec:dsigma_qq}. 
        Consequently, we normalise the resulting expected sensitivity to the values found by using \pyth~8.3 to generate the meson distributions shown in figure~\ref{fig:ternary}.
        The resulting sensitivity ratios are presented in figure~\ref{fig:ternary_comparison} for the proposed DUNE-ND and SHiP experiments.
        Both experiments are expected to be more sensitive when considering the empirical meson production cross sections. 
        We attribute this to the fact that \pyth estimates harder $p_T$ meson spectra than observed in data (c.f. figure~\ref{fig:DX_cc_fit} right) leading to more HNLs going out of acceptance.
        This effect is less pronounced in the baseline SHiP case, as the large angular coverage means that few HNLs go out of acceptance even in the \pyth case.
        This effect is especially pronounced in the $\tau$-philic case, as the production through intermediate $\tau$ leptons exacerbate this effect.
        Furthermore, it is important to note that the standard deviation in $1-U^2_\mathrm{AlpiFIT}\,/\,U^2_\mathrm{Pythia}$ is of the order of few to several percent with respect to the mean for both experiments.\footnote{Indeed, this is the case for all experiments presented in this work.} 
        Therefore, the impact of the meson spectra on the expected sensitivity can be viewed to first order as a coupling independent effect.
        This is especially true in the regime where $\tau$ coupling is minimal, as here kinematically similar channels dominate in the mass regimes relevant to beam-dump searches.
        
    \subsection{Impact on the sensitivity landscape}

        As the example case presented in section~\ref{sec:Impact_CHARM_HNL_spectra} shows, the choice of heavy meson (differential) production cross sections greatly impacts the sensitivity expectation of beam-dump style experiments to HNLs. 
        In the absence of perfectly perfectly reliable experimental data for these cross sections, a comparison between experiments with the same underlying biases is warranted. 
        \alpi is uniquely suited for this task as it  
        \begin{enumerate}[\roman{enumi}.]
            \setlength{\itemsep}{0pt}
            \setlength{\parskip}{0pt}
            \item reliably reproduces existing experimental bounds (see section~\ref{sec:ALPINIST_validation}),
            \item easily adapts underlying meson distributions,
            \item analyses the different experiments in the same underlying theory framework.
        \end{enumerate}
        In this section we present the collective past exclusions, current parameter reaches and future projected sensitivities to the most commonly considered HNL benchmark cases (BC6:$e$-philic, BC7:$\mu$-philic, BC8:$\tau$-philic) of beam-dump-style experiments derived under the assumptions established in this work.
        For this, we compare the past BEBC~\cite{WA66:1985mfx}, CHARM~\cite{CHARM:1985nku}, NuTeV~\cite{NuTeV:1999kej}, the ongoing NA62~\cite{NA62:2023qyn,NA62:2023nhs} in beam-dump mode (NA62-bd) and DarkQuest Phase-I~\cite{Batell:2020vqn}, and future DarkQuest Phase-II~\cite{Batell:2020vqn}, DUNE near detector~\cite{Berryman:2019dme,DUNE:2021tad}, and SHiP~\cite{SHiP2023} experiments.

\begin{figure}[h]
            \centering
            \includegraphics[width=.95\textwidth]{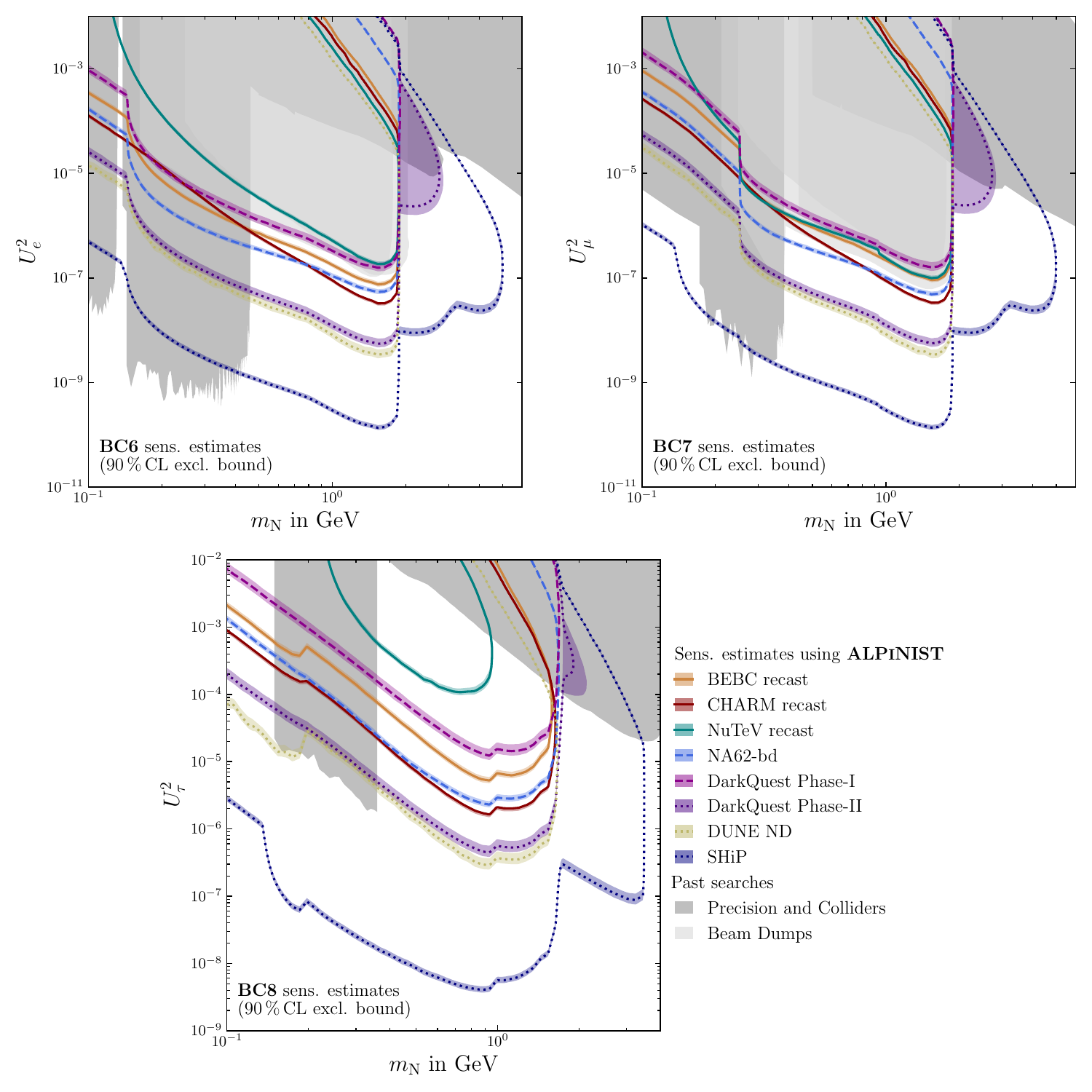}
            \caption{Sensitivities ($90\,\%\,\mathrm{CL}$ excl. bound)  of various past (solid lines) present (dashed lines) and future (dotted lines) beam-dump style experiments to Benchmark Case scenario HNLs (electron-philic top left, muon-philic top right, tau-philic bottom). 
            The underlying meson distributions are generated using \pyth~8.3, with quark level cross sections $\sigma_{q\bar q}$ as given in table~\ref{tab:qqbar_production_crosssection}. 
            The coloured bands surrounding the lines correspond to varying $\sigma_{q\bar q}$s within the $68\%\,\mathrm{CL}$ bounds.
            The grey shaded area is parameter space excluded by light meson decay precision experiments~\cite{PS191:1987ek,E949:2014gsn,KEK:1982wu,KEK:1984sj,PIENU:2019usb,T2K:2019jwa,NA62:2020mcv,NA62:2021bji}, as well as collider based experiments~\cite{DELPHI:1996qcc,BESIII:2019oef,CMS:2022fut,CMS:2024ake,CMS:2024ita}.
            The light grey shaded regions represent the parameter space excluded by beam-dump searches~\cite{CHARM:1985nku,BEBCWA66:1986err,NuTeV:1999kej}.
            For further details see text. 
            }
            \label{fig:pythia_HNLbd_overview}
        \end{figure}

\begin{figure}[h]
            \centering
            \includegraphics[width=.95\textwidth]{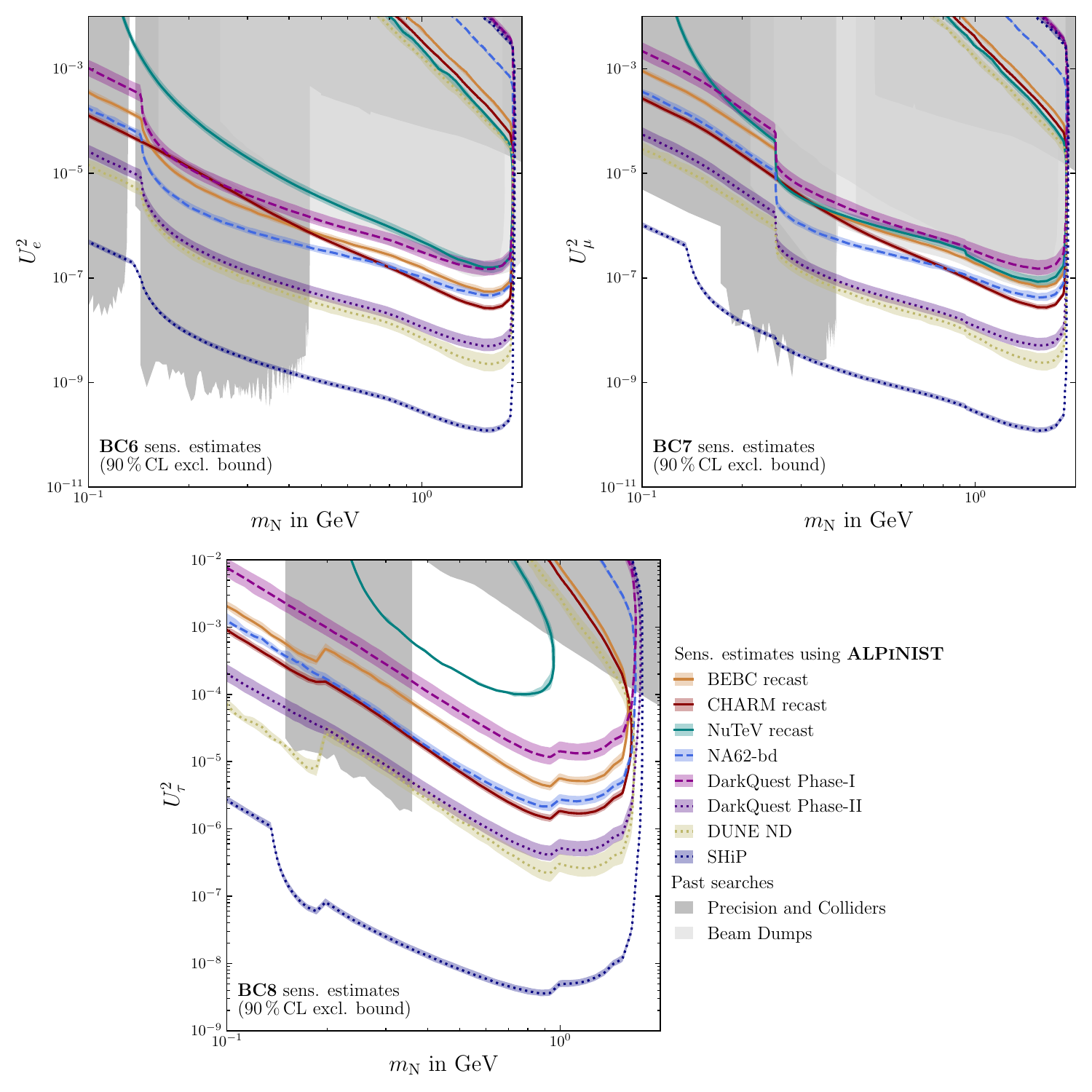}
            \caption{Sensitivities ($90\,\%\,\mathrm{CL}$ excl. bound) of various past (solid lines) present (dashed lines) and future (dotted lines) beam-dump style experiments to Benchmark Case scenario HNLs (electron-philic top left, muon-philic top right, tau-philic bottom). 
            The underlying meson distributions are generated using the parametrisation eq.~(\ref{eq:diff_meson_parametrisation}) with parameters as given in table~\ref{tab:DX_cc_fit}, and quark level cross sections $\sigma_{c\bar c}$ as given in table~\ref{tab:qqbar_production_crosssection}.
            The coloured bands surrounding the lines correspond to varying $\sigma_{c\bar c}$, $b_{c\bar c}$, and $n_{c\bar c}$ within the $68\%\,\mathrm{CL}$ bounds.
            The limitation to charmed meson production is due to the lack of experimental beauty meson data.
            The grey shaded area is parameter space excluded by light meson decay precision experiments~\cite{PS191:1987ek,E949:2014gsn,KEK:1982wu,KEK:1984sj,PIENU:2019usb,T2K:2019jwa,NA62:2020mcv,NA62:2021bji}, as well as collider based experiments~\cite{DELPHI:1996qcc,BESIII:2019oef,CMS:2022fut,CMS:2024ake,CMS:2024ita}.
            The light grey shaded regions represent the parameter space excluded by beam-dump searches~\cite{CHARM:1985nku,BEBCWA66:1986err,NuTeV:1999kej}.
            For further details see text.
            }
            \label{fig:FIT_HNLbd_overview}
        \end{figure}

        Figures~\ref{fig:pythia_HNLbd_overview} and \ref{fig:FIT_HNLbd_overview} show such comparisons of the above experiments for different meson input assumptions.
        These figures show 
        \begin{itemize}
            \setlength{\itemsep}{0pt} 
            \setlength{\parskip}{0pt} 
            \item parameter space which is presently excluded by light meson decay precision experiments~\cite{PS191:1987ek,E949:2014gsn,KEK:1982wu,KEK:1984sj,PIENU:2019usb,T2K:2019jwa,NA62:2020mcv,NA62:2021bji} and collider based experiments~\cite{DELPHI:1996qcc,BESIII:2019oef,CMS:2022fut,CMS:2024ake,CMS:2024ita} as grey shaded regions,
            \item parameter exclusion coming from beam-dump searches for HNLs~\cite{CHARM:1985nku,BEBCWA66:1986err,NuTeV:1999kej} as light grey shaded regions,
            \item \alpi recasts of these concluded beam-dump experiment searches with no significant signal events (i.e. already excluded parameter space within the experimental sensitivity) as solid lines,
            \item \alpi expectations for sensitivities of running experiments as dashed lines,
            \item \alpi sensitivity predictions for future experiments based on the respective baseline scenarios  represented as dotted lines.
        \end{itemize}
        
        We present all curves as the $90\%\,\mathrm{CL}$ exclusion limits, which for most experiments corresponds to an expected number of $n_{90\%\,\mathrm{CL}}=2.3$ events (i.e. background free assumption). 
        The exception to this are the DarkQuest and DUNE experiments which estimate $n_{90\%\,\mathrm{CL}}=10$~\cite{Batell:2020vqn,Berryman:2019dme}, and BEBC in the BC7 case where $n_{90\%\,\mathrm{CL}}=3.45$ due to observed events~\cite{WA66:1985mfx}.
        The shaded regions surrounding the curves of the same colour correspond to a variation of input assumptions within $68\%\,\mathrm{CL}$ limits as described below.
        
        Figure~\ref{fig:pythia_HNLbd_overview} presents these sensitivities with \pyth~8.309 in standard settings as a generator for the heavy mesons, while the overall meson production cross sections are the fit results on experimental data taken from section~\ref{sec:sigma_qq} (summarised in table~\ref{tab:qqbar_production_crosssection}) without cascade production factors.
        The uncertainties, represented by bands of the same colour as the lines, correspond to the propagated variation of the production cross sections within the $68\%\,\mathrm{CL}$ fit limits. 
        To the authors' knowledge, for the first time we show the NuTeV sensitivities in the BC6 and BC8 plots.
        This is possible as -- even though NuTeV was only sensitive to final states containing muons and therefore mostly BC7 -- the sensitivity to BC6 and BC8 stems from the final states $\nu \mu\mu$ being available through neutral current interactions, as well as $\nu e\mu$ also being a charged current mediated final state of BC6.
        For the NuTeV experiment we also note that using the charmed meson production as shown in table~\ref{tab:qqbar_production_crosssection} lowers the exclusion power in the case of BC7 with respect to the value used in the original search~\cite{NuTeV:1999kej}, but also relative to the BEBC and CHARM searches. 
        This is also due to, as already shown in recent recasts~\cite{Barouki:2022bkt,PhysRevD.104.095019}, the parameter exclusion in all mixing scenarios by the BEBC and CHARM experiments being stronger than originally published~\cite{WA66:1985mfx, CHARM:1985nku}.\footnote{As we are not able to fully reproduce the published CHARM limits with the assumptions given in the paper (c.f. appendix~\ref{sec:CHARM_validation}) the related exclusion limits should be taken with a grain of salt.} 
        We find the general hierarchy between the planned future experiments, as well as the order of improvement with respect to present and past experiments to be similar to what is commonly depicted in literature. 

        Figure~\ref{fig:FIT_HNLbd_overview}, on the other hand, shows the same sensitivities with the same cross sections, but with the heavy meson differential distributions sampled from the empirical distribution eq.~(\ref{eq:diff_meson_parametrisation}) with experimentally motivated parameters as established in section \ref{sec:dsigma_qq}. 
        The presented uncertainties represent the propagated variation of both the production cross section and the parameters entering eq.~(\ref{eq:diff_meson_parametrisation}) within the respective fit uncertainties.
        The multiplicities of the different open meson species per $q\bar q$ event are taken from \pyth~8.309 due to a lack of data.
        Notably, in this procedure, beauty meson contributions to the experimental sensitivities are not included as there is no experimental data available for differential cross sections at the relevant centre-of-mass energies (c.f. section~\ref{sec:dsigma_bb_experimental}).
        In the remaining sensitive parameter space, the relative differences between sensitivities are similar to those in figure~\ref{fig:pythia_HNLbd_overview}. 
        Meanwhile, we observe a slight sensitivity improvement on a global scale similar to the values presented in figure~\ref{fig:ternary_comparison} shifting the lower bounds by factors around $\sim 0.8$.

\section{Conclusions} \label{sec:conc}

    Heavy Neutral Leptons are an excellently motivated candidate for BSM physics that address a number of shortcomings in the Standard Model.
    The search for them is thus highly motivated and a number of proposals have been put forward and are poised to scan significant portions of the HNL parameter space~\cite{Berryman:2019dme,SHiP2023,Gorkavenko:2023nbk,Blondel:2022qqo}. 
    A promising way to find or exclude HNLs are proton beam-dumps. 
    In these configurations, HNLs are typically produced from decays of heavy mesons produced in the forward direction of the proton interaction.
    However, there is little experimental data available, for heavy meson production in this kinematic range. 
    For a credible search program, especially for those with data already on tape~\cite{NA62:2023qyn, NA62:2023nhs, gori}, it is thus mandatory to study possible differences in existing data and MCs typically employed for projected sensitivities.
    
    In this paper we have systematically studied the impact viable assumptions on meson distributions can have on the predicted sensitivity of an experiment. 
    To this effect, we have identified the underlying meson production parameters and attempted to collect the available experimental data in a unified set-up.
    We have used the updated implementation of the \alpi code to simulate HNLs from their production up to the detector response to the decay products in a simplified MC framework.
    Varying the meson inputs within motivated assumptions, we demonstrated that the experimental sensitivity can shift by up to an order of magnitude in the coupling suppression. 
    This effect was shown to be largely independent of the HNL coupling structure to the standard model.  
    Finally, for the first time, we have compared the sensitivities of different beam-dump experiments on equal footing both in terms input and also model parameters, albeit in a simplified MC set-up. 
    
    This work also highlights the need for further experimental constraints on the heavy meson spectra in proton nucleus interactions. 
    We would like to emphasise that ongoing experiments like DsTau~\cite{DsTauNA65:2023ogo} and SHiP-charm~\cite{SHiP:2024oua} at CERN may already take a large step towards a better understanding of the charm production.
    However, especially for beauty mesons, where experimental input is virtually non-existent, experimental validation of theory based simulations is needed.
    On the theory side, implementations of heavy quark generation mechanisms beyond $gg\to q \bar q$ and $q^\prime \bar q^\prime\to q\bar q$ in simulation frameworks like \pyth will likely be helpful in describing the meson spectra more accurately. 
    Steps toward this have already been taken (c.f. appendix~\ref{sec:leading_particle_effect}).

    Another important point to raise is that HNLs are special among the commonly considered FIPs in that they are produced in large numbers in open charmed decays. 
    For other FIPs like Axion-Like-Particles and Dark Photons, especially the beauty mesons are relevant in some production scenarios~\cite{Beacham:2019nyx}. 
    Yet, for these FIPs other channels like dark bremsstrahlung and mixing dominate, making accurate $B$-production prescriptions a secondary matter to inconsistencies in the commonly used radiation approximations~\cite{Foroughi-Abari:2021zbm} and mixing formalism~\cite{LoChiatto:2024guj}.
    In the case of the Dark Scalar, however, beauty meson decays are a dominant production mechanism~\cite{Beacham:2019nyx}, highlighting again the relevance of accurate beauty meson spectra. 
    In summary, especially beam-dump experiments that hope to exploit beauty meson decays to search for Feebly Interacting Particles should be very interested in exploring the relevant meson spectra from an experimental side.

\section*{Acknowledgements}
We would like to thank I. Abt and T. Hebbeker for discussions on the CHARM HNL limits and to M. U. Ashraf for discussions on the Pythia tunes. 
In addition, we gratefully acknowledge discussions with M. Ovchynnikov on the heavy meson spectra and FIP phenomenology. 
The authors acknowledge very useful conversations within the NA62 experiment. 
This study is performed independently of any experiment.
This work is supported by the European Research Council under grant ERC-2018-StG-802836 (AxScale project).
We also acknowledge discussions within COSMIC WISPers CA21106, supported by COST (European Cooperation in Science and Technology).

\appendix
\section{Heavy Neutral Lepton interactions}\label{sec:HNL_interactions}

    As already highlighted in section~\ref{sec:hnl_pheno}, the HNL is produced and decays in $U$-suppressed weak interactions. 
    The relevant width-formulae are well established~\cite{Gorbunov:2007ak, Atre:2009rg, Bondarenko:2018ptm}.
    In case of minor discord in the literature, we rely on the calculations by Bodarenko et al.~\cite{Bondarenko:2018ptm}. 
    As their review is very exhaustive, we would like to point the interested reader there.
    In the following we will only give a very brief summary, focusing on the points where our implementation diverges.
    
    \subsection{Production}

\begin{figure}
            \centering
            \begin{tikzpicture}
\begin{feynman}
  \vertex (q) {\(q_u\)};
  \path (q) ++ (1.25,-0.25) node[vertex] (v1);
  \vertex [right of=q](dummy1);
  \path (dummy1) ++ (0,-0.5) node[vertex] (dummy2);
  
  \vertex [left of= dummy2](qbar) {\(\bar q_d\)};
  \vertex [right of = v1] (v2){\(U_\alpha\)};
  \vertex [above right=of v2] (f1) {\(\mathrm N\)};
  \vertex [below right=of v2] (f2) {\(\overline \ell_\alpha\)};

  \diagram* {
    (q) -- [fermion, out = 0, in = 135] (v1) -- [fermion, out=225, in=0] (qbar),
    (v1) -- [boson, edge label'=\(W\), swap] (v2),
     (f2) -- [fermion] (v2) -- [plain, orange, very thick] (f1),
  };
 \draw (v1) node[dot, fill, draw=black] {};
 \draw [decoration={brace}, decorate] (qbar.south west) -- (q.north west) node [pos=0.5, left] {\(h\)};
% \draw (v2) node[dot, fill, draw=black] {};
\end{feynman}
\end{tikzpicture}
\hspace{4em}
\begin{tikzpicture}
\begin{feynman}
  \vertex (q) {\( q_u\)};
  \vertex[right of = q] (v1);      
 
  \path (v1) ++ (0,0.5) node[vertex] (v1dummy);
   \vertex [left of = v1dummy] (qin) {\( \bar q_d\)};
    \path (v1dummy) ++ (0.66,0) node[vertex] (dummyprime);
   \vertex[right = of dummyprime] (qout) {\( \bar q_d\)};
  
    \path (v1) ++ (0.66,0) node[vertex] (dummy);
   \vertex [right  =of dummy] (qprime){\( q_d ^\prime\)};
   \vertex [below right=of v1](v2) {\(U_\alpha\)};
   \vertex [ right=of v2] (f1) {\(\mathrm N\)};
   \vertex [below right=of v2] (f2) {\(\overline \ell_\alpha\)};

  \diagram* {
    (qout) -- [fermion]  (v1dummy)--  [fermion] (qin),
    (q) -- [fermion]  (v1)--  [fermion] (qprime),
    (v1) -- [boson, edge label'=\(W\)] (v2),
    (f2) -- [fermion] (v2) -- [plain, orange, very thick] (f1),
  };
 \draw (v1) node[dot, fill, draw=black] {};
 \draw [decoration={brace}, decorate] (q.south west) -- (qin.north west) node [pos=0.5, left] {\(h\)};
 \draw [decoration={brace}, decorate]  (qout.north east) -- (qprime.south east) node [pos=0.5, right] {\(h / h_V\)};
% \draw (v2) node[dot, fill, draw=black] {};

\end{feynman}
\end{tikzpicture}
            \caption{Dominant HNL production channels in the $\mathcal O \left(\SIrange[]{100}{1000}{MeV}\right)$ mass regime.
            In case the HNL is a Majorana particle, the charge conjugated process is allowed.}
            \label{fig:HNL_production_channels}
        \end{figure}
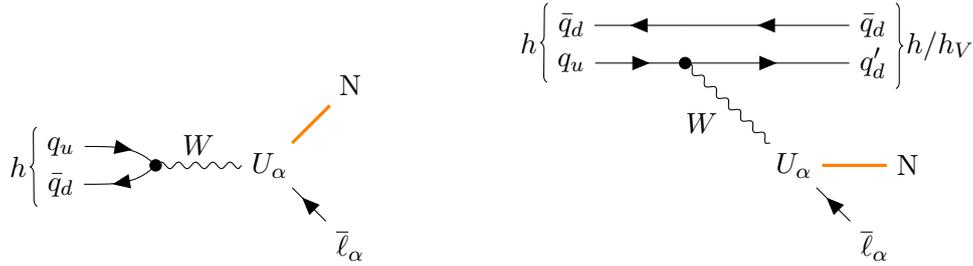

        The main production channels for HNLs in the $\mathcal O \left(\SIrange[]{100}{1000}{MeV}\right)$ range are the two and three-body decays of pseudoscalar mesons as sketched in figure~\ref{fig:HNL_production_channels}. 
        In particular, the decays of open charmed and beauty mesons are relevant for the production at energies typical for beam-dump experiments~\cite{Bondarenko:2018ptm}.
        The bulk of HNLs are produced in their (semi-) leptonic decays with up to a single pseudoscalar or vector meson in the final state.\footnote{In the massless HNL limit, where the neglected contributions are maximal, they make up around $\SI{20}{\%}$ of the total possible yield.}
        This corresponds to signatures such as $h_P\to \ell \mathrm{N},\,h_P\to h^\prime_P \ell \mathrm{N},\,\text{and }h_P\to h_V \ell \mathrm{N}$, where we focus on the dominant contributions with $h_P\in(\pi, K, D)$ and $h_V \in (K^*, D_s^*)$.
        
\begin{figure}
            \centering
            \includegraphics[width=.995\textwidth]{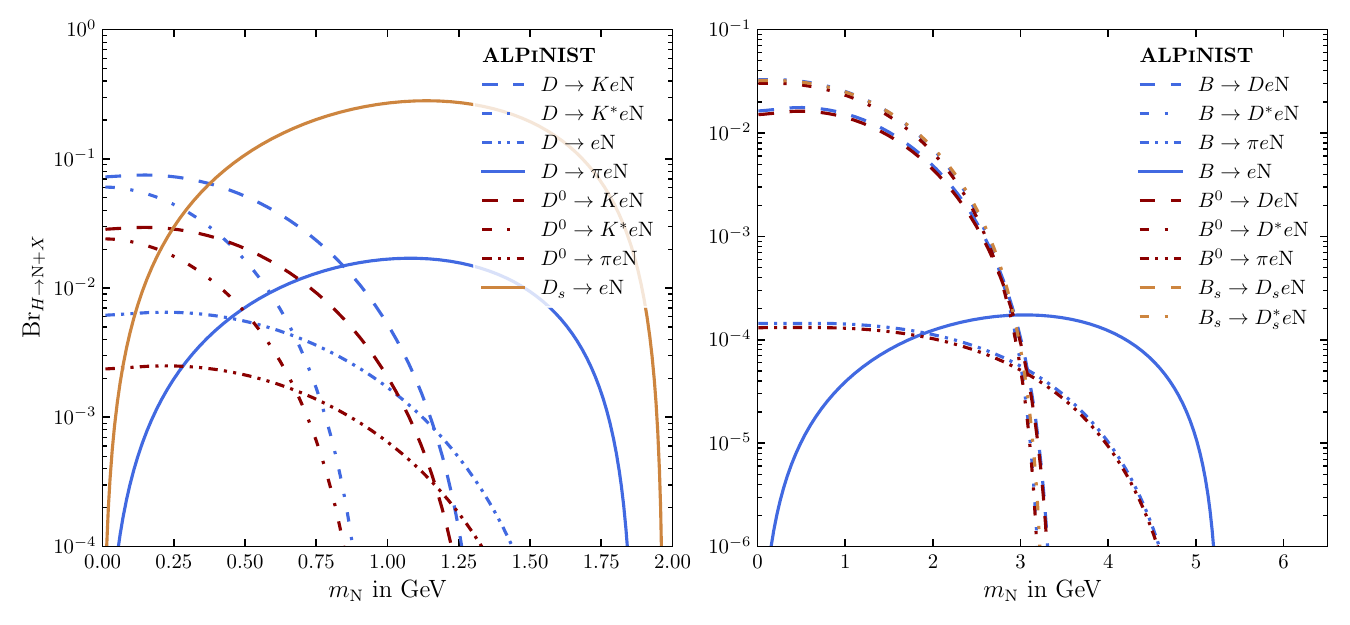}
            \caption{Branching ratios of open charm (left) and beauty (right) mesons to an electronphilic HNL $\mathrm{N}$ with coupling $U_e^2$ normalised to $1$.}
            \label{fig:BC6_heavyProdBrs}
        \end{figure}

        For two-body decays of the $h_P\to \ell \mathrm{N}$ type, the relevant pseudoscalar decay constants have been taken from PDG22 and where necessary FLAG21 review tables~\cite{ParticleDataGroup:2022pth, FLAG:2021npn}. 
        For three-body decays the underlying tabulated values are taken from the structure functions, where $\mathrm{N}$ is replaced with the SM neutrino.
        The underlying parameters have been updated to match best fit values presented in the FLAG21 review~\cite{FLAG:2021npn} for pseudoscalar meson decays into other pseudoscalars ($h_P\to h_P^\prime \ell \mathrm{N}$), while the decays into vector mesons have been matched to best fit results by the HPQCD collaboration~\cite{Harrison:2017fmw,Harrison:2021tol} for the decay of beauty mesons ($B\to h_V \ell \mathrm{N}$).
        For the decay $D\to K^*\ell\mathrm{N}$ we follow Bondarenko et al.~\cite{Bondarenko:2018ptm} using the form factors listed by Melikhov and Stech~\cite{Melikhov:2000yu}.
        The resulting branching ratios for production in open charmed and beauty meson decays considered in the \texttt{EXO\_decay} module of \alpi are presented in figure~\ref{fig:BC6_heavyProdBrs} for a coupling ratio $(U^2_e:U^2_\mu:U^2_\tau)= (1:0:0)$.
        
        We finally note, that HNLs can also be produced in the decays of light pseudoscalar mesons~\cite{Bondarenko:2018ptm}. 
        However, unlike for the mixing other FIPs undergo, the reabsorption rate of mesons and the propagation distance from the primary interaction point at decay time is strongly modulated by the composition and geometry of an experiments specific dump ($K_S$ being a borderline case).
        In the spirit of generalisability, we leave the simulation of these effects to experiments' dedicated full MC set-ups. 
        We would like to further point out, that light meson precision experiments routinely surpass beam-dumps in terms of sensitivity in this mass regime due to favourable scaling in terms of coupling suppression (c.f. figures~\ref{fig:pythia_HNLbd_overview} and~\ref{fig:FIT_HNLbd_overview}).

    \subsection{Decay}\label{sec:HNL_decay_width}
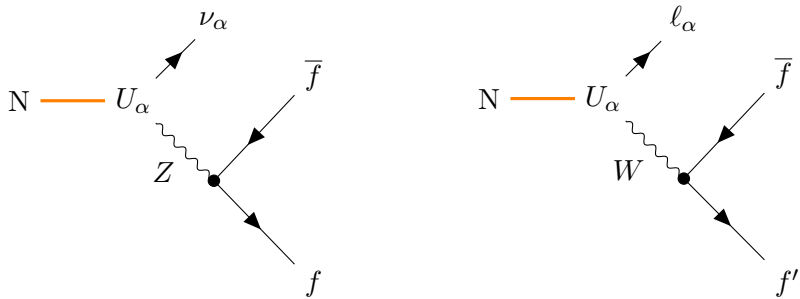
\begin{figure}
        \centering
        \begin{tikzpicture}
  \begin{feynman}
    \vertex (a) {\(\mathrm{N}\)};
    \vertex [right of = a] (b) {\(U_\alpha\)};
    \vertex [below right=of b] (B);
    \vertex [above right=of b] (f1) {\( \nu_\alpha\)};
    \vertex [above right=of B] (f2) {\(\overline f\)};
    \vertex [below right=of B] (f3) {\(f\)};

    \diagram* {
      (a) -- [plain, orange, very thick ] (b) -- [fermion] (f1),
      (b) -- [boson, edge label'=\(Z\)] (B),
      (f2) --[fermion] (B) -- [fermion] (f3),
    };
 \draw (B) node[dot, fill, draw=black] {};
  \end{feynman}
\end{tikzpicture}
\hspace{4em}
\begin{tikzpicture}
    \begin{feynman}
      \vertex (a) {\(\mathrm{N}\)};
      \vertex [right of = a] (b) {\(U_\alpha\)};
      \vertex [above right=of b] (f1) {\(\ell_{\alpha}\)};
      \vertex [below right=of b] (B);
      \vertex [above right=of B] (f2) {\(\overline f\)};
      \vertex [below right=of B] (f3)  {\( f^\prime\)};
  
      \diagram* {
        (a) -- [plain, orange, very thick] (b) -- [fermion] (f1),
        (b) -- [boson, edge label'=\(W\)] (B),
	    (f2) --[fermion] (B) -- [fermion] (f3),,
      };
    \draw (B) node[dot, fill, draw=black] {};
    \end{feynman}
  \end{tikzpicture}
        \caption{Leading order HNL decay channels in the $\mathcal O \left(\SIrange[]{100}{1000}{MeV}\right)$ mass regime.
        In case the HNL is a Majorana particle, the charge conjugated process is allowed.}
        \label{fig:HNL_decay_channels}
    \end{figure}

    The HNL with a coupling as described by eq.~(\ref{eq:Lagrangian_HNL_int}) most commonly decays into three fermions as schematically presented in figure~\ref{fig:HNL_decay_channels}.
    In the $Z$-mediated case, $f$ can either represent a lepton or a quark, while in the $W$-mediated case $f^\prime \bar f$ is either a neutrino anti-lepton pair or a up- anti-down-like pair of quarks.
    The leptonic decay channels of the HNL depend on the exact ratio of its mixing angles, but are of the structure $\nu\ell \bar\ell^{(\prime)}$ with $\ell=e,\mu,\tau$ and also include the invisible $3\nu$ final state. 
    The most relevant hadronic decay channels are of the kind $\ell h_P,\,\ell h_V$ for charged pseudoscalar (vector) mesons $h_P$ ($h_V$) and their neutral counterparts $\nu h_P,\,\nu h_V$. 
    Of particular note are the decays to $\ell\pi$ and $\ell\rho$ as they leave a clear signature in a detector. However, for the relevant calculations we also include the above combinations with $h_P\in (K,\eta,\eta^\prime,D,D_s,J/\psi)$ and $h_V\in(\rho,\omega,\phi,D_s^*)$.
    
\begin{figure}
        \centering
        \includegraphics[width = 0.595\textwidth]{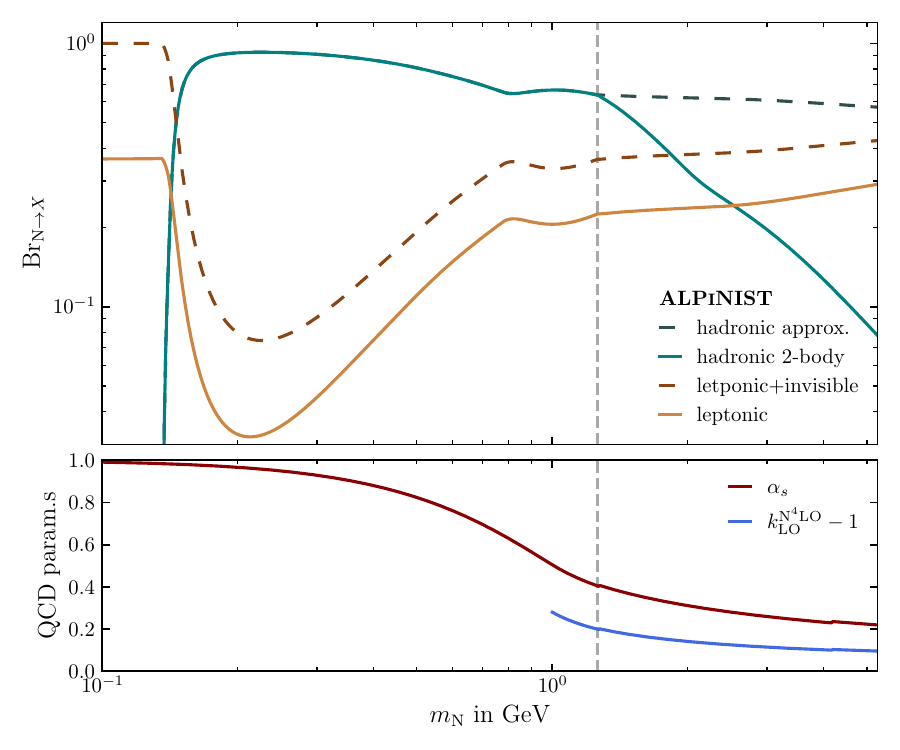}
        \includegraphics[width = 0.395\textwidth]{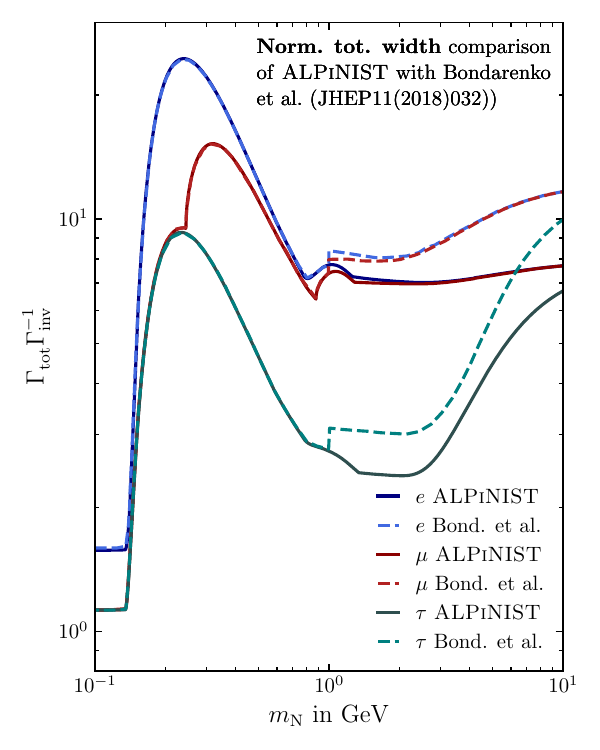}
        \caption{Left: Branching ratios of hadronic and leptonic decay channels of an electronphilic HNL (top), with solid lines indicating the observable components and dashed lines the contributions to the total width. 
        Underlying $\mathrm{QCD}$ related parameters strong coupling $\alpha_s$ and scaling ratio $k^\mathrm{N^4LO}_\mathrm{LO}$ (bottom).
        Right: Comparison of total decay widths normalised to the decay width to invisible ($3\nu$ final state) for the different coupling dominance scenarios ($e$-BC6, $\mu$-BC7, and $\tau$-BC8) between the implementation in \alpi and Bondarenko et al.~\cite{Bondarenko:2018ptm}.
        For further information see text.
        }
        \label{fig:semileptonic_running}
    \end{figure}

    In case of a semileptonic decay, the QCD interactions of the produced quarks play a major role in the width of the HNL.
    In case the quarks form a stable meson, the decay can be well approximated as a two-body decay including the appropriate form factors. 
    However, for HNLs heavier than twice the pion mass, also multibody bound QCD final states become possible. 
    This is of particular importance when calculating the total width of the HNL.
    Following Bodarenko et al.~\cite{Bondarenko:2018ptm}, we estimate their importance using the hadronic width ratio of the $\tau$ lepton $R_{\tau,h}=\Gamma (\nu_\tau + \mathrm{had}) \Gamma^{-1} (\nu_\tau + e \bar \nu_e) $, which is known up to $\mathrm{N^4LO}$ in $\alpha_S(m_\tau)\pi^{-1}$~\cite{Baikov:2008jh,Deur:2023dzc}.
    We then promote $R_{\tau,h}$ to relate the approximated full hadronic width of the HNL to its charged and neutral current decays with $u$ and/or $d$ quarks in the final state
    with $\alpha_S = \alpha_S(m_\mathrm{N})$ as\footnote{This formulation is the $\mathrm{N^4LO}$ approximation of $R_{\tau,h}$ in $\overline{\mathrm{MS}}$ at $\alpha_s = \alpha_S(m_\tau) $~\cite{Baikov:2008jh,Deur:2023dzc}, not including the number of quark colours and CKM suppression, as these are included in the reference widths.}
    \begin{equation}
            k^\mathrm{N^4LO}_\mathrm{LO} = 1.019 \left(0.994 \vphantom{\left(\frac{\alpha_S}{\pi}\right)^5}
            + \frac{\alpha_S}{\pi} + 1.640\left(\frac{\alpha_S}{\pi}\right)^2 + 6.371\left(\frac{\alpha_S}{\pi}\right)^3  + 49.076\left(\frac{\alpha_S}{\pi}\right)^4 \right).
    \end{equation}
    Using the $\beta$ evolution for the running of the strong coupling~\cite{Deur:2023dzc} up to $\beta_3$, and connecting it below $q<\SI{1}{GeV}$ with a third order polynomial ensuring that $\alpha_s(q)$ run smoothly to $\alpha_s(q=0)=1$, we get an approximation ratio as shown in the bottom left panel of figure~\ref{fig:semileptonic_running}.
    
    For HNL masses above $\SI{1}{GeV}$ we then calculate the semileptonic HNL width as the maximum between the thus calculated approximate width and the possible two-body semileptonic decays considered.
    This ensures, that $\alpha_s \pi^{-1}$ is already in a perturbative regime justifying the $R_{\tau,h}$ approach.
    The resulting branching ratios for (visible) leptonic and (approximated) hadronic decays of an electrophilic HNL are shown in the top left panel of figure~\ref{fig:semileptonic_running}.
    Typically, the semileptonic width is then approximated for masses just above the $\SI{1}{GeV}$ threshold, as indicated for this case by the grey dashed line.
    Notably, this results in a total HNL width that differs from the prescription by Bondarenko et al.~\cite{Bondarenko:2018ptm} in all benchmark cases for HNL masses above $\SI{1}{GeV}$ (see figure~\ref{fig:semileptonic_running} right). 

\section{Leading particle effects on meson distributions}\label{sec:leading_particle_effect}

    The leading particle effect describes an asymmetry in hadron collisions, where the $x_F$ and $p_T$ spectra of secondary particles sharing a valence quark with the beam particle are harder with respect to their charge conjugate partners that do not have such a valence quark in common.
    From the theory side, many ideas have been put forward to derive this experimentally observed asymmetry ranging from intrinsic charm coalescence~\cite{BRODSKY1980451, Vogt:1995fsa}, charm excitation~\cite{Combridge:1978kx}, over-production in colour strings (used in the standard configuration of \pyth)~\cite{Andersson:143966, Andersson:1983ia, Norrbin:1998bw}, to heavy recombination~\cite{Braaten:2002yt} and light fragmentation~\cite{Gao:2007ht}.
    In all these models, any asymmetry in the hard $x_F$ region of open charm/beauty meson production can be attributed to the valence quarks of the beam particle either as remnants involved in the recombination or as initial states to the scattering process. 
    In the following we will give some insight in the experimental validation of these ideas at beam-dump energies. 
    
    \subsection{Open charmed mesons}
        \label{sec:leading_particle_cc}

\begin{figure}[ht]
            \centering
            \includegraphics[width = 0.95\textwidth]{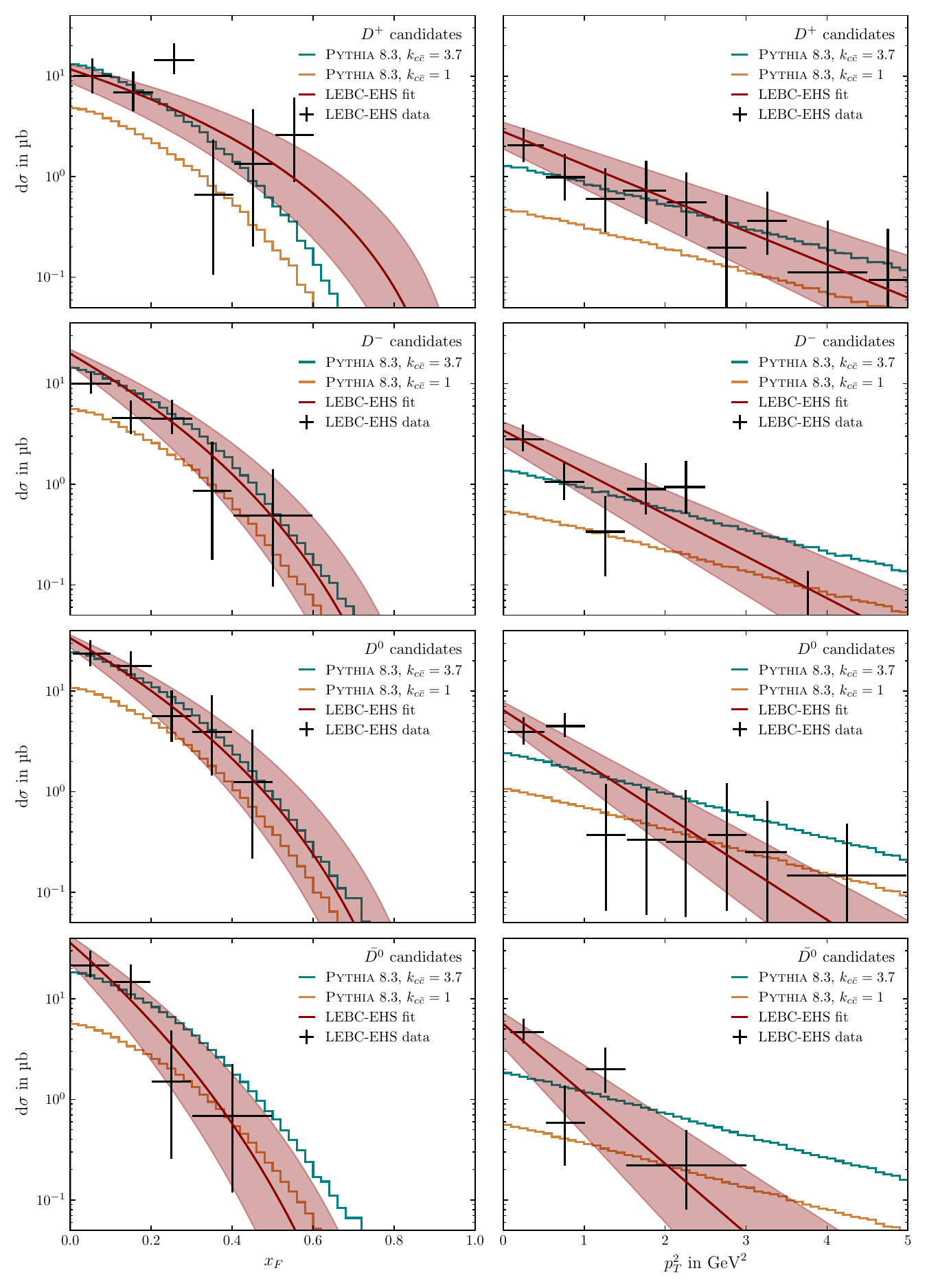}
            \vspace{-1em}
            \caption{Comparison of $x_F$ and $p_T^2$ spectra for different open charmed mesons as measured by the LEBC-EHS collaboration~\cite{LEBC-EHS:1988oic} compared to results from \pyth~8.3 with \pyth suggested ($k_{c\bar c}=1$, orange) and adjusted ($k_{c\bar c}=3.7$, teal) cross section  $\sigma_{cc} $.}
            \vspace{-1em}
            \label{fig:LEBC-EHS_Pythia_comparison}
        \end{figure}

        %about leading particles 
        Leading particle effects were observed experimentally in the differential spectra of open charmed mesons in $\pi^- N_\mathrm{C/Ti}$~\cite{E791:1997eip} as well as $\Sigma^- N_\mathrm{Be}$ scattering~\cite{WA89:1998wdl}. 
        However, in $\SI{400}{GeV}$ protons scattering on $N_\mathrm{H_2}$, the LEBC-EHS collaboration (NA27) observed an asymmetry preferring hard $D^+$ and $D^0$ mesons over their charge conjugate partners in the high $x_F$ regime~\cite{LEBC-EHS:1988oic}.
        A comparison of these results with \pyth~8.3 is shown in figure~\ref{fig:LEBC-EHS_Pythia_comparison}, where both the differential cross sections from \pyth ($k_{c\bar c}=1$) and that scaled to match the cross section found by LEBC-EHS ($k_{c\bar c}=3.7$) are shown. A tension between expected and measured values is present.
        Data from $pN_\mathrm{Cu}$ scattering by the WA82 experiment~\cite{Adamovich:1992cv} at a similar beam energy of $\SI{370}{GeV}$ is in better agreement with theoretical models.\footnote{As presented in table~\ref{tab:NA27_results}, the tension is somewhat less significant in our reevaluation of the spectral data for the case of $D^\pm$, however still persists for $\overset{\textbf{\fontsize{1pt}{1pt}\selectfont(---)}}{D_0}$.\par
        Employing a different \pyth reconnection scheme based on QCD colour of the beam remnants~\cite{Christiansen:2015yqa} results in overall harder $x_F$ meson spectra while the $p_T$ spectra are softer, which also releases the tension.
        However, we still observe a major discrepancy in the $p_T$ spectra.\par
        We found the Ropewalk mechanism~\cite{Bierlich:2014xba,Bierlich:2023fmh} to have little impact on the resulting spectra.}
        We would like to highlight that the intrinsic charm model, further substantiated by the recent findings using LHCb data~\cite{Ball:2022qks,NNPDF:2023tyk}, could lead to asymmetries specific to $pN$ scattering that would not occur in $\pi^\pm N$.
        It is, however, beyond the scope of this work to establish a new theoretical framework for describing the production mechanisms of open charmed mesons.
        Therefore, we limited ourselves to using the LEBC-EHS and other measured differential cross sections as inputs for a subsequent FIP simulation.

\begin{figure}
            \centering
            \includegraphics[width =.995\textwidth]{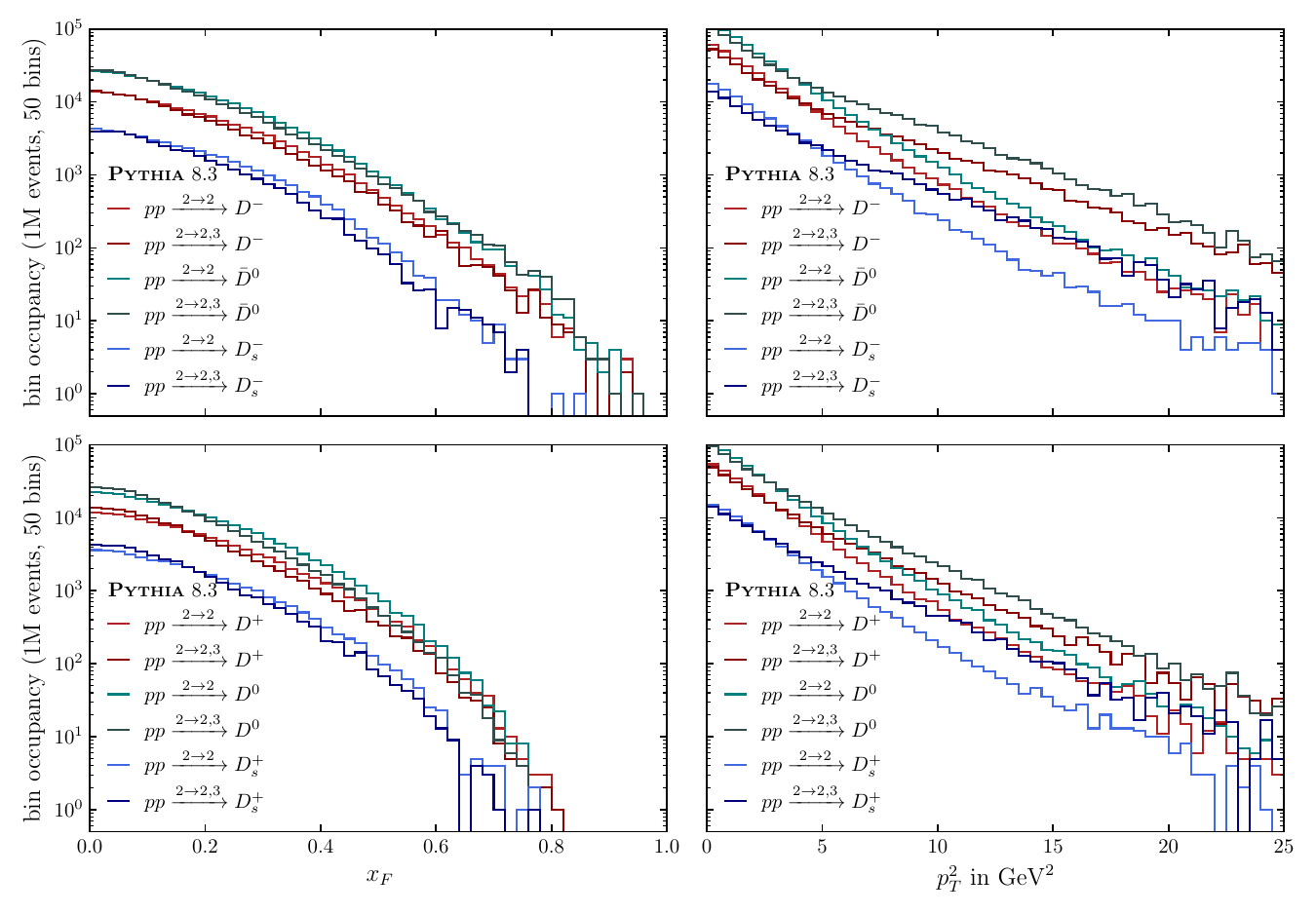}
            \caption{Comparison of $x_F$ and $p_T^2$ spectra considering \textsc{Pythia}'s \texttt{HardQCD:hardccbar} ($2\to 2$) and a weighted combination including \texttt{HardQCD:3parton} with \texttt{HardQCD:nQuarkNew=4} ($2\to 2,3$) for open charm and anti charm mesons in the top and bottom panels respectively in $\sqrt{s}=\SI{27.43}{GeV}$ $pp$ scattering.}
            \label{fig:23cc_ppScatter}
        \end{figure}

        In recent versions of \pyth~8 a new option for simulating \texttt{HardQCD} processes was added, namely the \texttt{HardQCD:3parton} functionality. 
        This allows the simulation including $2\to 3$ diagrams describing parton level scattering.
        In principle, this introduces production channels of open charmed mesons, where a charmed quark can combine with a parton involved in the scattering process, as required by heavy recombination~\cite{Braaten:2002yt}.
        This is in contrast to the standard colour string scheme of \pyth, where the combination only occurs with beam remnant parts.
        However, as of the writing of this paper, the \texttt{HardQCD:3parton} processes are only implemented with amplitudes assuming massless quarks. 
        This is especially concerning, as these processes can scale explicitly with the quark mass~\cite{Braaten:2002yt}.
        We verified that in the standard \pyth production of open charmed meson production in $2\to 2$ processes, no significant difference in $x_F$ or $p_T$ spectra was observed between considering the charmed quark massive or massless.
        Even though the extension of this check to $2\to 3$ processes is far from sound, we present in figure~\ref{fig:23cc_ppScatter} the impact on the spectra of charmed and anti-charmed quarks when considering the three-parton processes and weighting the contributions between two- and three-parton processes with the associated cross section estimated by \pyth.
        The $x_F$ spectra for open anti-charmed mesons (upper left) are roughly comparable when considering only $2\to 2$ compared to $2\to 2,3$ processes, while the open charmed meson $x_F$ spectra (bottom left) are slightly softer when also considering $2\to 3$.
        The $p_T$ spectra are significantly harder for all mesons when including the $2\to 3$ processes.
        This is in worse agreement with experimental data (c.f. figure~\ref{fig:DX_cc_fit}).
        
        \begin{table}[h]
            \centering
            \begin{tabular}{lcc|cc}
                   & \multicolumn{2}{c}{LEBC-EHS}& \multicolumn{2}{c}{Refits}\\
                \hline
                  & $n$ & \multicolumn{1}{c}{$b$} & \multicolumn{1}{c}{$n$} & $b$ \\
                 \hline
                 $D^+$      &  $3.1\pm 0.8$ & $0.75\pm 0.14$ & $4.94 \pm 2.54$ & $0.67 \pm 0.09$ \\
                 $D^-$      &  $5.4\pm 1.2$ & $0.93\pm 0.18$ & $5.1  \pm 1.07$ & $1.24 \pm 0.43$\\
                 $D^0$      &  $5.4\pm 1.1$ & $1.04\pm 0.19$ & $5.47 \pm 0.8 $ & $1.97 \pm 0.68$\\  
                 $\bar D^0$ &  $8.1\pm 1.9$ & $1.49\pm 0.32$ & $9.25 \pm 2.84$ & $1.87 \pm 1.04$\\
                 \hline
            \end{tabular}
            \caption{LEBC-EHS parametrisations of individual open charmed mesons~\cite{LEBC-EHS:1988oic} for differential production cross sections as given by eq.~(\ref{eq:diff_meson_parametrisation}). 
            Also presented are the results of refits performed on the published LEBC-EHS data points, assuming the presented uncertainties as standard deviations in a orthogonal distance reduction fit.}
            \label{tab:NA27_results}
        \end{table}
        
    \subsection{Open beauty mesons}
        \label{sec:leading_particle_bb}
\begin{figure}
            \centering
            \includegraphics[width =.995\textwidth]{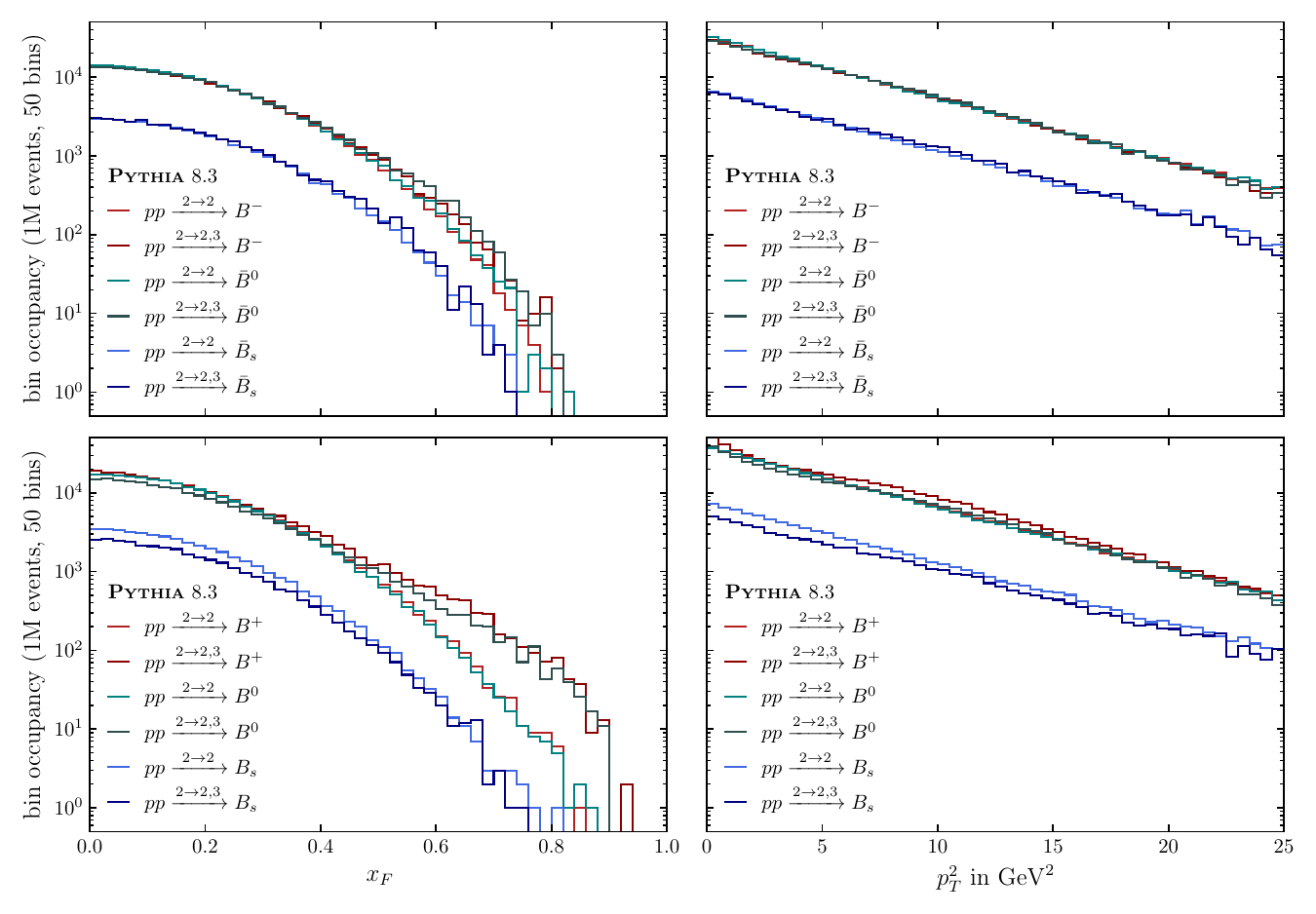}
            \caption{Comparison of $x_F$ and $p_T^2$ spectra considering \textsc{Pythia}'s \texttt{HardQCD:hardbbbar} ($2\to 2$) and a weighted combination including \texttt{HardQCD:3parton} with \texttt{HardQCD:nQuarkNew=5} ($2\to 2,3$) for open beauty and anti beauty mesons in the bottom and top panels respectively in $\sqrt{s}=\SI{27.43}{GeV}$ $pp$ scattering.}
            \label{fig:bb_ppScatter}
        \end{figure}

        %a word on B mesons
        In principle, one would expect the leading particle effect to also impact the spectra of beauty mesons. 
        However, given the lack of experimental data we limit ourselves here purely to analysing the impacts with \pyth.
        Figure~\ref{fig:bb_ppScatter} shows the $B$ meson differential distribution, using \pyth's \texttt{HardQCD:3parton} with \texttt{HardQCD:nQuarkNew=5}. 
        The spectra for anti-beauty mesons generated with only $2\to 2$ processes do not show sizeable differences to those generated including also $2 \to 3$ parton-level processes. 
        For beauty mesons, the $x_F$ spectra for the $B^+$ and $B^0$ are significantly harder when also including $2 \to 3$ parton level processes, while the corresponding $B_s$ spectrum does not change significantly. 
        In all cases, the $p_T$ spectra show little discrepancy between the two approaches.
        We observe, however, that comparing the $2\to 2 $ process \texttt{HardQCD:hardbbbar} using a massive $b$ quark and the massless equivalent of  \texttt{HardQCD:qqbar2qqbarNew} combined with \texttt{HardQCD:gg2qqbar} and \texttt{HardQCD:nQuarkNew=5}, the spectra differ significantly.
       Moreover, the transverse effects due to $2\to 3$ diagrams scale with the quark mass~\cite{Braaten:2002yt}.
        Therefore, it questionable whether the $2\to 2,3$ spectra shown in figure~\ref{fig:bb_ppScatter} paint a more accurate picture than those of only the $2\to 2$ processes.
        
\section{Fit details}\label{sec:fit_details}

    The underlying framework for all the fits presented in section~\ref{sec:distributions} is the PyMC Gaussian Process~\cite{AbrilPla2023} which we use to perform a Bayesian fit to the data. 
    The heart of the fit is the Mat\'{e}rn covariance function with $\nu = 5/2$ as the Gaussian Process' kernel function.
    To this we assign a gamma distributed length scale with variable $\alpha_\mathrm{ls}$ and $\beta_\mathrm{ls}$ scales, and a half normally distributed amplitude with variable scale $\sigma_\mathrm{amp}$.
    The models are assumed normally distributed at a given $x$ ($\sqrt{s}$) using a theory motivated prior with expectation value $\mu_\mathrm{prior}$ and standard deviation $\sigma_\mathrm{prior}$.
    The posterior distribution is then generated by marginal likelihood minimisation at new $x$ values (120 points from $x^\mathrm{new}_{\min}$ to $x^\mathrm{new}_{\max}$). 
    The (initial) values used for the specific fits are summarised in table~\ref{tab:fit_details}.
    This choice of covariance function and model building is motivated by its ability to fit data points with a cohesive function, while the uncertainty is modulated by the data point uncertainties, data point multiplicities, and the notion of $x$ distance to data.
    The initial values for the covariance function were then derived by a parameter scan with the goal of allowing a dynamic enough $x$-dependence to match the data in terms of overall likelihood optimisation, while avoiding divergences and strong oscillations (overfitting) on qualitative grounds.
    
    \begin{table}[ht]
        \centering
        \begin{tabular}{c|c c c c}
            fit    & $\sigma_{c \bar c}$ & $\sigma_{b \bar b}$ & $n_{c \bar c}$  & $b_{c \bar c}$  \\
             \hline
             $x$ scale & lin. & log. & lin. & lin.\\
             $y$ scale & log. & log. & lin. & lin.\\
             $\sigma_\mathrm{amp}$ & 0.4 & 5 & 10 & 10\\
             $\alpha_\mathrm{ls}$  & 10  & 3 & 10 & 10\\
             $\beta_\mathrm{ls}$   & 1   & 1 & 1   & 1 \\
             $x^\mathrm{new}_{\min}$ in $\mathrm{GeV}$ & 5   & 10    & 5  & 5 \\
             $x^\mathrm{new}_{\max}$ in $\mathrm{GeV}$ & 48  & 20000 & 48 & 48 \\
             {expected LPD} & -8.9 & -12.8 & -15.1 & -2.8 \\
        \end{tabular}
        \caption{Model details for the fits presented in section~\ref{sec:distributions}. 
        For details see text.}
        \label{tab:fit_details}
    \end{table}
    
    All fits were conducted in 4 independent chains using 1000 samples each. 
    Table~\ref{tab:fit_details} also gives their respective expected log pointwise predictive density (LPD)~\cite{gelman2013understanding} as evaluated in Pareto-smoothed importance sampling leave-one-out cross-validation~\cite{Vehtari_2016,vehtari2024pareto}, even though a reliable objective evaluation is limited due to the lack of data to compare predictions against.
    It also details which optimisations occurred considering the data at either a linear or logarithmic $x$ and $y$ ($\sigma_{q \bar q}$, $n_{c \bar c}$, and $b_{c \bar c}$) scales.

\section{Framework overview}

    The \alpi simulation framework used to generate the FIP sensitivity estimates in this work has already been described in significant detail in previous publications~\cite{Jerhot:2022chi,Afik:2023mhj} and theses~\cite{Jerhot:2023web}. 
    For the reader's convenience, we summarise the basic assumptions here in appendix~\ref{sec:factorisation}, followed by a brief (updated) overview of the simulation chain in appendix~\ref{sec:implementation}, and finally a comment on the details specific to the simulation of HNLs in appendix~\ref{sec:HNL_sim}.

    \subsection{Factorisation assumptions}\label{sec:factorisation}

        The first underlying assumption of the \alpi simulation set-up is, that the number of detected Feebly Interacting Particles as presented in eq.~(\ref{eq:N_detections}) can be reinterpreted as 
        \begin{equation}\label{eq:Ndet_proddec_factorisation}
            N_\mathrm{det} \left(m_X, g_X\right) = \int \mathrm{d}\theta_X \mathrm{d}E_X\frac{\mathrm{d}^2N^\mathrm{prod}\left(m_X,\mathbf{g}_X\right)}{\mathrm{d}\theta_X\mathrm{d}E_X}\chi^\mathrm{det}\left(m_X,\mathbf{g}_X,\Gamma_X,\theta_X,E_X\right),
        \end{equation}
        where ${\mathrm{d}^2N^\mathrm{prod}\left(m_X,\mathbf{g}_X\right)}/{\mathrm{d}\theta_X\mathrm{d}E_X}$ is the spectrum of a FIP with mass $m_X$, SM couplings $\mathbf{g}_X$, and width $\Gamma_X$, with respect to the FIP energy $E_X$ and its emission angle from the target with respect to the beam axis $\theta_X$ in the laboratory frame. 
        The second component of this integral describes the detection probability $\chi^\mathrm{det}$ of such a FIP with a given $\left(E_X,\theta_X\right)$.
        In other words, we assume, that FIP production and detection factorise, which is motivated by the macroscopic on-shell propagation of the FIP from its production to the point of its decay.\footnote{For a nearly degenerate HNL pair this assumption does not hold, as HNL oscillations determine the breaking or conservation of Lepton number. However, in the approximation of a single HNL as presented in section~\ref{sec:hnl_pheno}, no oscillations occur, and the assumption when implemented as described in appendix \ref{sec:HNL_sim}.
        We leave the implementation of oscillation phenomena to future works.}

        Having successfully disentangled FIP production and decay, we now observe that both production and associated decay probability can be expressed as the sum over individual channels, i.e. that the integrand of eq.~(\ref{eq:Ndet_proddec_factorisation}) can be written as
        \begin{equation}\label{eq:spectrum_factorisation}
            \sum_{i}  \frac{\mathrm{d}^2N^\mathrm{prod}_i\left(m_X,\mathbf{g}_X\right)}{\mathrm{d}\theta_X\mathrm{d}E_X} \times \sum_{f} \chi^\mathrm{det}_f\left( m_X,\mathbf{g}_X,\Gamma_X,\theta_X,E_X\right),
        \end{equation}
        where $i$ denotes the $i$th production channel and $f$ the $f$th decay channel of $X$.
        The second assumption of the simulation is now that both $\mathrm{d}N_i^\mathrm{prod}$ and $\chi_f^\mathrm{det}$ further factorise in such a way that it is possible to write 
        \begin{align}
                \frac{\mathrm{d}^2N^\mathrm{prod}_i\left(m_X,\mathbf{g}_X\right)}{\mathrm{d}\theta_X\mathrm{d}E_X} &= \lambda^\mathrm{prod}_i\left(\mathbf{g}_X,\mathbf{g}_X^\mathrm{ref}\right)\frac{\mathrm{d}^2N^\mathrm{prod}_i\left(m_X,\mathbf{g}_X^\mathrm{ref}\right)}{\mathrm{d}\theta_X\mathrm{d}E_X}\\
                \chi_f^\mathrm{det}\left( m_X,\mathbf{g}_X,\Gamma_X,\theta_X,E_X\right) &= \lambda^\mathrm{dec}_f\left(\mathbf{g}_X,\mathbf{g}_X^\mathrm{ref}\right)\chi_f^\mathrm{det}\left( m_X,\mathbf{g}_X^\mathrm{ref},\Gamma_X,\theta_X,E_X\right).
        \end{align}
        This is to say, that the model dependence introduced by an arbitrarily chosen set of $\mathbf{g}_X$ can be encapsulated in rescaling functions $\lambda$ for a set of appropriately chosen reference couplings $\mathbf{g}_X^\mathrm{ref}$.\footnote{Note that $\lambda$ generally relates the relevant branching fraction to the reference coupling. This also ensures that the rescaled $\chi^\mathrm{det}$ can never exceed $1$.} 
        Thus, we can write 
        \begin{equation} \label{eq:Ndet_model_factorisation}
            \begin{split}
                N_\mathrm{det}  =  \sum_{i,f} &\overbrace{\lambda^\mathrm{prod}_i\left(\mathbf{g}_X,\mathbf{g}_X^\mathrm{ref}\right) \lambda^\mathrm{dec}_f\left(\mathbf{g}_X,\mathbf{g}_X^\mathrm{ref}\right)}^{\text{model-dependent}}\\
            &\times\underbrace{\int \mathrm{d}\theta_X \mathrm{d}E_X \frac{\mathrm{d}^2N^\mathrm{prod}_i\left(m_X,\mathbf{g}_X^\mathrm{ref}\right)}{\mathrm{d}\theta_X\mathrm{d}E_X} \chi_f^\mathrm{det}\left( m_X,\mathbf{g}_X^\mathrm{ref},\Gamma_X,\theta_X,E_X\right)}_{\text{kinematics and geometry}}.
            \end{split}
        \end{equation}
        In this way, if the mass $m_X$ and the total width $\Gamma_X$ are treated as free parameters, $N_\mathrm{det}$ can be factorised into a model-dependent part, and a second part determined only by the kinematics of $X$ and the experimental geometry.

    \subsection{Structural layout} \label{sec:implementation}
    
\begin{figure}
            \centering
            % \begin{flushleft}
\begin{footnotesize}
\vspace*{-8em}
\hspace*{2em}
\begin{tikzpicture}[line cap=round,line join=round,>=triangle 45,x=.72cm,y=.72cm]
\clip(3.2524247492428073,-4.8026854573384945) rectangle (26.281169335211395,15.337292745132187);
\fill[line width=2pt,fill=black,fill opacity=0.1] (4.8,6) -- (4.8,4) -- (8,4) -- (8,6) -- cycle;
\fill[line width=2pt,fill=black,fill opacity=0.1] (11,6) -- (11,4) -- (14.2,4) -- (14.2,6) -- cycle;
\fill[line width=2pt,fill=black,fill opacity=0.1] (17.2,6) -- (17.2,4) -- (20.4,4) -- (20.4,6) -- cycle;
\fill[line width=2pt,fill=black,fill opacity=0.1] (9.2,5.4) -- (9.2,4.6) -- (9.8,5) -- cycle;
\fill[line width=2pt,fill=black,fill opacity=0.1] (15.4,5.4) -- (15.4,4.6) -- (16,5) -- cycle;
\fill[line width=2pt,fill=black,fill opacity=0.1] (8,9) -- (8,8) -- (6.4,8) -- (6.4,9) -- cycle;
\fill[line width=2pt,fill=black,fill opacity=0.1] (6.8,7.2) -- (7.6,7.2) -- (7.2,6.6) -- cycle;
\fill[line width=2pt,fill=black,fill opacity=0.1] (5.2,10.2) -- (6,10.2) -- (5.6,9.6) -- cycle;
\fill[line width=2pt,fill=black,fill opacity=0.1] (13,10) -- (13,8) -- (16.2,8) -- (16.2,10) -- cycle;
\fill[line width=2pt,fill=black,fill opacity=0.1] (13.2,7.2) -- (14,7.2) -- (13.6,6.6) -- cycle;
\fill[line width=2pt,fill=black,fill opacity=0.1] (17.4,7.2) -- (18.2,7.2) -- (17.8,6.6) -- cycle;
\fill[line width=2pt,fill=black,fill opacity=0.1] (18.3,3.2) -- (18.7,2.6) -- (19.1,3.2) -- cycle;
\fill[line width=2pt,fill=black,fill opacity=0.1] (17.9,2) -- (17.9,1) -- (19.5,1) -- (19.5,2) -- cycle; 
\draw (5.1,5.6) node[anchor=north west] {\texttt{Exotic}};
\draw (5.1,5.1) node[anchor=north west] {\texttt{production}};
\draw (11.3,5.6) node[anchor=north west] {\texttt{Exotic}};
\draw (11.3,5.1) node[anchor=north west] {\texttt{decay}};
\draw (17.5,5.6) node[anchor=north west] {\texttt{Exotic}};
\draw (17.5,5.1) node[anchor=north west] {\texttt{rescale}};
\draw [rounded corners] (4.8,6) -- (4.8,4) -- (8,4) -- (8,6)-- cycle;
\draw [rounded corners] (11,6)-- (11,4)-- (14.2,4) -- (14.2,6) -- cycle;
\draw [rounded corners] (17.2,6)-- (17.2,4) -- (20.4,4) -- (20.4,6)-- cycle;
\draw (8.25,4.65) node[anchor=north west] {/tab\_prod};
\draw (14.35,4.65) node[anchor=north west] {/tab\_decay};
\draw (9.2,5.4)-- (9.2,4.6) -- (9.8,5) -- cycle;
\draw (8,5)-- (9.2,5);
\draw (9.8,5)-- (11,5);
\draw (15.4,5.4)-- (15.4,4.6) -- (16,5) -- cycle;
\draw (14.2,5)-- (15.4,5);
\draw (16,5)-- (17.2,5);
\draw (6.0,10.3) node[anchor=north west] {/tab\_mesons};
\draw (7.5,7.35) node[anchor=north west] {/tab\_gammas};
\draw [rounded corners] (8,9)-- (8,8) -- (6.4,8) -- (6.4,9) -- cycle;
\draw (6.35,9.05) node[anchor=north west] {decay};
\draw (6.35,8.5) node[anchor=north west] {mesons};
\draw (6.8,7.2) -- (7.6,7.2)-- (7.2,6.6) -- cycle;
\draw (5.2,10.2)-- (6,10.2) -- (5.6,9.6) -- cycle;
\draw (5.6,9.6)-- (5.6,9.3);
\draw [rounded corners] (5.6,9.3) -- (7.2,9.3) -- (7.2,9);
\draw (7.2,8)-- (7.2,7.2);
\draw (7.2,6.6)-- (7.2,6);
\draw (5.6,9.3)-- (5.6,6);
\draw [rounded corners] (13,10)-- (13,8) -- (16.2,8) -- (16.2,10) -- cycle;
\draw (13.2,7.2) -- (14,7.2) -- (13.6,6.6) -- cycle; 
\draw (17.4,7.2)-- (18.2,7.2) -- (17.8,6.6) -- cycle;
\draw (13.6,7.2)-- (13.6,8);
\draw (13.6,6.6)-- (13.6,6);
\draw [rounded corners] (16.2,9)-- (17.8,9) -- (17.8,7.2);
\draw (17.8,6.6) -- (17.8,6);
\draw (13.1,9.95) node[anchor=north west] {decay};
\draw (13.1,9.4) node[anchor=north west] {width};
\draw (13.1,8.85) node[anchor=north west] {calculation};
\draw (11.6,7.35) node[anchor=north west] {/Dalitz};
\draw (14.9,7.35) node[anchor=north west] {/integrated};

\draw (18.7,4.) -- (18.7,3.2);
\draw (18.3,3.2) -- (18.7,2.6) -- (19.1,3.2) -- cycle;
\draw (18.7,2.6) -- (18.7,2);
\draw [rounded corners] (17.9,2) -- (17.9,1) -- (19.5,1) -- (19.5,2) -- cycle; 

\draw (15.7,3.3) node[anchor=north west] {/tab\_toPlot};
\draw (18,1.8) node[anchor=north west] {plots};

\draw (4.8,12.) node[anchor=north west] {External};
\draw (4.8,11.6) node[anchor=north west] {generator};
\draw [dashed,->] (5.6,11)-- (5.6,10.2);

\draw (13.3,12) node[anchor=north west] {Lagrangian};
\draw [dashed,->, rounded corners] (15.8,11.65) -- (19.6,11.65) -- (19.6,6);
\draw [dashed,->] (14.5,11.3) -- (14.5,10);
% \draw [dashed,->] (17.71179758998091,11.218756487487265)-- (19.6,6);
\end{tikzpicture}

\end{footnotesize}
% \end{flushleft}
\vspace*{-13em}
\hspace*{-2em}
            \caption{Updated schematic overview of the \alpi simulation framework.  }
            \label{fig:code_schematic}
        \end{figure}
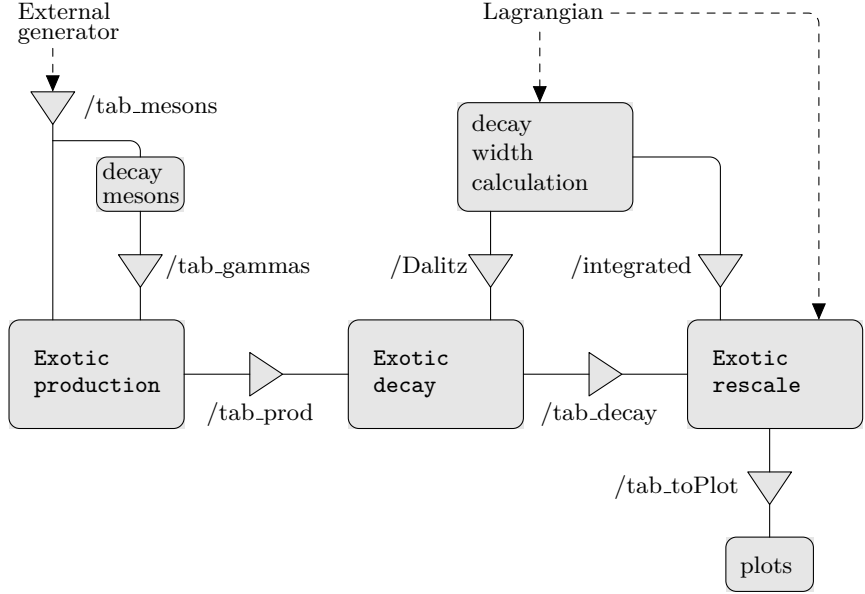

        The \alpi framework is built around the idea that the number of detectable events of Beyond Standard Model decays at a beam-dump experiments factorises into a kinematic and a model-dependent part as presented in section~\ref{sec:factorisation}.
        It can be summarised as presented in figure~\ref{fig:code_schematic}, with boxes representing code elements and triangles the tables for intermediate data storage with the orientation indicating the direction of data flow.\footnote{Given the appropriate inputs, FIPs of alternative origins can also be to studied skipping the production module in the simulation chain presented in figure~\ref{fig:code_schematic}.}
        The three major modular simulation components are related to the components of eq.~(\ref{eq:Ndet_model_factorisation}) as given in table~\ref{tab:module_functionlaties}.
        \begin{table}[h]
            \centering
            \begin{tabular}{p{.185\textwidth}p{.765\textwidth}}
            \texttt{EXO\_production} & calculates the spectrum $\tfrac{\mathrm{d}^2N_i}{\mathrm{d}\theta_X\mathrm{d}E_X}$ for a given production channel $i$;\\
            \texttt{EXO\_decay} & calculates yields for decay channels $f$ for each production spectrum $\mathrm{d}^2N_i$;\\
            \texttt{EXO\_rescale} & performs the appropriate rescaling $\lambda^\mathrm{prod}_i\left(\mathbf{g}_X,\mathbf{g}_X^\mathrm{ref}\right) \lambda^\mathrm{dec}_f\left(\mathbf{g}_X,\mathbf{g}_X^\mathrm{ref}\right)$ for the relevant channels and consequently sums over the individual yields.
        \end{tabular}
            \caption{Summary of the main \alpi module functionalities.}
            \label{tab:module_functionlaties}
        \end{table}
        The individual parts and their function will be described in more detail below.
        
        The \texttt{EXO\_production} module is written in \textsc{Python}~3 and uses functionalities of the \textsc{ROOT} framework~\cite{ROOT_FW}. 
        It emulates the beam particle's interaction with the target material leading to the production of a feebly interacting particle. 
        As FIP production can occur in primary or secondary interactions, for some production channels intermediate Standard Model states are required. 
        These are typically light or heavy mesons which mix with or decay into the FIP. 
        Their momentum distributions can be calculated using either the directly interfaced \textsc{Pythia}~8~\cite{Bierlich:2022pfr}, from implemented empirical distributions relying on experimental data, or any other an external generator when provided as a data table.
        For several configurations of beam energy and underlying theoretical description in \textsc{Pythia8}, pregenerated meson tables are available under~\texttt{/tab\_mesons}. 
        On-shell photons are generated in a separate module using \textsc{ROOT} to simulate the $2\gamma$ decays of light mesons ($\pi^0,\eta,\eta^\prime$) and consequently stored under~\texttt{/tab\_gammas}.
        The resulting FIP spectra are then tabulated on a regular grid in terms of the FIP mass $m_X$ in $\si{MeV}$, its energy $E_X$ in $\si{GeV}$ and emission angle with respect to the beam axis $\theta_X$ in $\si{rad}$ under~\texttt{/tab\_prod}.
        
        The \texttt{EXO\_decay} module is written in {C\nolinebreak[4]\hspace{-.05em}\raisebox{.4ex}{\tiny\bf ++}} using the \textsc{ROOT} framework. 
        It simulates the detector arrays response to the decay of the FIP using the spectra generated in \texttt{EXO\_production}. 
        For three-body decays, model specific Dalitz densities can be requested, to accurately represent the decay kinematics. 
        These densities are calculated%\footnote{Due to the non-trivial normalization of the decay weights in \textsc{ROOT}'s \texttt{TGenPhaseSpace} class, c.f. \href{https://github.com/root-project/root/issues/14765}{https://github.com/root-project/root/issues/14765} , the implementation of this module is preferred over using this existing solution in \textsc{ROOT}.} 
         in an external module using \textsc{Mathetmatica} with \textsc{FeynCalc}~\cite{MERTIG1991345} and stored under~\texttt{/Dalitz}.\footnote{At fixed-target experiments the FIP decay products are typically highly boosted, therefore, unless a very strong momentum cut is applied at the experimental or analysis level to some of the decay products, the change to the yield by assuming a flat phase space of the FIP decay is marginal compared to other experimental or theoretical uncertainties. The same may not be the case for a FIP production in three-body decays due to a typically large distance between the target and the detector.} 
        The yield of a decay channel in a given production spectrum is then scaled with the number of beam particles on target and the appropriate cross sections.
        This yield is tabulated on a regular grid in terms of FIP mass $m_X$ in $\si{MeV}$, and its total width $\Gamma_X$ in $\si{MeV}$ under~\texttt{/tab\_decay}.
        
        The \texttt{EXO\_rescale} module finally introduces the model specific FIP parameters.
        Up until this point in the simulation pipeline, the only model-dependent assumptions about the simulated FIP are the relative branching ratios involved in the FIP's production, and the Dalitz distributions in three-body decays of and into the FIP.
        This module, written in \textsc{Python3}, can be used to appropriately rescale the outputs of the \texttt{EXO\_decay} module into tables of expected yield in terms of mass and coupling strength, sensitivity plots in various FIP parameter representations, and the respective tabulated contours.

    \subsection{Simulating Heavy Neutral Leptons}\label{sec:HNL_sim}

        The \alpi simulation method relies on the underlying assumption that the production and decay of the Feebly Interacting Particle can be treated independent of one another (c.f. appendix~\ref{sec:factorisation}). 
        For a degenerate pair of HNLs undergoing oscillations, this is not necessarily the case.
        But even for a single HNL described by the phenomenological seesaw interactions (see section~\ref{sec:hnl_pheno}), this is not always the case. 
        In \alpi, at the kinematic level, we treat electronphilic, muonphilic and tauphilic mixing as separate production channels.
        This allows us to simulate decay channels of the HNL based on the three production spectra.
        The overall yields for this channel is then the incoherent sum of the three yields based on the HNL's coupling ratios. 
        For two-body decays this approach is fully justified, as the coupling factorises from the kinematic description of the decays. 
        However, for three-body decays of the type $\mathrm{N}\to\nu\ell\ell^{(\prime)}$, the exact kinematic distribution of the decay products depends on the coupling structure (i.e. the Dalitz plot of the tauphilic decay into $\nu_e e\bar e$ differs from that of the electronphilic decay into the related channel $\nu_\tau e\bar e$).
        This is due to the HNL being able to couple both via neutral or charged current as shown in figure~\ref{fig:HNL_decay_channels} and the contributions possibly adding coherently. 
        Consequently, one would have to simulate not only $\nu ee$ for the three different production spectra, but $\nu_e e\bar e$ and $\nu_{\mu/\tau} e\bar e$, increasing the total number of combinations to three-body leptonic decays with only $e$s and $\mu$s in the final state from $9$ to $18$.
    
        In practise, however, we find that the exact kinematic distribution of the three-body HNL decays has little influence on the final sensitivities.
        In order to reduce the amount of necessary simulations, we therefore limit ourselves to simulating the decays using the Dalitz distributions according to the active coupling production spectrum.\footnote{In \alpi, the simulations can alternatively be performed using flat Dalitz distributions instead by invoking the \texttt{--flat-dalitz} flag.}
        This ensures, that the trivial benchmark cases are simulated with the exact kinematics, while the error resulting for admixtures of HNLs is comparable to the difference presented in previous works~\cite{Jerhot:2023web} (often below statistical fluctuations).

\section{Further sensitivity comparisons}

\subsection{NA62 experiment in beam-dump mode} \label{sec:NA62_prediction}
    A commonly cited reference for the prospect sensitivity of the NA62 sensitivity in beam-dump mode is the study by  Drewes et al. \cite{Drewes:2018gkc}. 
    This study employed a toy MC with decay products required to be within acceptance of the charged hodoscope.
    Moreover, the study of \cite{Drewes:2018gkc} states that the composition of the shower and the kinematics of the produced c- and b-hadrons were obtained by simulating the 400 GeV proton beam on a thick ($\sim$ 11 interaction lengths) high-Z target with Pythia 6.4. Our results of figure~\ref{fig:pythia_HNLbd_overview} suggest weaker sensitivity for the NA62 experiment in beam-dump mode than estimated in \cite{Drewes:2018gkc}. 
    We would like to note however, that using the differential cross sections of charmed mesons measured by the LEBC-EHS collaboration~\cite{LEBC-EHS:1987evz}, we reproduce results similar to those by Drewes et al.~\cite{Drewes:2018gkc}. 
    Similarly to the analyses of di-lepton searches~\cite{NA62:2023qyn,NA62:2023nhs}, the NA62-BD data can also be analysed for the possibility of HNL decays.
    This would make both projections obsolete in the future.

\subsection{SHiP experiment}\label{sec:SHiP_validation}
\begin{figure}
        \centering
        \includegraphics[width=.95\textwidth]{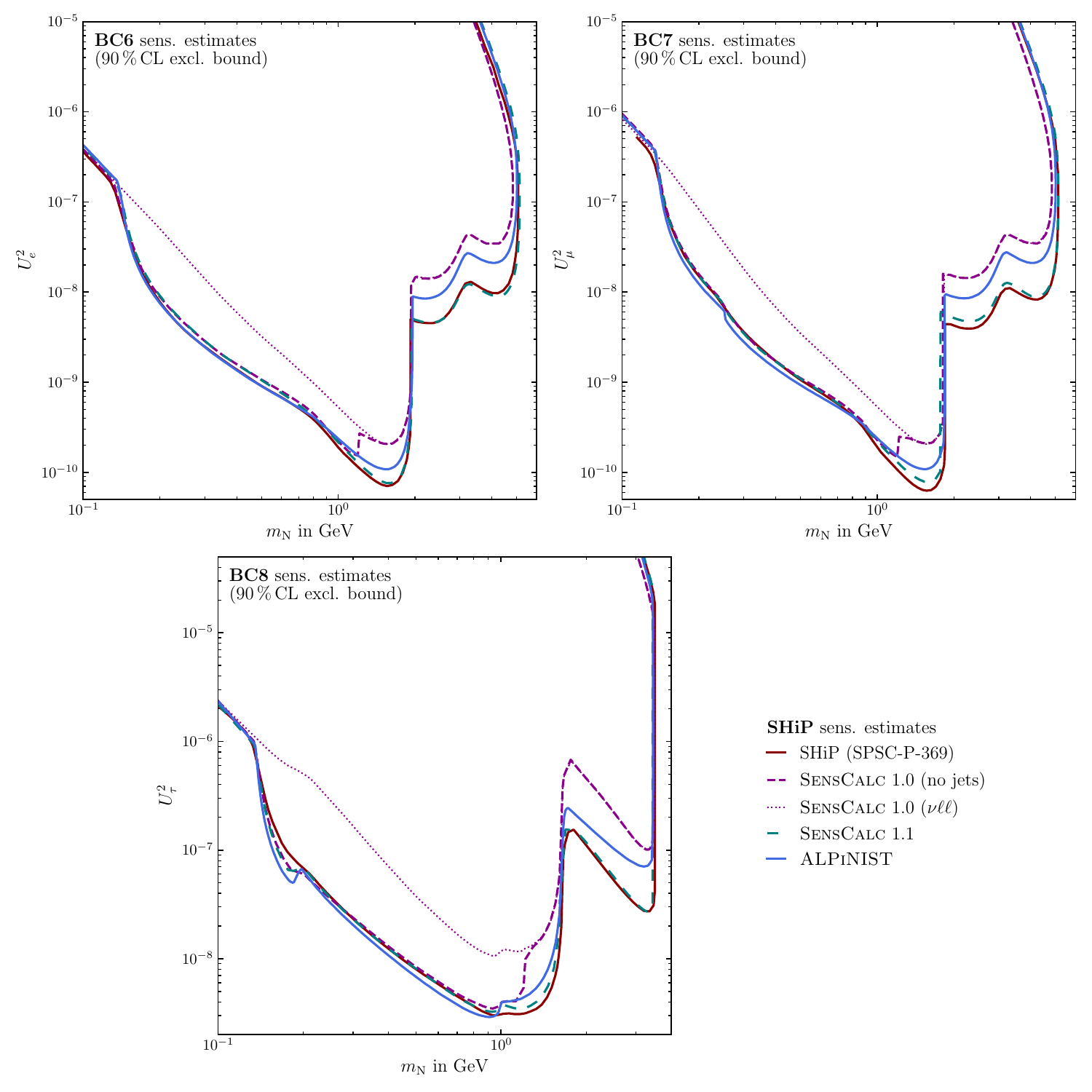}
        \caption{Predicted sensitivity ($90\,\%\,\mathrm{CL}$ excl. bound) for a electronphilic (top left), muonphilic (top right) and tauphilic (bottom) HNLs of the SHiP experiment at ECN3.
        The red lines show the published prediction by the BDF/SHiP collaboration~\cite{SHiP2023}, relying on the semi-analytic light-weight-MC \textsc{SensCalc}~\cite{Ovchynnikov:2023cry}.
        Using the same tool, but not including hadronising open-quark-final-states (hadronic final states) as signal results in the purple dashed (dotted) line.
        The dashed teal curves show the expectation using the latest version of \textsc{SensCalc} available as of the writing of this paper, considering all available final states as signal.
        The blue solid lines show the \alpi prediction using the same meson inputs as \textsc{SensCalc}~\cite{CERN-SHiP-NOTE-2015-009}.}
        \label{fig:SHiP_comparison}
    \end{figure}

    Since its first inception, the concept design of the SHiP experiment has undergone major changes.
    In this section we present the \alpi sensitivity estimates for the latest implementation, following the latest proposal by the BDF/SHiP collaboration~\cite{SHiP2023}, the key parameters of which are summarised in table~\ref{tab:experiment_geometries}.
    We include in our comparisons estimates of the semi-analytic light-weight-MC \textsc{SensCalc}~\cite{Ovchynnikov:2023cry} which has been widely employed in phenomenological studies of SHiP sensitivities.\footnote{We used v1.0.6.1 to reproduce the BDF/SHiP results~\cite{SHiP2023}.
    We adjusted the mass step sizes with respect to the standard settings of this version resulting in slight differences between contours. 
    This, as well as improved calculation of the decay products acceptance are implemented in versions after v1.1.2.1, leading to the dashed teal sensitivity curves presented in figure~\ref{fig:SHiP_comparison}.\label{fn:SC_binning}}
    
    Important components of the SHiP experiment are a $4\times 6 \,\mathrm{m^2}$ spectrometer system just behind the decay volume (DV), featuring a $\SI{162.5}{mT}\times \SI{4}{m}$ magnet.
    Further downstream, the ${5.3\times 6.7}\,\mathrm{m^2}$ electromagnetic calorimeter, and a subsequent ${5.2 \times 6.6}\,\mathrm{m^2}$ muon/hadronic calorimeter are foreseen. 
    Beyond the resulting geometrical cuts, we use the signal cuts as prescribed by BDF/SHiP~\cite{SHiP2023} on final state kinematics
    \begin{enumerate}
        \setlength\itemsep{0em}
        \item $\SI{1}{GeV\, c^{-1}}< p_\mathrm{track}$,
        \item $\min |\Vec{r}_\mathrm{vtx2} - \Vec{r}_\mathrm{DV wall}| > \SI{5}{cm}$,
    \end{enumerate}
    which demands that all observable final state particles have a momentum greater than $\SI{1}{GeV\, c^{-1}}$ and the reconstructed decay vertex be $\SI{5}{cm}$ away from the inner wall of the decay volume.
    These are complemented for decay channels including a neutrino by the additional requirements
    \begin{enumerate}
        \setlength\itemsep{0em}
        \setcounter{enumi}{2}
        \item $ z_\mathrm{vtx2} - z_\mathrm{DV} > \SI{1}{m}$,
        \item $b_\mathrm{target} < \SI{2.5}{m}$.
    \end{enumerate}
    Here, $b_\mathrm{target}$ is the impact parameter of the reconstructed FIP track with respect to the beam axis at the interaction point.

    The resulting parameter sensitivities for electronphilic, muonphilic, and tauphilic HNLs are presented in figure~\ref{fig:SHiP_comparison}.
    In all three scenarios, the estimates are in reasonable agreement with the literature values~\cite{SHiP2023}.
    Just above the $\SI{1}{GeV}$ threshold the predictions including all possible final states of \textsc{SensCalc} and \alpi diverge in all Benchmark Cases.
    This is due to the fact, that in this regime semileptonic decays including multiple hadrons begin to dominate the total width. 
    \textsc{SensCalc} models these as decays into quarks with subsequent hadronisation, while \alpi considers only the final states with a single meson as signal, taking multi-hadron final states into account only as a contribution to the total HNL decay width as presented in section~\ref{sec:HNL_decay_width}.
    Above this threshold, the purple dashed lines showcase the sensitivity to only fully leptonic decays and are in reasonable agreement with the \alpi generated curves.
    Another point of disagreement is the sensitivity just below $m_\mathrm{N}=\SI{200}{MeV}$ in the tauphilic case, where the decay chain $D_s\to \tau (\mathrm{N} \to \pi\nu)$ gives the dominant contribution to the sensitivity. 
    This disagreement is less significant with our evaluation using \textsc{SensCalc}.\footref{fn:SC_binning}

\subsection{CHARM experiment} \label{sec:CHARM_validation}

    \begin{figure}
        \centering
        \includegraphics[width=.95\textwidth]{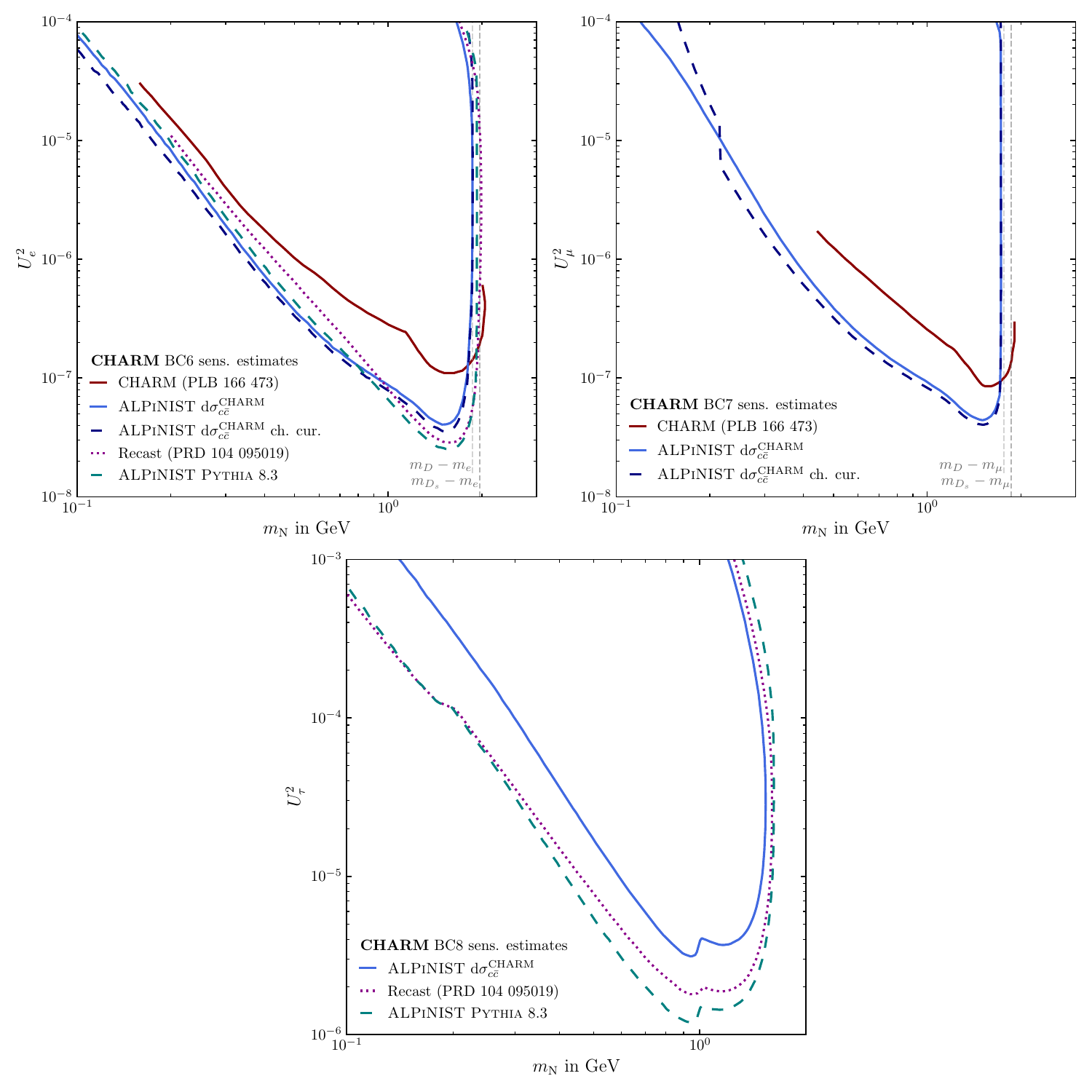}
        \caption{Comparison of the CHARM experiments HNL sensitivity estimates ($90\,\%\,\mathrm{CL}$ excl. bound) using \alpi to those by the CHARM collaboration~\cite{CHARM:1985nku} and a more recent recast of the search~\cite{PhysRevD.104.095019}. 
        Note that the \alpi curves using the $\mathrm{d}\sigma_{c\bar c}^\mathrm{CHARM}$ differential production cross sections do not contain contributions from $D_s$ mesons~\cite{CHARM:1985nku} relevant above $m_\mathrm{N}\simeq\SI{600}{MeV}$ for BC6 and BC7, and at all masses for BC8.} 
        \label{fig:CHARM_literature_comparison}
    \end{figure}

    In the scenario of electron and muon-coupling dominance, one of the strongest HNL exclusions is set by a search performed by the CHARM experiment.\footnote{These results are complemented by a re-interpretation of the CHARM limits in terms of $\tau$ couplings  through a recast~\cite{PhysRevD.104.095019}.}
    Around 2$\times 10^{18}$ protons of $\SI{400}{GeV}$ were dumped onto a thick copper beam-dump. 
    A search was performed looking for visible decays with electrons/muons in the final state in the 35 m long decay volume~\cite{CHARM:1985nku}.
    %with a spectrometer of around 2.4$\times$2.4 $m^2$ cross section 
    In order to validate \alpi against this limit, some understanding of the shape and overall numbers of D-mesons assumed in the analysis is needed.
    In the final version of the search~\cite{CHARM:1985nku}, the input assumption for $N_D$ (the number of $D$ mesons produced in the dump) is mentioned only as scaling with another measured quantity, not as an absolute number.
    The scaling procedure is explained in an earlier publication by CHARM from 1983~\cite{BERGSMA1983361}. 
    It takes as a reference the number of prompt single-muon events observed in the CHARM calorimeter in order to infer $N_D$. 
    In terms of differential $D$ meson distribution, CHARM quote eq.~(\ref{eq:diff_meson_parametrisation}) with $a=2$, $b=0$, and where in the earlier paper~\cite{BERGSMA1983361}, the $x_F$ exponent is taken to be 4, while in ref.~\cite{CHARM:1985nku}, 5 is assumed.
    
    While we are able to re-reproduce the overall shape of the CHARM exclusion\footnote{We would like to observe that the quality of the printed contour limit at large masses ($\sim\SI{2}{GeV}$) in figures~2,3 of \cite{CHARM:1985nku}, including the BEBC contour, do not exactly facilitate the contour validation. Partially, limits seem to have been drawn beyond $\sim\SI{2}{GeV}$, in contrast to the performed measurements and what is kinematically possible.}, the absolute value of the contour is not reproducible with our understanding of the d$\sigma_{cc}$ assumed.
    We attribute this to the fact that we are unable to fully follow the scaling procedure proposed in ref.~\cite{BERGSMA1983361}, given the 
     very satisfactory reproduction of the BEBC limit with \alpi.
     
    Figure~\ref{fig:CHARM_literature_comparison} shows a comparison between the sensitivity estimate for the CHARM experiment to HNLs using \alpi and literature.
    The differential production cross section used  in an attempt to replicate the data published by the CHARM collaboration match those presented in said publication~\cite{CHARM:1985nku}. 
    The blue \alpi curves are presented using production exclusively from $D^\pm$ and $D^0$ mesons with only charged current (ch. cur.), as well as also including neutral current mediated decays of the HNL. 
    To compare to a more modern recast of the experiment's results~\cite{PhysRevD.104.095019}, we also present the \alpi curves as given by \pyth~8.3 including production from $D_s$ mesons.

\bibliographystyle{JHEP_improved}
\bibliography{biblio}

\providecommand{\href}[2]{#2}\begingroup\raggedright\begin{thebibliography}{100}

\bibitem{Beacham:2019nyx}
J.~Beacham et~al., \href{http://dx.doi.org/10.1088/1361-6471/ab4cd2}{{\it
  {Physics Beyond Colliders at CERN: Beyond the Standard Model Working Group
  Report}}, } {\em J. Phys. G} {\bf 47} (2020), no.~1 010501,
  [\href{http://arxiv.org/abs/1901.09966}{{\tt 1901.09966}}].

\bibitem{Antel:2023hkf}
C.~Antel et~al., \href{http://dx.doi.org/10.1140/epjc/s10052-023-12168-5}{{\it
  {Feebly Interacting Particles: FIPs 2022 workshop report}}, } {\em Eur. Phys.
  J. C} {\bf 83} (2023) 1122, [\href{http://arxiv.org/abs/2305.01715}{{\tt
  2305.01715}}].

\bibitem{babyiaxo}
{\bf IAXO Collaboration}, A.~Abeln et~al.,
  \href{http://dx.doi.org/10.1007/JHEP05(2021)137}{{\it {Conceptual design of
  BabyIAXO, the intermediate stage towards the International Axion
  Observatory}}, } {\em JHEP} {\bf 05} (2021) 137,
  [\href{http://arxiv.org/abs/2010.12076}{{\tt 2010.12076}}].

\bibitem{Dutta:2024kuj}
B.~Dutta, W.-C. Huang, D.~Kim, J.~L. Newstead, J.-C. Park, et~al., {\it
  {Exciting Prospects for Dark Matter at Large-Volume Neutrino Detectors}},
  \href{http://arxiv.org/abs/2402.04184}{{\tt 2402.04184}}.

\bibitem{FASER:2018eoc}
{\bf FASER Collaboration}, A.~Ariga et~al.,
  \href{http://dx.doi.org/10.1103/PhysRevD.99.095011}{{\it
  {FASER\textquoteright{}s physics reach for long-lived particles}}, } {\em
  Phys. Rev. D} {\bf 99} (2019), no.~9 095011,
  [\href{http://arxiv.org/abs/1811.12522}{{\tt 1811.12522}}].

\bibitem{Gorkavenko:2023nbk}
V.~Gorkavenko, B.~Jashal, V.~Kholoimov, Y.~Kyselov, D.~Mendoza, et~al., {\it
  {LHCb potential to discover long-lived new physics particles with lifetimes
  above 100 ps}},  \href{http://arxiv.org/abs/2312.14016}{{\tt 2312.14016}}.

\bibitem{MATHUSLA:2018bqv}
{\bf MATHUSLA Collaboration}, C.~Alpigiani et~al., {\it {A Letter of Intent for
  MATHUSLA: A Dedicated Displaced Vertex Detector above ATLAS or CMS.}},
  \href{http://arxiv.org/abs/1811.00927}{{\tt 1811.00927}}.

\bibitem{SHiP2023}
{\bf SHiP Collaboration}, R.~Albanese, J.~Alt, A.~Alexandrov, S.~Aoki,
  D.~Aritunov, et~al., \href{https://cds.cern.ch/record/2878604}{{\it {BDF/SHiP
  at the ECN3 high-intensity beam facility}}, } tech. rep., CERN, Geneva, 2023.

\bibitem{SHADOWS2023}
{\bf SHADOWS Collaboration}, M.~Alviggi, S.~Bachmann, W.~Baldini, A.~Balla,
  M.~Barth, et~al., \href{https://cds.cern.ch/record/2878470}{{\it {SHADOWS
  Technical Proposal}}, } tech. rep., CERN, Geneva, 2023.

\bibitem{HIKE2023}
{\bf HIKE Collaboration}, M.~U. Ashraf et~al., {\it {High Intensity Kaon
  Experiments (HIKE) at the CERN SPS Proposal for Phases 1 and 2}},
  \href{http://arxiv.org/abs/2311.08231}{{\tt 2311.08231}}.

\bibitem{Giffin:2022rei}
P.~Giffin, S.~Gori, Y.-D. Tsai, and D.~Tuckler,
  \href{http://dx.doi.org/10.1007/JHEP04(2023)046}{{\it {Heavy neutral leptons
  at beam dump experiments of future lepton colliders}}, } {\em JHEP} {\bf 04}
  (2023) 046, [\href{http://arxiv.org/abs/2206.13745}{{\tt 2206.13745}}].

\bibitem{CHARM:1980yym}
{\bf CHARM Collaboration}, M.~Jonker et~al.,
  \href{http://dx.doi.org/10.1016/0370-2693(81)91123-0}{{\it {Experimental
  Study of Neutral Current and Charged Current Neutrino Cross-Sections}}, }
  {\em Phys. Lett. B} {\bf 99} (1981) 265. [Erratum: Phys.Lett.B 100, 520
  (1981), Erratum: Phys.Lett.B 103, 469 (1981)].

\bibitem{CCFRNuTeV:1996vbm}
{\bf CCFR/NuTeV Collaboration}, A.~Romosan et~al.,
  \href{http://dx.doi.org/10.1103/PhysRevLett.78.2912}{{\it {A High statistics
  search for muon-neutrino (anti-muon-neutrino) --\ensuremath{>}
  electron-neutrino (anti-electron-neutrino) oscillations in the small mixing
  angle regime}}, } {\em Phys. Rev. Lett.} {\bf 78} (1997) 2912--2915,
  [\href{http://arxiv.org/abs/hep-ex/9611013}{{\tt hep-ex/9611013}}].

\bibitem{DUNE:2020fgq}
{\bf DUNE Collaboration}, B.~Abi et~al.,
  \href{http://dx.doi.org/10.1140/epjc/s10052-021-09007-w}{{\it {Prospects for
  beyond the Standard Model physics searches at the Deep Underground Neutrino
  Experiment}}, } {\em Eur. Phys. J. C} {\bf 81} (2021), no.~4 322,
  [\href{http://arxiv.org/abs/2008.12769}{{\tt 2008.12769}}].

\bibitem{NA62:2017rwk}
{\bf NA62 Collaboration}, E.~Cortina~Gil et~al.,
  \href{http://dx.doi.org/10.1088/1748-0221/12/05/P05025}{{\it {The Beam and
  detector of the NA62 experiment at CERN}}, } {\em JINST} {\bf 12} (2017),
  no.~05 P05025, [\href{http://arxiv.org/abs/1703.08501}{{\tt 1703.08501}}].

\bibitem{PIONEER:2022yag}
{\bf PIONEER Collaboration}, W.~Altmannshofer et~al., {\it {PIONEER: Studies of
  Rare Pion Decays}},  \href{http://arxiv.org/abs/2203.01981}{{\tt
  2203.01981}}.

\bibitem{Minkowski:1977sc}
P.~Minkowski, \href{http://dx.doi.org/10.1016/0370-2693(77)90435-X}{{\it {$\mu
  \to e\gamma$ at a Rate of One Out of $10^{9}$ Muon Decays?}}, } {\em Phys.
  Lett. B} {\bf 67} (1977) 421--428.

\bibitem{Gell-Mann:1979vob}
M.~Gell-Mann, P.~Ramond, and R.~Slansky, {\it {Complex Spinors and Unified
  Theories}},  {\em Conf. Proc. C} {\bf 790927} (1979) 315--321,
  [\href{http://arxiv.org/abs/1306.4669}{{\tt 1306.4669}}].

\bibitem{Shaposhnikov:2006nn}
M.~Shaposhnikov, \href{http://dx.doi.org/10.1016/j.nuclphysb.2006.11.003}{{\it
  {A Possible symmetry of the nuMSM}}, } {\em Nucl. Phys. B} {\bf 763} (2007)
  49--59, [\href{http://arxiv.org/abs/hep-ph/0605047}{{\tt hep-ph/0605047}}].

\bibitem{Dodelson:1993je}
S.~Dodelson and L.~M. Widrow,
  \href{http://dx.doi.org/10.1103/PhysRevLett.72.17}{{\it {Sterile-neutrinos as
  dark matter}}, } {\em Phys. Rev. Lett.} {\bf 72} (1994) 17--20,
  [\href{http://arxiv.org/abs/hep-ph/9303287}{{\tt hep-ph/9303287}}].

\bibitem{Shi:1998km}
X.-D. Shi and G.~M. Fuller,
  \href{http://dx.doi.org/10.1103/PhysRevLett.82.2832}{{\it {A New dark matter
  candidate: Nonthermal sterile neutrinos}}, } {\em Phys. Rev. Lett.} {\bf 82}
  (1999) 2832--2835, [\href{http://arxiv.org/abs/astro-ph/9810076}{{\tt
  astro-ph/9810076}}].

\bibitem{Abazajian:2001nj}
K.~Abazajian, G.~M. Fuller, and M.~Patel,
  \href{http://dx.doi.org/10.1103/PhysRevD.64.023501}{{\it {Sterile neutrino
  hot, warm, and cold dark matter}}, } {\em Phys. Rev. D} {\bf 64} (2001)
  023501, [\href{http://arxiv.org/abs/astro-ph/0101524}{{\tt
  astro-ph/0101524}}].

\bibitem{Asaka:2006nq}
T.~Asaka, M.~Laine, and M.~Shaposhnikov,
  \href{http://dx.doi.org/10.1088/1126-6708/2007/01/091}{{\it {Lightest sterile
  neutrino abundance within the nuMSM}}, } {\em JHEP} {\bf 01} (2007) 091,
  [\href{http://arxiv.org/abs/hep-ph/0612182}{{\tt hep-ph/0612182}}]. [Erratum:
  JHEP 02, 028 (2015)].

\bibitem{Boyarsky:2009ix}
A.~Boyarsky, O.~Ruchayskiy, and M.~Shaposhnikov,
  \href{http://dx.doi.org/10.1146/annurev.nucl.010909.083654}{{\it {The Role of
  sterile neutrinos in cosmology and astrophysics}}, } {\em Ann. Rev. Nucl.
  Part. Sci.} {\bf 59} (2009) 191--214,
  [\href{http://arxiv.org/abs/0901.0011}{{\tt 0901.0011}}].

\bibitem{Fukugita:1986hr}
M.~Fukugita and T.~Yanagida,
  \href{http://dx.doi.org/10.1016/0370-2693(86)91126-3}{{\it {Baryogenesis
  Without Grand Unification}}, } {\em Phys. Lett. B} {\bf 174} (1986) 45--47.

\bibitem{Akhmedov:1998qx}
E.~K. Akhmedov, V.~A. Rubakov, and A.~Y. Smirnov,
  \href{http://dx.doi.org/10.1103/PhysRevLett.81.1359}{{\it {Baryogenesis via
  neutrino oscillations}}, } {\em Phys. Rev. Lett.} {\bf 81} (1998) 1359--1362,
  [\href{http://arxiv.org/abs/hep-ph/9803255}{{\tt hep-ph/9803255}}].

\bibitem{Asaka:2005pn}
T.~Asaka and M.~Shaposhnikov,
  \href{http://dx.doi.org/10.1016/j.physletb.2005.06.020}{{\it {The $\nu$MSM,
  dark matter and baryon asymmetry of the universe}}, } {\em Phys. Lett. B}
  {\bf 620} (2005) 17--26, [\href{http://arxiv.org/abs/hep-ph/0505013}{{\tt
  hep-ph/0505013}}].

\bibitem{Klaric:2021cpi}
J.~Klari\'c, M.~Shaposhnikov, and I.~Timiryasov,
  \href{http://dx.doi.org/10.1103/PhysRevD.104.055010}{{\it {Reconciling
  resonant leptogenesis and baryogenesis via neutrino oscillations}}, } {\em
  Phys. Rev. D} {\bf 104} (2021), no.~5 055010,
  [\href{http://arxiv.org/abs/2103.16545}{{\tt 2103.16545}}].

\bibitem{Drewes:2021nqr}
M.~Drewes, Y.~Georis, and J.~Klari\'c,
  \href{http://dx.doi.org/10.1103/PhysRevLett.128.051801}{{\it {Mapping the
  Viable Parameter Space for Testable Leptogenesis}}, } {\em Phys. Rev. Lett.}
  {\bf 128} (2022), no.~5 051801, [\href{http://arxiv.org/abs/2106.16226}{{\tt
  2106.16226}}].

\bibitem{HNL}
S.~N. N.~L. Constraints. \url{www.sterile-neutrino.org}.

\bibitem{Ruf:2115534}
{\bf SHiPcollaboration Collaboration}, T.~Ruf and H.~Dijkstra,
  \href{https://cds.cern.ch/record/2115534}{{\it {Heavy Flavour Cascade
  Production in a Beam Dump}}, }.

\bibitem{Gorkavenko:2021mpj}
V.~M. Gorkavenko, Y.~R. Borysenkova, and M.~S. Tsarenkova,
  \href{http://dx.doi.org/10.1088/1361-6471/ac1394}{{\it {Production of
  GeV-scale heavy neutral leptons in three-body decays. Comparison with the
  PYTHIA approach}}, } {\em J. Phys. G} {\bf 48} (2021), no.~10 105001,
  [\href{http://arxiv.org/abs/2103.11494}{{\tt 2103.11494}}].

\bibitem{Dobrich:2019dxc}
B.~D\"obrich, J.~Jaeckel, and T.~Spadaro,
  \href{http://dx.doi.org/10.1007/JHEP05(2019)213}{{\it {Light in the beam dump
  - ALP production from decay photons in proton beam-dumps}}, } {\em JHEP} {\bf
  05} (2019) 213, [\href{http://arxiv.org/abs/1904.02091}{{\tt 1904.02091}}].
  [Erratum: JHEP 10, 046 (2020)].

\bibitem{Lourenco:2006vw}
C.~Lourenco and H.~K. Wohri,
  \href{http://dx.doi.org/10.1016/j.physrep.2006.05.005}{{\it {Heavy flavour
  hadro-production from fixed-target to collider energies}}, } {\em Phys.
  Rept.} {\bf 433} (2006) 127--180,
  [\href{http://arxiv.org/abs/hep-ph/0609101}{{\tt hep-ph/0609101}}].

\bibitem{SHiP:2024oua}
{\bf SHiP Collaboration}, C.~Ahdida et~al.,
  \href{http://dx.doi.org/10.1140/epjc/s10052-024-12655-3}{{\it {Reconstruction
  of 400 GeV/c proton interactions with the SHiP-charm project}}, } {\em Eur.
  Phys. J. C} {\bf 84} (2024), no.~6 562,
  [\href{http://arxiv.org/abs/2406.04730}{{\tt 2406.04730}}].

\bibitem{Jerhot:2022chi}
J.~Jerhot, B.~D\"obrich, F.~Ertas, F.~Kahlhoefer, and T.~Spadaro,
  \href{http://dx.doi.org/10.1007/JHEP07(2022)094}{{\it {ALPINIST: Axion-Like
  Particles In Numerous Interactions Simulated and Tabulated}}, } {\em JHEP}
  {\bf 07} (2022) 094, [\href{http://arxiv.org/abs/2201.05170}{{\tt
  2201.05170}}].

\bibitem{Afik:2023mhj}
Y.~Afik, B.~D\"obrich, J.~Jerhot, Y.~Soreq, and K.~Tobioka,
  \href{http://dx.doi.org/10.1103/PhysRevD.108.055007}{{\it {Probing long-lived
  axions at the KOTO experiment}}, } {\em Phys. Rev. D} {\bf 108} (2023), no.~5
  055007, [\href{http://arxiv.org/abs/2303.01521}{{\tt 2303.01521}}].

\bibitem{Buonocore:2018xjk}
L.~Buonocore, C.~Frugiuele, F.~Maltoni, O.~Mattelaer, and F.~Tramontano,
  \href{http://dx.doi.org/10.1007/JHEP05(2019)028}{{\it {Event generation for
  beam dump experiments}}, } {\em JHEP} {\bf 05} (2019) 028,
  [\href{http://arxiv.org/abs/1812.06771}{{\tt 1812.06771}}].

\bibitem{Ovchynnikov:2023cry}
M.~Ovchynnikov, J.-L. Tastet, O.~Mikulenko, and K.~Bondarenko,
  \href{http://dx.doi.org/10.1103/PhysRevD.108.075028}{{\it {Sensitivities to
  feebly interacting particles: Public and unified calculations}}, } {\em Phys.
  Rev. D} {\bf 108} (2023), no.~7 075028,
  [\href{http://arxiv.org/abs/2305.13383}{{\tt 2305.13383}}].

\bibitem{Domingo:2023dew}
F.~Domingo, J.~G\"unther, J.~S. Kim, and Z.~S. Wang, {\it {A C++ program for
  estimating detector sensitivities to long-lived particles: Displaced Decay
  Counter}},  \href{http://arxiv.org/abs/2308.07371}{{\tt 2308.07371}}.

\bibitem{Dobrich:2024ajq}
B.~Dobrich, \href{http://dx.doi.org/10.22323/1.457.0012}{{\it {Exotic particle
  searches at beam-dumps \textendash{} dos and don\textquoteright{}ts}}, } {\em
  PoS} {\bf WIFAI2023} (2024) 012.

\bibitem{Abdullahi:2022jlv}
A.~M. Abdullahi et~al., \href{http://dx.doi.org/10.1088/1361-6471/ac98f9}{{\it
  {The present and future status of heavy neutral leptons}}, } {\em J. Phys. G}
  {\bf 50} (2023), no.~2 020501, [\href{http://arxiv.org/abs/2203.08039}{{\tt
  2203.08039}}].

\bibitem{Bondarenko:2018ptm}
K.~Bondarenko, A.~Boyarsky, D.~Gorbunov, and O.~Ruchayskiy,
  \href{http://dx.doi.org/10.1007/JHEP11(2018)032}{{\it {Phenomenology of
  GeV-scale Heavy Neutral Leptons}}, } {\em JHEP} {\bf 11} (2018) 032,
  [\href{http://arxiv.org/abs/1805.08567}{{\tt 1805.08567}}].

\bibitem{Shrock:1980ct}
R.~E. Shrock, \href{http://dx.doi.org/10.1103/PhysRevD.24.1232}{{\it {General
  Theory of Weak Leptonic and Semileptonic Decays. 1. Leptonic Pseudoscalar
  Meson Decays, with Associated Tests For, and Bounds on, Neutrino Masses and
  Lepton Mixing}}, } {\em Phys. Rev. D} {\bf 24} (1981) 1232.

\bibitem{Shrock:1981wq}
R.~E. Shrock, \href{http://dx.doi.org/10.1103/PhysRevD.24.1275}{{\it {General
  Theory of Weak Processes Involving Neutrinos. 2. Pure Leptonic Decays}}, }
  {\em Phys. Rev. D} {\bf 24} (1981) 1275.

\bibitem{Johnson:1997cj}
L.~M. Johnson, D.~W. McKay, and T.~Bolton,
  \href{http://dx.doi.org/10.1103/PhysRevD.56.2970}{{\it {Extending sensitivity
  for low mass neutral heavy lepton searches}}, } {\em Phys. Rev. D} {\bf 56}
  (1997) 2970--2981, [\href{http://arxiv.org/abs/hep-ph/9703333}{{\tt
  hep-ph/9703333}}].

\bibitem{Gorbunov:2007ak}
D.~Gorbunov and M.~Shaposhnikov,
  \href{http://dx.doi.org/10.1088/1126-6708/2007/10/015}{{\it {How to find
  neutral leptons of the $\nu$MSM?}}, } {\em JHEP} {\bf 10} (2007) 015,
  [\href{http://arxiv.org/abs/0705.1729}{{\tt 0705.1729}}]. [Erratum: JHEP 11,
  101 (2013)].

\bibitem{Atre:2009rg}
A.~Atre, T.~Han, S.~Pascoli, and B.~Zhang,
  \href{http://dx.doi.org/10.1088/1126-6708/2009/05/030}{{\it {The Search for
  Heavy Majorana Neutrinos}}, } {\em JHEP} {\bf 05} (2009) 030,
  [\href{http://arxiv.org/abs/0901.3589}{{\tt 0901.3589}}].

\bibitem{Ruchayskiy:2011aa}
O.~Ruchayskiy and A.~Ivashko,
  \href{http://dx.doi.org/10.1007/JHEP06(2012)100}{{\it {Experimental bounds on
  sterile neutrino mixing angles}}, } {\em JHEP} {\bf 06} (2012) 100,
  [\href{http://arxiv.org/abs/1112.3319}{{\tt 1112.3319}}].

\bibitem{PhysRevD.104.095019}
I.~Boiarska, A.~Boyarsky, O.~Mikulenko, and M.~Ovchynnikov,
  \href{https://link.aps.org/doi/10.1103/PhysRevD.104.095019}{{\it Constraints
  from the charm experiment on heavy neutral leptons with tau mixing}, } {\em
  Phys. Rev. D} {\bf 104} (Nov, 2021) 095019.

\bibitem{CERN-SHiP-NOTE-2015-009}
{\bf SHiP Collaboration}, T.~Ruf and H.~Dijkstra,
  \href{https://cds.cern.ch/record/2115534}{{\it {Heavy Flavour Cascade
  Production in a Beam Dump}}, }.

\bibitem{Moghaddam:2022tac}
{\bf DUNE Collaboration}, Z.~G. Moghaddam, {\it {Sensitivity to Heavy Neutral
  Leptons with the SAND detector at the DUNE ND complex}},
  \href{http://arxiv.org/abs/2209.01899}{{\tt 2209.01899}}.

\bibitem{Fieg:2023kld}
M.~Fieg, F.~Kling, H.~Schulz, and T.~Sj\"ostrand,
  \href{http://dx.doi.org/10.1103/PhysRevD.109.016010}{{\it {Tuning pythia for
  forward physics experiments}}, } {\em Phys. Rev. D} {\bf 109} (2024), no.~1
  016010, [\href{http://arxiv.org/abs/2309.08604}{{\tt 2309.08604}}].

\bibitem{Vogt:2007aw}
R.~Vogt, \href{http://dx.doi.org/10.1140/epjst/e2008-00603-5}{{\it {The Total
  charm cross-section}}, } {\em Eur. Phys. J. ST} {\bf 155} (2008) 213--222,
  [\href{http://arxiv.org/abs/0709.2531}{{\tt 0709.2531}}].

\bibitem{Cacciari:1998it}
M.~Cacciari, M.~Greco, and P.~Nason,
  \href{http://dx.doi.org/10.1088/1126-6708/1998/05/007}{{\it {The $p_T$
  spectrum in heavy-flavour hadroproduction.}}, } {\em JHEP} {\bf 05} (1998)
  007, [\href{http://arxiv.org/abs/hep-ph/9803400}{{\tt hep-ph/9803400}}].

\bibitem{Cacciari:2001td}
M.~Cacciari, S.~Frixione, and P.~Nason,
  \href{http://dx.doi.org/10.1088/1126-6708/2001/03/006}{{\it {The p(T)
  spectrum in heavy flavor photoproduction}}, } {\em JHEP} {\bf 03} (2001) 006,
  [\href{http://arxiv.org/abs/hep-ph/0102134}{{\tt hep-ph/0102134}}].

\bibitem{Buckley:2014ana}
A.~Buckley, J.~Ferrando, S.~Lloyd, K.~Nordstr\"om, B.~Page, et~al.,
  \href{http://dx.doi.org/10.1140/epjc/s10052-015-3318-8}{{\it {LHAPDF6: parton
  density access in the LHC precision era}}, } {\em Eur. Phys. J. C} {\bf 75}
  (2015) 132, [\href{http://arxiv.org/abs/1412.7420}{{\tt 1412.7420}}].

\bibitem{NNPDF:2021njg}
{\bf NNPDF Collaboration}, R.~D. Ball et~al.,
  \href{http://dx.doi.org/10.1140/epjc/s10052-022-10328-7}{{\it {The path to
  proton structure at 1\% accuracy}}, } {\em Eur. Phys. J. C} {\bf 82} (2022),
  no.~5 428, [\href{http://arxiv.org/abs/2109.02653}{{\tt 2109.02653}}].

\bibitem{Frixione:1994nb}
S.~Frixione, M.~L. Mangano, P.~Nason, and G.~Ridolfi,
  \href{http://dx.doi.org/10.1016/0550-3213(94)90213-5}{{\it {Charm and bottom
  production: Theoretical results versus experimental data}}, } {\em Nucl.
  Phys. B} {\bf 431} (1994) 453--483.

\bibitem{AbrilPla2023}
O.~Abril-Pla, V.~Andreani, C.~Carroll, L.~Dong, C.~J. Fonnesbeck, et~al.,
  \href{http://dx.doi.org/10.7717/peerj-cs.1516}{{\it Pymc: a modern, and
  comprehensive probabilistic programming framework in python}, } {\em PeerJ
  Computer Science} {\bf 9} (Sept., 2023) e1516.

\bibitem{SVD-2:2017ovj}
{\bf SVD-2 Collaboration}, A.~Aleev et~al.,
  \href{http://dx.doi.org/10.1140/epja/i2017-12230-9}{{\it {Charmed particles
  production in pA -interactions at $\surd s = 11.8$ GeV}}, } {\em Eur. Phys.
  J. A} {\bf 53} (2017), no.~3 45.

\bibitem{ACCMOR:1988pxc}
{\bf ACCMOR Collaboration}, S.~Barlag et~al.,
  \href{http://dx.doi.org/10.1007/BF01555972}{{\it {Production of D, D* and
  D(s) Mesons in 200-GeV/c pi-, K- and p Si Interactions}}, } {\em Z. Phys. C}
  {\bf 39} (1988) 451.

\bibitem{E769:1996jqf}
{\bf E769 Collaboration}, G.~A. Alves et~al.,
  \href{http://dx.doi.org/10.1103/PhysRevLett.77.2388}{{\it {Forward
  cross-sections for production of D+, D0, D(s), D*+ and Lambda(c) in 250-GeV
  pi+-, K+-, and p - nucleon interactions}}, } {\em Phys. Rev. Lett.} {\bf 77}
  (1996) 2388--2391. [Erratum: Phys.Rev.Lett. 81, 1537 (1998)].

\bibitem{Adamovich:1992cv}
M.~Adamovich et~al., \href{http://dx.doi.org/10.1016/0920-5632(92)90053-U}{{\it
  {Results on charm physics from WA82}}, } {\em Nucl. Phys. B Proc. Suppl.}
  {\bf 27} (1992) 212--218.

\bibitem{LEBC-EHS:1983llu}
{\bf LEBC-EHS Collaboration}, M.~Aguilar-Benitez et~al.,
  \href{http://dx.doi.org/10.1016/0370-2693(84)90489-1}{{\it {$D$ Meson
  Branching Ratios and Hadronic Charm Production Cross-sections}}, } {\em Phys.
  Lett. B} {\bf 135} (1984) 237--242.

\bibitem{LEBC-EHS:1988oic}
{\bf LEBC-EHS Collaboration}, M.~Aguilar-Benitez et~al.,
  \href{http://dx.doi.org/10.1007/BF01548848}{{\it {Charm Hadron Properties in
  400-GeV/c p p Interactions}}, } {\em Z. Phys. C} {\bf 40} (1988) 321.

\bibitem{Ammar:1988ta}
R.~Ammar et~al., \href{http://dx.doi.org/10.1103/PhysRevLett.61.2185}{{\it
  {D-Meson Production in 800-GeV/c p Pinteractions}}, } {\em Phys. Rev. Lett.}
  {\bf 61} (1988) 2185--2188.

\bibitem{FermilabE653:1991vmo}
{\bf Fermilab E653 Collaboration}, K.~Kodama et~al.,
  \href{http://dx.doi.org/10.1016/0370-2693(91)90506-L}{{\it {Charm Meson
  Production in 800-GeV/c Proton - Emulsion Interactions}}, } {\em Phys. Lett.
  B} {\bf 263} (1991) 573--578.

\bibitem{HERA-B:2007rfd}
{\bf HERA-B Collaboration}, I.~Abt et~al.,
  \href{http://dx.doi.org/10.1140/epjc/s10052-007-0427-z}{{\it {Measurement of
  D0, D+, D+(s) and D*+ Production in Fixed Target 920-GeV Proton-Nucleus
  Collisions}}, } {\em Eur. Phys. J. C} {\bf 52} (2007) 531--542,
  [\href{http://arxiv.org/abs/0708.1443}{{\tt 0708.1443}}].

\bibitem{E789:1994nhc}
{\bf E789 Collaboration}, M.~J. Leitch et~al.,
  \href{http://dx.doi.org/10.1103/PhysRevLett.72.2542}{{\it {Nuclear dependence
  of neutral D meson production by 800-GeV/c protons}}, } {\em Phys. Rev.
  Lett.} {\bf 72} (1994) 2542--2545.

\bibitem{Jansen:1994bz}
D.~M. Jansen et~al., \href{http://dx.doi.org/10.1103/PhysRevLett.74.3118}{{\it
  {Measurement of the bottom quark production cross-section in 800-GeV/c proton
  - gold collisions}}, } {\em Phys. Rev. Lett.} {\bf 74} (1995) 3118--3121.

\bibitem{HERA-B:2005tnp}
{\bf HERA-B Collaboration}, I.~Abt et~al.,
  \href{http://dx.doi.org/10.1103/PhysRevD.73.052005}{{\it {Improved
  measurement of the b-anti-b production cross section in 920-GeV fixed-target
  proton-nucleus collisions}}, } {\em Phys. Rev. D} {\bf 73} (2006) 052005,
  [\href{http://arxiv.org/abs/hep-ex/0512030}{{\tt hep-ex/0512030}}].

\bibitem{PHENIX:2018dwt}
{\bf PHENIX Collaboration}, C.~Aidala et~al.,
  \href{http://dx.doi.org/10.1103/PhysRevD.99.072003}{{\it {Measurements of
  $\mu\mu$ pairs from open heavy flavor and Drell-Yan in $p+p$ collisions at
  $\sqrt{s}=200$ GeV}}, } {\em Phys. Rev. D} {\bf 99} (2019), no.~7 072003,
  [\href{http://arxiv.org/abs/1805.02448}{{\tt 1805.02448}}].

\bibitem{PHENIX:2020iaj}
{\bf PHENIX Collaboration}, U.~Acharya et~al.,
  \href{http://dx.doi.org/10.1103/PhysRevD.102.092002}{{\it {Production of
  $b\bar{b}$ at forward rapidity in $p$+$p$ collisions at $\sqrt{s}=510$ GeV}},
  } {\em Phys. Rev. D} {\bf 102} (2020), no.~9 092002,
  [\href{http://arxiv.org/abs/2005.14276}{{\tt 2005.14276}}].

\bibitem{UA1:1990vvp}
{\bf UA1 Collaboration}, C.~Albajar et~al.,
  \href{http://dx.doi.org/10.1016/0370-2693(91)90228-I}{{\it {Beauty production
  at the CERN p anti-p collider}}, } {\em Phys. Lett. B} {\bf 256} (1991)
  121--128. [Erratum: Phys.Lett.B 262, 497 (1991)].

\bibitem{ALICE:2012acz}
{\bf ALICE Collaboration}, B.~Abelev et~al.,
  \href{http://dx.doi.org/10.1016/j.physletb.2013.01.069}{{\it {Measurement of
  electrons from beauty hadron decays in $pp$ collisions at $\sqrt{s}=7$ TeV}},
  } {\em Phys. Lett. B} {\bf 721} (2013) 13--23,
  [\href{http://arxiv.org/abs/1208.1902}{{\tt 1208.1902}}]. [Erratum:
  Phys.Lett.B 763, 507--509 (2016)].

\bibitem{LHCb:2010wqx}
{\bf LHCb Collaboration}, R.~Aaij et~al.,
  \href{http://dx.doi.org/10.1016/j.physletb.2010.10.010}{{\it {Measurement of
  $\sigma(pp \to b \bar{b} X)$ at $\sqrt{s}=7~\rm{TeV}$ in the forward
  region}}, } {\em Phys. Lett. B} {\bf 694} (2010) 209--216,
  [\href{http://arxiv.org/abs/1009.2731}{{\tt 1009.2731}}].

\bibitem{LHCb:2016qpe}
{\bf LHCb Collaboration}, R.~Aaij et~al.,
  \href{http://dx.doi.org/10.1103/PhysRevLett.118.052002}{{\it {Measurement of
  the $b$-quark production cross-section in 7 and 13 TeV $pp$ collisions}}, }
  {\em Phys. Rev. Lett.} {\bf 118} (2017), no.~5 052002,
  [\href{http://arxiv.org/abs/1612.05140}{{\tt 1612.05140}}]. [Erratum:
  Phys.Rev.Lett. 119, 169901 (2017)].

\bibitem{ParticleDataGroup:2022pth}
{\bf Particle Data Group Collaboration}, R.~L. Workman et~al.,
  \href{http://dx.doi.org/10.1093/ptep/ptac097}{{\it {Review of Particle
  Physics}}, } {\em PTEP} {\bf 2022} (2022) 083C01.

\bibitem{LEBC-EHS:1987evz}
{\bf LEBC-EHS Collaboration}, M.~Aguilar-Benitez et~al.,
  \href{http://dx.doi.org/10.1016/0370-2693(87)90663-0}{{\it {$D$ Meson
  Production From 400 GeV/$c p p$ Interactions}}, } {\em Phys. Lett. B} {\bf
  189} (1987) 476. [Erratum: Phys.Lett.B 208, 530 (1988)].

\bibitem{Bergsma:1987br}
{\bf CHARM Collaboration}, F.~Bergsma, {\it {Charm production measured in a
  400-GeV proton copper beam dump experiment}},  {\em Annals N. Y. Acad. Sci.}
  {\bf 535} (1988) 506--515.

\bibitem{Boyanovsky:2014una}
D.~Boyanovsky, \href{http://dx.doi.org/10.1103/PhysRevD.90.105024}{{\it {Nearly
  degenerate heavy sterile neutrinos in cascade decay: mixing and
  oscillations}}, } {\em Phys. Rev. D} {\bf 90} (2014), no.~10 105024,
  [\href{http://arxiv.org/abs/1409.4265}{{\tt 1409.4265}}].

\bibitem{Tastet:2019nqj}
J.-L. Tastet and I.~Timiryasov,
  \href{http://dx.doi.org/10.1007/JHEP04(2020)005}{{\it {Dirac vs. Majorana
  HNLs (and their oscillations) at SHiP}}, } {\em JHEP} {\bf 04} (2020) 005,
  [\href{http://arxiv.org/abs/1912.05520}{{\tt 1912.05520}}].

\bibitem{Mikulenko:2023iqq}
O.~Mikulenko, K.~Bondarenko, A.~Boyarsky, and O.~Ruchayskiy, {\it {Unveiling
  new physics with discoveries at Intensity Frontier}},
  \href{http://arxiv.org/abs/2312.05163}{{\tt 2312.05163}}.

\bibitem{jan_jerhot_2022_5844011}
J.~Jerhot, B.~D\"{o}brich, E.~Ertas, F.~Kahlhoefer, and T.~Spadaro,
  \href{https://doi.org/10.5281/zenodo.5844011}{{\it {ALPINIST: v1.0.0}}, }.

\bibitem{Harigel:160549}
G.~G. Harigel, {\em {List of publications covering BEBC experiments}}.
\newblock CERN Yellow Reports: Monographs. CERN, Geneva, 1985.

\bibitem{BEBCWA66:1986err}
{\bf BEBC WA66 Collaboration}, H.~Grassler et~al.,
  \href{http://dx.doi.org/10.1016/0550-3213(86)90246-4}{{\it {Prompt Neutrino
  Production in 400-{GeV} Proton Copper Interactions}}, } {\em Nucl. Phys. B}
  {\bf 273} (1986) 253--274.

\bibitem{FOETH1980203}
H.~Foeth,
  \href{https://www.sciencedirect.com/science/article/pii/0029554X8090703X}{{\it
  The internal picket fence for bebc}, } {\em Nuclear Instruments and Methods}
  {\bf 176} (1980), no.~1 203--206.

\bibitem{Wittgenstein:1972zz}
F.~Wittgenstein, {\it {Preliminary Test Results of BEBC Superconducting
  Magnet}},  {\em eConf} {\bf C720919} (1972) 295.

\bibitem{Sakumoto:1990py}
W.~K. Sakumoto et~al.,
  \href{http://dx.doi.org/10.1016/0168-9002(90)91832-V}{{\it {Calibration of
  the CCFR Target Calorimeter}}, } {\em Nucl. Instrum. Meth. A} {\bf 294}
  (1990) 179--192.

\bibitem{King:1991gs}
B.~J. King et~al., \href{http://dx.doi.org/10.1016/0168-9002(91)90408-I}{{\it
  {Measuring Muon Momenta with the CCFR Neutrino Detector}}, } {\em Nucl.
  Instrum. Meth. A} {\bf 302} (1991) 254--260.

\bibitem{NuTeV:1999kej}
{\bf NuTeV, E815 Collaboration}, A.~Vaitaitis et~al.,
  \href{http://dx.doi.org/10.1103/PhysRevLett.83.4943}{{\it {Search for neutral
  heavy leptons in a high-energy neutrino beam}}, } {\em Phys. Rev. Lett.} {\bf
  83} (1999) 4943--4946, [\href{http://arxiv.org/abs/hep-ex/9908011}{{\tt
  hep-ex/9908011}}].

\bibitem{gori}
W.~would like to thank S. Gori for providing us with the PDF of~this
  APS~contribution. \url{https://meetings.aps.org/Meeting/APR24/Session/D14.5}.

\bibitem{vallee}
C.~P. study~group presentation.
  \url{https://indico.cern.ch/event/1369776/contributions/5760315/attachments/2825823/4936322/PBC_introduction.pdf}.

\bibitem{WA66:1985mfx}
{\bf WA66 Collaboration}, A.~M. Cooper-Sarkar et~al.,
  \href{http://dx.doi.org/10.1016/0370-2693(85)91493-5}{{\it {Search for Heavy
  Neutrino Decays in the {BEBC} Beam Dump Experiment}}, } {\em Phys. Lett. B}
  {\bf 160} (1985) 207--211.

\bibitem{Barouki:2022bkt}
R.~Barouki, G.~Marocco, and S.~Sarkar,
  \href{http://dx.doi.org/10.21468/SciPostPhys.13.5.118}{{\it {Blast from the
  past II: Constraints on heavy neutral leptons from the BEBC WA66 beam dump
  experiment}}, } {\em SciPost Phys.} {\bf 13} (2022) 118,
  [\href{http://arxiv.org/abs/2208.00416}{{\tt 2208.00416}}].

\bibitem{CHARM:1985nku}
{\bf CHARM Collaboration}, F.~Bergsma et~al.,
  \href{http://dx.doi.org/10.1016/0370-2693(86)91601-1}{{\it {A Search for
  Decays of Heavy Neutrinos in the Mass Range 0.5-{GeV} to 2.8-{GeV}}}, } {\em
  Phys. Lett. B} {\bf 166} (1986) 473--478.

\bibitem{BERGSMA1983361}
F.~Bergsma, J.~Dorenbosch, M.~Jonker, C.~Nieuwenhuis, J.~Allaby, et~al.,
  \href{https://www.sciencedirect.com/science/article/pii/0370269383902757}{{\it
  A search for decays of heavy neutrinos}, } {\em Physics Letters B} {\bf 128}
  (1983), no.~5 361--366.

\bibitem{Carvalho:2003pza}
J.~Carvalho, \href{http://dx.doi.org/10.1016/S0375-9474(03)01597-5}{{\it
  {Compilation of cross sections for proton nucleus interactions at the HERA
  energy}}, } {\em Nucl. Phys. A} {\bf 725} (2003) 269--275.

\bibitem{ROPP2022}
P.~D. Group, R.~L. Workman, V.~D. Burkert, V.~Crede, E.~Klempt, et~al.,
  \href{https://doi.org/10.1093/ptep/ptac097}{{\it {Review of Particle
  Physics}}, } {\em Progress of Theoretical and Experimental Physics} {\bf
  2022} (08, 2022) 083C01,
  [\href{http://arxiv.org/abs/https://academic.oup.com/ptep/article-pdf/2022/8/083C01/49175539/ptac097.pdf}{{\tt
  https://academic.oup.com/ptep/article-pdf/2022/8/083C01/49175539/ptac097.pdf}}].

\bibitem{Berryman:2019dme}
J.~M. Berryman, A.~de~Gouvea, P.~J. Fox, B.~J. Kayser, K.~J. Kelly, et~al.,
  \href{http://dx.doi.org/10.1007/JHEP02(2020)174}{{\it {Searches for Decays of
  New Particles in the DUNE Multi-Purpose Near Detector}}, } {\em JHEP} {\bf
  02} (2020) 174, [\href{http://arxiv.org/abs/1912.07622}{{\tt 1912.07622}}].

\bibitem{DUNE:2021tad}
{\bf DUNE Collaboration}, V.~Hewes et~al.,
  \href{http://dx.doi.org/10.3390/instruments5040031}{{\it {Deep Underground
  Neutrino Experiment (DUNE) Near Detector Conceptual Design Report}}, } {\em
  Instruments} {\bf 5} (2021), no.~4 31,
  [\href{http://arxiv.org/abs/2103.13910}{{\tt 2103.13910}}].

\bibitem{NA62:2023qyn}
{\bf NA62 Collaboration}, E.~Cortina~Gil et~al.,
  \href{http://dx.doi.org/10.1007/JHEP09(2023)035}{{\it {Search for dark photon
  decays to $\mu^+\mu^-$ at NA62}}, } {\em JHEP} {\bf 09} (2023) 035,
  [\href{http://arxiv.org/abs/2303.08666}{{\tt 2303.08666}}].

\bibitem{NA62:2023nhs}
{\bf NA62 Collaboration}, E.~Cortina~Gil et~al., {\it {Search for leptonic
  decays of the dark photon at NA62}},
  \href{http://arxiv.org/abs/2312.12055}{{\tt 2312.12055}}.

\bibitem{Batell:2020vqn}
B.~Batell, J.~A. Evans, S.~Gori, and M.~Rai,
  \href{http://dx.doi.org/10.1007/JHEP05(2021)049}{{\it {Dark Scalars and Heavy
  Neutral Leptons at DarkQuest}}, } {\em JHEP} {\bf 05} (2021) 049,
  [\href{http://arxiv.org/abs/2008.08108}{{\tt 2008.08108}}].

\bibitem{PS191:1987ek}
G.~Bernardi et~al., \href{http://dx.doi.org/10.1016/0370-2693(88)90563-1}{{\it
  {FURTHER LIMITS ON HEAVY NEUTRINO COUPLINGS}}, } {\em Phys. Lett. B} {\bf
  203} (1988) 332--334.

\bibitem{E949:2014gsn}
{\bf E949 Collaboration}, A.~V. Artamonov et~al.,
  \href{http://dx.doi.org/10.1103/PhysRevD.91.052001}{{\it {Search for heavy
  neutrinos in $K^+\to\mu^+\nu_H$ decays}}, } {\em Phys. Rev. D} {\bf 91}
  (2015), no.~5 052001, [\href{http://arxiv.org/abs/1411.3963}{{\tt
  1411.3963}}]. [Erratum: Phys.Rev.D 91, 059903 (2015)].

\bibitem{KEK:1982wu}
R.~S. Hayano et~al., \href{http://dx.doi.org/10.1103/PhysRevLett.49.1305}{{\it
  {HEAVY NEUTRINO SEARCH USING K(mu2) DECAY}}, } {\em Phys. Rev. Lett.} {\bf
  49} (1982) 1305.

\bibitem{KEK:1984sj}
T.~Yamazaki et~al., {\it {Search for Heavy Neutrinos in Kaon Decay}},  {\em
  Conf. Proc. C} {\bf 840719} (1984) 262.

\bibitem{PIENU:2019usb}
{\bf PIENU Collaboration}, A.~Aguilar-Arevalo et~al.,
  \href{http://dx.doi.org/10.1016/j.physletb.2019.134980}{{\it {Search for
  heavy neutrinos in $\pi \to \mu\nu$ decay}}, } {\em Phys. Lett. B} {\bf 798}
  (2019) 134980, [\href{http://arxiv.org/abs/1904.03269}{{\tt 1904.03269}}].

\bibitem{T2K:2019jwa}
{\bf T2K Collaboration}, K.~Abe et~al.,
  \href{http://dx.doi.org/10.1103/PhysRevD.100.052006}{{\it {Search for heavy
  neutrinos with the T2K near detector ND280}}, } {\em Phys. Rev. D} {\bf 100}
  (2019), no.~5 052006, [\href{http://arxiv.org/abs/1902.07598}{{\tt
  1902.07598}}].

\bibitem{NA62:2020mcv}
{\bf NA62 Collaboration}, E.~Cortina~Gil et~al.,
  \href{http://dx.doi.org/10.1016/j.physletb.2020.135599}{{\it {Search for
  heavy neutral lepton production in K+ decays to positrons}}, } {\em Phys.
  Lett. B} {\bf 807} (2020) 135599,
  [\href{http://arxiv.org/abs/2005.09575}{{\tt 2005.09575}}].

\bibitem{NA62:2021bji}
{\bf NA62 Collaboration}, E.~Cortina~Gil et~al.,
  \href{http://dx.doi.org/10.1016/j.physletb.2021.136259}{{\it {Search for
  $K^+$ decays to a muon and invisible particles}}, } {\em Phys. Lett. B} {\bf
  816} (2021) 136259, [\href{http://arxiv.org/abs/2101.12304}{{\tt
  2101.12304}}].

\bibitem{DELPHI:1996qcc}
{\bf DELPHI Collaboration}, P.~Abreu et~al.,
  \href{http://dx.doi.org/10.1007/s002880050370}{{\it {Search for neutral heavy
  leptons produced in Z decays}}, } {\em Z. Phys. C} {\bf 74} (1997) 57--71.
  [Erratum: Z.Phys.C 75, 580 (1997)].

\bibitem{BESIII:2019oef}
{\bf BESIII Collaboration}, M.~Ablikim et~al.,
  \href{http://dx.doi.org/10.1103/PhysRevD.99.112002}{{\it {Search for heavy
  Majorana neutrino in lepton number violating decays of $D\to K \pi e^+
  e^+$}}, } {\em Phys. Rev. D} {\bf 99} (2019), no.~11 112002,
  [\href{http://arxiv.org/abs/1902.02450}{{\tt 1902.02450}}].

\bibitem{CMS:2022fut}
{\bf CMS Collaboration}, A.~Tumasyan et~al.,
  \href{http://dx.doi.org/10.1007/JHEP07(2022)081}{{\it {Search for long-lived
  heavy neutral leptons with displaced vertices in proton-proton collisions at
  $ \sqrt{\mathrm{s}} $ =13 TeV}}, } {\em JHEP} {\bf 07} (2022) 081,
  [\href{http://arxiv.org/abs/2201.05578}{{\tt 2201.05578}}].

\bibitem{CMS:2024ake}
{\bf CMS Collaboration}, A.~Hayrapetyan et~al., {\it {Search for long-lived
  heavy neutral leptons decaying in the CMS muon detectors in proton-proton
  collisions at $\sqrt{s}$ = 13 TeV}},
  \href{http://arxiv.org/abs/2402.18658}{{\tt 2402.18658}}.

\bibitem{CMS:2024ita}
{\bf CMS Collaboration}, A.~Hayrapetyan et~al.,
  \href{http://dx.doi.org/10.1007/JHEP06(2024)183}{{\it {Search for long-lived
  heavy neutrinos in the decays of B mesons produced in proton-proton
  collisions at $\sqrt{s}$ = 13 TeV}}, } {\em JHEP} {\bf 06} (2024) 183,
  [\href{http://arxiv.org/abs/2403.04584}{{\tt 2403.04584}}].

\bibitem{Blondel:2022qqo}
A.~Blondel et~al., \href{http://dx.doi.org/10.3389/fphy.2022.967881}{{\it
  {Searches for long-lived particles at the future FCC-ee}}, } {\em Front. in
  Phys.} {\bf 10} (2022) 967881, [\href{http://arxiv.org/abs/2203.05502}{{\tt
  2203.05502}}].

\bibitem{DsTauNA65:2023ogo}
{\bf DsTau (NA65) Collaboration}, S.~Aoki et~al.,
  \href{http://dx.doi.org/10.1088/1748-0221/18/10/P10008}{{\it {Development of
  proton beam irradiation system for the NA65/DsTau experiment}}, } {\em JINST}
  {\bf 18} (2023), no.~10 P10008, [\href{http://arxiv.org/abs/2303.13070}{{\tt
  2303.13070}}].

\bibitem{Foroughi-Abari:2021zbm}
S.~Foroughi-Abari and A.~Ritz,
  \href{http://dx.doi.org/10.1103/PhysRevD.105.095045}{{\it {Dark sector
  production via proton bremsstrahlung}}, } {\em Phys. Rev. D} {\bf 105}
  (2022), no.~9 095045, [\href{http://arxiv.org/abs/2108.05900}{{\tt
  2108.05900}}].

\bibitem{LoChiatto:2024guj}
P.~Lo~Chiatto and F.~Yu, {\it {Consistent Electroweak Phenomenology of a Nearly
  Degenerate $Z'$ Boson}},  \href{http://arxiv.org/abs/2405.03396}{{\tt
  2405.03396}}.

\bibitem{FLAG:2021npn}
{\bf Flavour Lattice Averaging Group (FLAG) Collaboration}, Y.~Aoki et~al.,
  \href{http://dx.doi.org/10.1140/epjc/s10052-022-10536-1}{{\it {FLAG Review
  2021}}, } {\em Eur. Phys. J. C} {\bf 82} (2022), no.~10 869,
  [\href{http://arxiv.org/abs/2111.09849}{{\tt 2111.09849}}].

\bibitem{Harrison:2017fmw}
{\bf HPQCD Collaboration}, J.~Harrison, C.~Davies, and M.~Wingate,
  \href{http://dx.doi.org/10.1103/PhysRevD.97.054502}{{\it {Lattice QCD
  calculation of the ${{B}_{(s)}\to D_{(s)}^{*}\ell{\nu}}$ form factors at zero
  recoil and implications for ${|V_{cb}|}$}}, } {\em Phys. Rev. D} {\bf 97}
  (2018), no.~5 054502, [\href{http://arxiv.org/abs/1711.11013}{{\tt
  1711.11013}}].

\bibitem{Harrison:2021tol}
{\bf HPQCD Collaboration}, J.~Harrison and C.~T.~H. Davies,
  \href{http://dx.doi.org/10.1103/PhysRevD.105.094506}{{\it
  {Bs\textrightarrow{}Ds* form factors for the full q2 range from lattice
  QCD}}, } {\em Phys. Rev. D} {\bf 105} (2022), no.~9 094506,
  [\href{http://arxiv.org/abs/2105.11433}{{\tt 2105.11433}}].

\bibitem{Melikhov:2000yu}
D.~Melikhov and B.~Stech,
  \href{http://dx.doi.org/10.1103/PhysRevD.62.014006}{{\it {Weak form-factors
  for heavy meson decays: An Update}}, } {\em Phys. Rev. D} {\bf 62} (2000)
  014006, [\href{http://arxiv.org/abs/hep-ph/0001113}{{\tt hep-ph/0001113}}].

\bibitem{Baikov:2008jh}
P.~A. Baikov, K.~G. Chetyrkin, and J.~H. Kuhn,
  \href{http://dx.doi.org/10.1103/PhysRevLett.101.012002}{{\it {Order
  alpha**4(s) QCD Corrections to Z and tau Decays}}, } {\em Phys. Rev. Lett.}
  {\bf 101} (2008) 012002, [\href{http://arxiv.org/abs/0801.1821}{{\tt
  0801.1821}}].

\bibitem{Deur:2023dzc}
A.~Deur, S.~J. Brodsky, and C.~D. Roberts,
  \href{http://dx.doi.org/10.1016/j.ppnp.2023.104081}{{\it {QCD running
  couplings and effective charges}}, } {\em Prog. Part. Nucl. Phys.} {\bf 134}
  (2024) 104081, [\href{http://arxiv.org/abs/2303.00723}{{\tt 2303.00723}}].

\bibitem{BRODSKY1980451}
S.~Brodsky, P.~Hoyer, C.~Peterson, and N.~Sakai,
  \href{https://www.sciencedirect.com/science/article/pii/0370269380903640}{{\it
  The intrinsic charm of the proton}, } {\em Physics Letters B} {\bf 93}
  (1980), no.~4 451--455.

\bibitem{Vogt:1995fsa}
R.~Vogt and S.~J. Brodsky,
  \href{http://dx.doi.org/10.1016/0550-3213(96)00380-X}{{\it {Charmed hadron
  asymmetries in the intrinsic charm coalescence model}}, } {\em Nucl. Phys. B}
  {\bf 478} (1996) 311--334, [\href{http://arxiv.org/abs/hep-ph/9512300}{{\tt
  hep-ph/9512300}}].

\bibitem{Combridge:1978kx}
B.~L. Combridge, \href{http://dx.doi.org/10.1016/0550-3213(79)90449-8}{{\it
  {Associated Production of Heavy Flavor States in p p and anti-p p
  Interactions: Some QCD Estimates}}, } {\em Nucl. Phys. B} {\bf 151} (1979)
  429--456.

\bibitem{Andersson:143966}
B.~Andersson, H.~U. Bengtsson, and G.~Gustafson,
  \href{https://cds.cern.ch/record/143966}{{\it {Charm production and the
  confining force field}}, } tech. rep., Lund Univ., Lund, 1983.

\bibitem{Andersson:1983ia}
B.~Andersson, G.~Gustafson, G.~Ingelman, and T.~Sjostrand,
  \href{http://dx.doi.org/10.1016/0370-1573(83)90080-7}{{\it {Parton
  Fragmentation and String Dynamics}}, } {\em Phys. Rept.} {\bf 97} (1983)
  31--145.

\bibitem{Norrbin:1998bw}
E.~Norrbin and T.~Sjostrand,
  \href{http://dx.doi.org/10.1016/S0370-2693(98)01244-1}{{\it {Production
  mechanisms of charm hadrons in the string model}}, } {\em Phys. Lett. B} {\bf
  442} (1998) 407--416, [\href{http://arxiv.org/abs/hep-ph/9809266}{{\tt
  hep-ph/9809266}}].

\bibitem{Braaten:2002yt}
E.~Braaten, Y.~Jia, and T.~Mehen,
  \href{http://dx.doi.org/10.1103/PhysRevLett.89.122002}{{\it {The Leading
  particle effect from heavy quark recombination}}, } {\em Phys. Rev. Lett.}
  {\bf 89} (2002) 122002, [\href{http://arxiv.org/abs/hep-ph/0205149}{{\tt
  hep-ph/0205149}}].

\bibitem{Gao:2007ht}
P.~Gao and B.-Q. Ma,
  \href{http://dx.doi.org/10.1140/epjc/s10052-007-0227-5}{{\it {The Leading
  particle effect from light quark fragmentation in charm hadroproduction}}, }
  {\em Eur. Phys. J. C} {\bf 50} (2007) 603--608,
  [\href{http://arxiv.org/abs/hep-ph/0703133}{{\tt hep-ph/0703133}}].

\bibitem{E791:1997eip}
{\bf E791 Collaboration}, E.~M. Aitala et~al.,
  \href{http://dx.doi.org/10.1016/S0370-2693(97)00952-0}{{\it {Asymmetries
  between the production of $D_s^-$ and $D_s^+$ mesons from 500-Gev/c $\pi^{-}$
  nucleon interactions as functions of x(F) and $p-transverse^{2}$}}, } {\em
  Phys. Lett. B} {\bf 411} (1997) 230--236,
  [\href{http://arxiv.org/abs/hep-ex/9708040}{{\tt hep-ex/9708040}}].

\bibitem{WA89:1998wdl}
{\bf WA89 Collaboration}, M.~I. Adamovich et~al.,
  \href{http://dx.doi.org/10.1007/s100529900019}{{\it {Charge asymmetries for
  D, D(s) and Lambda(c) production in Sigma- nucleus interactions at
  340-GeV/c}}, } {\em Eur. Phys. J. C} {\bf 8} (1999) 593--601,
  [\href{http://arxiv.org/abs/hep-ex/9803021}{{\tt hep-ex/9803021}}].

\bibitem{Christiansen:2015yqa}
J.~R. Christiansen and P.~Z. Skands,
  \href{http://dx.doi.org/10.1007/JHEP08(2015)003}{{\it {String Formation
  Beyond Leading Colour}}, } {\em JHEP} {\bf 08} (2015) 003,
  [\href{http://arxiv.org/abs/1505.01681}{{\tt 1505.01681}}].

\bibitem{Bierlich:2014xba}
C.~Bierlich, G.~Gustafson, L.~L\"onnblad, and A.~Tarasov,
  \href{http://dx.doi.org/10.1007/JHEP03(2015)148}{{\it {Effects of Overlapping
  Strings in pp Collisions}}, } {\em JHEP} {\bf 03} (2015) 148,
  [\href{http://arxiv.org/abs/1412.6259}{{\tt 1412.6259}}].

\bibitem{Bierlich:2023fmh}
C.~Bierlich, P.~Ilten, T.~Menzo, S.~Mrenna, M.~Szewc, et~al.,
  \href{http://dx.doi.org/10.21468/SciPostPhys.16.5.134}{{\it {Reweighting
  Monte Carlo Predictions and Automated Fragmentation Variations in Pythia 8}},
  } {\em SciPost Phys.} {\bf 16} (2024) 134,
  [\href{http://arxiv.org/abs/2308.13459}{{\tt 2308.13459}}].

\bibitem{Ball:2022qks}
{\bf NNPDF Collaboration}, R.~D. Ball, A.~Candido, J.~Cruz-Martinez, S.~Forte,
  T.~Giani, et~al., \href{http://dx.doi.org/10.1038/s41586-022-04998-2}{{\it
  {Evidence for intrinsic charm quarks in the proton}}, } {\em Nature} {\bf
  608} (2022), no.~7923 483--487, [\href{http://arxiv.org/abs/2208.08372}{{\tt
  2208.08372}}].

\bibitem{NNPDF:2023tyk}
{\bf NNPDF Collaboration}, R.~D. Ball, A.~Candido, J.~Cruz-Martinez, S.~Forte,
  T.~Giani, et~al., {\it {The intrinsic charm quark valence distribution of the
  proton}},  \href{http://arxiv.org/abs/2311.00743}{{\tt 2311.00743}}.

\bibitem{gelman2013understanding}
A.~Gelman, J.~Hwang, and A.~Vehtari, {\it Understanding predictive information
  criteria for bayesian models},  2013.

\bibitem{Vehtari_2016}
A.~Vehtari, A.~Gelman, and J.~Gabry,
  \href{http://dx.doi.org/10.1007/s11222-016-9696-4}{{\it Practical bayesian
  model evaluation using leave-one-out cross-validation and waic}, } {\em
  Statistics and Computing} {\bf 27} (Aug., 2016) 1413–1432.

\bibitem{vehtari2024pareto}
A.~Vehtari, D.~Simpson, A.~Gelman, Y.~Yao, and J.~Gabry, {\it Pareto smoothed
  importance sampling},  2024.

\bibitem{Jerhot:2023web}
J.~Jerhot, {\em {Hidden sector searches with fixed-target experiments}}.
\newblock PhD thesis, Louvain U., Louvain U., CP3, 2023.

\bibitem{ROOT_FW}
R.~Brun, F.~Rademakers, P.~Canal, A.~Naumann, O.~Couet, et~al.,
  \href{https://doi.org/10.5281/zenodo.3895860}{{\it root-project/root:
  v6.18/02}, }.

\bibitem{Bierlich:2022pfr}
C.~Bierlich et~al., \href{http://dx.doi.org/10.21468/SciPostPhysCodeb.8}{{\it
  {A comprehensive guide to the physics and usage of PYTHIA 8.3}}, } {\em
  SciPost Phys. Codeb.} {\bf 2022} (2022) 8,
  [\href{http://arxiv.org/abs/2203.11601}{{\tt 2203.11601}}].

\bibitem{MERTIG1991345}
R.~Mertig, M.~B{\"o}hm, and A.~Denner,
  \href{https://www.sciencedirect.com/science/article/pii/001046559190130D}{{\it
  Feyn calc - computer-algebraic calculation of feynman amplitudes}, } {\em
  Computer Physics Communications} {\bf 64} (1991), no.~3 345--359.

\bibitem{Drewes:2018gkc}
M.~Drewes, J.~Hajer, J.~Klaric, and G.~Lanfranchi,
  \href{http://dx.doi.org/10.1007/JHEP07(2018)105}{{\it {NA62 sensitivity to
  heavy neutral leptons in the low scale seesaw model}}, } {\em JHEP} {\bf 07}
  (2018) 105, [\href{http://arxiv.org/abs/1801.04207}{{\tt 1801.04207}}].

\end{thebibliography}\endgroup

\end{document}